\documentclass[twocolumns,10pt,final]{IEEEtran}
\usepackage{latexsym}
\usepackage{cite}
\usepackage{amssymb}
\usepackage{bbm}
\usepackage{bm}
\usepackage[usenames,dvipsnames]{color}
\usepackage{dsfont}
\usepackage{amsmath, amssymb}
\usepackage[dvips]{graphics}
\usepackage{graphicx}
\usepackage{amsmath}
\usepackage{psfrag}
\usepackage{subfloat}
\usepackage{comment}
\usepackage{subfig}
\usepackage{setspace}
\usepackage{algorithm}
\usepackage{algpseudocode}
\usepackage{theorem}
\usepackage{float}
\usepackage{mathtools}
\usepackage{times}
\usepackage{epsfig,verbatim}
\usepackage{mathrsfs}
\usepackage{ifthen}
\usepackage{syntonly}
\usepackage{epstopdf}
\usepackage[font=footnotesize]{caption}
\usepackage{multirow}
\usepackage{stfloats}
\usepackage{latexsym}
\usepackage{cite}

 \allowdisplaybreaks

\newtheorem{definition}{Definition}

\newtheorem{theorem}{Theorem}
\newtheorem{lemma}[theorem]{Lemma}
\newtheorem{remark}[theorem]{Remark}

\newtheorem{corollary}[theorem]{Corollary}

\newcommand\numberthis{\addtocounter{equation}{1}\tag{\theequation}}
{ %<- modify value to suit your needs
\bibliographystyle{unsrt}
\bibliographystyle{IEEEtran}
\begin{document}

\makeatletter
\newcommand{\vasti}{\bBigg@{3}}
\newcommand{\vast}{\bBigg@{4}}
\newcommand{\Vast}{\bBigg@{5}}
\makeatother
\newcommand{\be}{\begin{equation}}
\newcommand{\ee}{\end{equation}}
\newcommand{\ba}{\begin{align}}
\newcommand{\ea}{\end{align}}
\newcommand{\baa}{\begin{align*}}
\newcommand{\eaa}{\end{align*}}
\newcommand{\bea}{\begin{eqnarray}}
\newcommand{\eea}{\end{eqnarray}}
\newcommand{\beaa}{\begin{eqnarray*}}
\newcommand{\eeaa}{\end{eqnarray*}}
\newcommand{\p}[1]{\left(#1\right)}
\newcommand{\pp}[1]{\left[#1\right]}
\newcommand{\ppp}[1]{\left\{#1\right\}}
\newcommand{\ber}{$\ \mbox{Ber}$}
\newcommand\aatop[2]{\genfrac{}{}{0pt}{}{#1}{#2}}

\title{Duality of a Source Coding Problem and the Semi-Deterministic Broadcast Channel with Rate-Limited Cooperation}

\author{Ziv Goldfeld, \emph{Student Member, IEEE}, Haim H. Permuter, \emph{Senior Member, IEEE}, and Gerhard Kramer, \emph{Fellow, IEEE}
\thanks{Manuscript received May 19, 2014; revised November 03, 2014; accepted November 26,
2015. Date of current version November 27, 2015. The work of Z. Goldfeld and H. H. Permuter was supported by the Israel Science Foundation (grant no. 684/11) and an ERC starting grant. The work of G. Kramer was supported by an Alexander von Humboldt Professorship endowed by the German Federal Ministry of Education and Research.
\newline This paper was presented in part at the 2014 IEEE International Symposium on Information Theory, Honolulu, HI, USA, in part at the 2014 IEEE 28-th Convention of Electrical and Electronics Engineers in Israel, Eilat, Israel, November, 2014, and in part at the 2015 IEEE Information Theory Workshop, Jerusalem, Israel.
\newline Z. Goldfeld and H. H. Permuter are with the Department of Electrical and Computer Engineering, Ben-Gurion University of the Negev, Beer-Sheva, Israel (gziv@post.bgu.ac.il, haimp@bgu.ac.il). G. Kramer is with the Institute for Communications Engineering, Technische Universit{\"a}t M{\"u}nchen, Munich D-80333, Germany (gerhard.kramer@tum.de).
\newline Copyright (c) 2014 IEEE. Personal use of this material is permitted. However, permission to use this material for any other purposes must be obtained from the IEEE by sending a request to pubs-permissions@ieee.org.}}
\maketitle

%%%%%%%%%%%%%%%%%%%%%%%%%%%%%%%%%%%%%%%%%%%%%%%%%%%%%%%%%%%%%%%%%%%%%%%%%%%%%%%%%%%%%%%%%%%%%%%%%%%%%%%%%%%%%%%%%%%
%%%%%%%%%%%%%%%%%%%%%%%%%%%%%%%%%%%%%%%%%%%%%%%%%%%%%%%%%%%%%%%%%%%%%%%%%%%%%%%%%%%%%%%%%%%%%%%%%%%%%%%%%%%%%%%%%%%
%%%%%%%%%%%%%%%%%%%%%%%%%%%%%%%%%%%%%%%%%%                         %%%%%%%%%%%%%%%%%%%%%%%%%%%%%%%%%%%%%%%%%%%%%%%%
%%%%%%%%%%%%%%%%%%%%%%%%%%%%%%%%%%%%%%%%%%         Abstract        %%%%%%%%%%%%%%%%%%%%%%%%%%%%%%%%%%%%%%%%%%%%%%%%
%%%%%%%%%%%%%%%%%%%%%%%%%%%%%%%%%%%%%%%%%%                         %%%%%%%%%%%%%%%%%%%%%%%%%%%%%%%%%%%%%%%%%%%%%%%%
%%%%%%%%%%%%%%%%%%%%%%%%%%%%%%%%%%%%%%%%%%%%%%%%%%%%%%%%%%%%%%%%%%%%%%%%%%%%%%%%%%%%%%%%%%%%%%%%%%%%%%%%%%%%%%%%%%%
%%%%%%%%%%%%%%%%%%%%%%%%%%%%%%%%%%%%%%%%%%%%%%%%%%%%%%%%%%%%%%%%%%%%%%%%%%%%%%%%%%%%%%%%%%%%%%%%%%%%%%%%%%%%%%%%%%%

\begin{abstract}
The Wyner-Ahlswede-K{\"o}rner (WAK) empirical-coordination problem where the encoders cooperate via a finite-capacity one-sided link is considered. The coordination-capacity region is derived by combining several source coding techniques, such as Wyner-Ziv (WZ) coding, binning and superposition coding. Furthermore, a semi-deterministic (SD) broadcast channel (BC) with one-sided decoder cooperation is considered. Duality principles relating the two problems are presented, and the capacity region for the SD-BC setting is derived. The direct part follows from an achievable region for a general BC that is tight for the SD scenario. A converse is established by using telescoping identities. The SD-BC is shown to be operationally equivalent to a class of relay-BCs (RBCs) and the correspondence between their capacity regions is established. The capacity region of the SD-BC is transformed into an equivalent region that is shown to be dual to the admissible region of the WAK problem in the sense that the information measures defining the corner points of both regions coincide. Achievability and converse proofs for the equivalent region are provided. For the converse, we use a probabilistic construction of auxiliary random variables that depends on the distribution induced by the codebook. Several examples illustrate the results.
\end{abstract}

%%%%%%%%%%%%%%%%%%%%%%%%%%%%%%%%%%%%%%%%%%%%%%%%%%%%%%%%%%%%%%%%%%%%%%%%%%%%%%%%%%%%%%%%%%%%%%%%%%%%%%%%%%%%%%%%%%%
%%%%%%%%%%%%%%%%%%%%%%%%%%%%%%%%%%%%%%%%%%%%%%%%%%%%%%%%%%%%%%%%%%%%%%%%%%%%%%%%%%%%%%%%%%%%%%%%%%%%%%%%%%%%%%%%%%%
%%%%%%%%%%%%%%%%%%%%%%%%%%%%%%%%%%%%%%%%%%%%%%                     %%%%%%%%%%%%%%%%%%%%%%%%%%%%%%%%%%%%%%%%%%%%%%%%
%%%%%%%%%%%%%%%%%%%%%%%%%%%%%%%%%%%%%%%%%%%%%%     keywords        %%%%%%%%%%%%%%%%%%%%%%%%%%%%%%%%%%%%%%%%%%%%%%%%
%%%%%%%%%%%%%%%%%%%%%%%%%%%%%%%%%%%%%%%%%%%%%%                     %%%%%%%%%%%%%%%%%%%%%%%%%%%%%%%%%%%%%%%%%%%%%%%%
%%%%%%%%%%%%%%%%%%%%%%%%%%%%%%%%%%%%%%%%%%%%%%%%%%%%%%%%%%%%%%%%%%%%%%%%%%%%%%%%%%%%%%%%%%%%%%%%%%%%%%%%%%%%%%%%%%%
%%%%%%%%%%%%%%%%%%%%%%%%%%%%%%%%%%%%%%%%%%%%%%%%%%%%%%%%%%%%%%%%%%%%%%%%%%%%%%%%%%%%%%%%%%%%%%%%%%%%%%%%%%%%%%%%%%%


%%%%%%%%%%%%%%%%%%%%%%%%%%%%%%%%%%%%%%%%%%%%%%%%%%%%%%%%%%%%%%%%%%%%%%%%%%%%%%%%%%%%%%%%%%%%%%%%%%%%%%%%%%%%%%%%%%%
%%%%%%%%%%%%%%%%%%%%%%%%%%%%%%%%%%%%%%%%%%%%%%%%%%%%%%%%%%%%%%%%%%%%%%%%%%%%%%%%%%%%%%%%%%%%%%%%%%%%%%%%%%%%%%%%%%%
%%%%%%%%%%%%%%%%%%%%%%%%%%%%%%%%%%%%%%%%%%%%                         %%%%%%%%%%%%%%%%%%%%%%%%%%%%%%%%%%%%%%%%%%%%%%
%%%%%%%%%%%%%%%%%%%%%%%%%%%%%%%%%%%%%%%%%%%%     Introduction        %%%%%%%%%%%%%%%%%%%%%%%%%%%%%%%%%%%%%%%%%%%%%%
%%%%%%%%%%%%%%%%%%%%%%%%%%%%%%%%%%%%%%%%%%%%                         %%%%%%%%%%%%%%%%%%%%%%%%%%%%%%%%%%%%%%%%%%%%%%
%%%%%%%%%%%%%%%%%%%%%%%%%%%%%%%%%%%%%%%%%%%%%%%%%%%%%%%%%%%%%%%%%%%%%%%%%%%%%%%%%%%%%%%%%%%%%%%%%%%%%%%%%%%%%%%%%%%
%%%%%%%%%%%%%%%%%%%%%%%%%%%%%%%%%%%%%%%%%%%%%%%%%%%%%%%%%%%%%%%%%%%%%%%%%%%%%%%%%%%%%%%%%%%%%%%%%%%%%%%%%%%%%%%%%%%

\section{Introduction}

\begin{figure}[!t]
\begin{center}
\begin{psfrags}
    \psfragscanon
    \psfrag{A}[][][1]{\ \ \ \ \ $\mathbf{X}_1$}
    \psfrag{B}[][][1]{\ \ \ \ \ $\mathbf{X}_2$}
    \psfrag{O}[][][1]{\ \ \ \ \ \ \ \ Encoder 1}
    \psfrag{D}[][][1]{\ \ \ \ \ \ \ \ Encoder 2}
    \psfrag{T}[][][1]{$\mspace{-7mu}T_{12}(\mathbf{X}_1)$}
    \psfrag{F}[][][1]{\ \ \ \ \ \ \ \ \ \ \ \ $T_2(T_{12},\mathbf{X}_2)$}
    \psfrag{E}[][][1]{\ \ \ \ \ \ \ $T_1(\mathbf{X}_1)$}
    \psfrag{H}[][][1]{$\mathbf{Y}$}
    \psfrag{G}[][][1]{\ \ \ \ \ \ \ Decoder}
\includegraphics[scale=0.4]{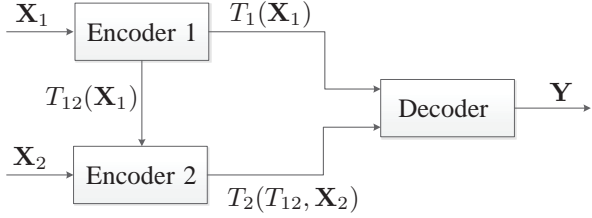}
\caption{The WAK source coding problem.} \label{fig_coordination_model}
\psfragscanoff
\end{psfrags}
\end{center}%\vspace{-7mm}
\end{figure}

Cooperation can substantially improve the performance of a network. A common form of cooperation permits information exchange between the transmitting and receiving ends via rate-limited links, generally referred to as \emph{conferencing} \cite{Willems83_cooperating}. In this work, conferencing is incorporated in a special case of the fundamental two-encoder multiterminal source coding problem (cf., e.g., \cite{Berger78} and \cite{multi_Tung1978}). Solutions for several special cases of the two-encoder source coding problem have been provided. Among these are the Slepian-Wolf (SW) \cite{Slepian_wolf_73_source_coding}, Wyner-Ziv (WZ) \cite{Wyner_ziv76_side_info_decoder}, Gaussian quadratic \cite{Wagner08_two_sources} and Wyner-Ahlswede-K{\"o}rner (WAK) \cite{Wyner_AK1974,Ahlswede-Korner75} problems. The last setting refers to two correlated sources that are separately compressed, and their compressed versions are conveyed to the decoder, which reproduces only one of the sources in a lossless manner. We consider the WAK problem with conferencing (Fig. \ref{fig_coordination_model}) in which a pair of correlated sources $(X_1^n,X_2^n)$ are compressed by two encoders that are connected via a one-sided rate-limited link that extends from the 1st encoder to the 2nd. The compressed versions are conveyed to the decoder that outputs an empirical coordination sequence $Y^n$ from which $X_1^n$ can be reproduced in a lossless manner.

\par Source coordination is an alternative formulation for lossy source coding. \emph{Strong coordination} was considered by Wyner \cite{Strong_Coord_Wyner75}, while \emph{empirical coordination} was studied in \cite{Kramer_Coord_ISIT2002,kramer_savari07,Anantharam_Brkar2007}. Cuff \emph{et al.} extended these results to the multiuser case \cite{Cuff_Permuter_Cover_Coordination}. Rather than sending data from one point to another with a fidelity constraint, in a coordination problem all network nodes should develop certain joint statistics. Moreover, it was shown in \cite{Cuff_Permuter_Cover_Coordination} that rate-distortion theory is a special case of source coordination. In this work, we consider empirical coordination, a problem in which the terminals, upon observing correlated sources, generate sequences with a desired empirical joint distribution. A closely related empirical coordination problem was presented by Bereyhi \emph{et al.} \cite{triangular_coord2013}, who considered a triangular multiterminal network. In this setting, each of the two terminals receives a different correlated source that it compresses and conveys to the decoder. The decoder outputs a sequence that achieves the desired coordination. Moreover, the encoders in \cite{triangular_coord2013} may share information via a one-sided cooperation link (see \cite{Successive_Refinemen_Permuter2013} and references therein for additional work involving cooperation in source coding problems). The main contributions of \cite{triangular_coord2013} comprise inner and outer bounds on the optimal rate region.

%%%%%%%%%%%%%%%%%%%%%%%%%%%%%%%%%%%%%%%%%%%%%%%%%%%%%%%%%%%%%%%%%%%%%%%%%%%%%%%%%%%%%%%%%%%%%%%%%%%%%%%%%%%%%%%%%%
%%%%%%%%%%%%%%%%%%%%%       Figure - AK coordination problem with cooperation     %%%%%%%%%%%%%%%%%%%%%%%%%%%%%%%%

%%%%%%%%%%%%%%%%%%%%%%%%%%%%%%%%%%%%%%%%%%%%%%%%%%%%%%%%%%%%%%%%%%%%%%%%%%%%%%%%%%%%%%%%%%%%%%%%%%%%%%%%%%%%%%%%%%

\par The WAK problem with cooperation considered here is a special case of the triangular multiterminal network in \cite{triangular_coord2013} where the sequence $X_1^n$ is losslessly reproduced from the output coordination sequence. We derive a single-letter characterization of the coordination-capacity region for this problem. The direct proof unifies several concepts in source coding by relying on WZ coding \cite{Wyner_ziv76_side_info_decoder}, binning \cite{Cover75SliepenWolf} and superposition coding \cite{Cover_BC_channel}. Note that in the classical WAK problem, where the encoders are non-cooperative, coordination of the output with the side information (i.e., the sequence $X_2^n$ in Fig. \ref{fig_coordination_model}) is achieved even though it is not required. Therefore, adding such a coordination constraint to the classic WAK problem does not alter its solution, which can be obtained as a special case of the rate region we give here. The non-cooperative version of the problem in Fig. \ref{fig_coordination_model}, i.e., where one of the sources is losslessly reproduced while coordination with the other source \emph{is} required, was studied by Berger and Yeung in \cite{Berger_Yeung89_one_distortion}.

%
%In the classical WAK problem, where the encoders are non-cooperative, the output sequence admits joint typicality with the side information (i.e., with the source sequence $X_2^n$ in Fig. \ref{fig_coordination_model}) even though this is not required. In the considered setting, however, the output sequence is coordinated with both source sequences. Nonetheless, since coordination is achieved by means of joint typicality of the sources and output with respect to the desired statistics, adding this typicality constraint to the WAK problem does not alter its solution. Indeed, under the proper assumptions, the rate region we give here reduces to the classical AK region. We note that the non-cooperative version of the problem in Fig. \ref{fig_coordination_model}, i.e., where there is no cooperation link available between the encoders, was studies by Berger and Yeung in \cite{Berger_Yeung89_one_distortion}. Thus, our result for the AK problem with cooperation also serves as a generalization of \cite{Berger_Yeung89_one_distortion}.
%
%
%%They show that the quantize-and-bin scheme \cite{Wagner08_two_sources} is optimal.
%%In \cite{wagner2011}, the authors showed that the scheme in \cite{Berger_Yeung89_one_distortion} is suboptimal with a distortion constraint because it does not fully exploit the correlation between the sources.

%%%%%%%%%%%%%%%%%%%%%%%%%%%%%%%%%%%%%%%%%%%%%%%%%%%%%%%%%%%%%%%%%%%%%%%%%%%%%%%%%%%%%%%%%%%%%%%%%%%%%%%%%%%%%%%%%%%
%%%%%%%%%%%%%%%%%%%%%%%     Figure - Semi-Deterministic BC with One-Sided Cooperation      %%%%%%%%%%%%%%%%%%%%%%%%

\begin{figure}[t!]
    \begin{center}
        \begin{psfrags}
            \psfragscanon
            
            \psfrag{I}[][][0.9]{$(M_1,M_2)$}
            \psfrag{J}[][][0.9]{\ \ \ \ \ \ \ \ Encoder}
            \psfrag{K}[][][0.9]{\ \ \ $\mathbf{X}$}
            \psfrag{L}[][][0.9]{\ \ \ \ \ \ \ \ \ \ $\mspace{3mu}\mathds{1}_{\{Y_1=f(X)\}}$}
            \psfrag{V}[][][0.9]{\ \ \ \ \ \ \ \ \ \ \ $\times P_{Y_2|X}$}
            \psfrag{S}[][][0.9]{\ \ \ \ \ \ \ \ \ \ Channel}
            \psfrag{M}[][][0.9]{\ \ \ \ $\mathbf{Y}_1$}
            \psfrag{N}[][][0.9]{\ \ \ \ $\mathbf{Y}_2$}
            \psfrag{O}[][][0.9]{\ \ \ \ \ \ \ \ Decoder 1}
            \psfrag{P}[][][0.9]{\ \ \ \ \ \ \ \ Decoder 2}
            \psfrag{Q}[][][0.9]{\ \ $\hat{M}_1$}
            \psfrag{R}[][][0.9]{\ \ $\hat{M}_2$}
            \psfrag{T}[][][0.9]{\ \ \ $M_{12}$}
            \includegraphics[scale = .34]{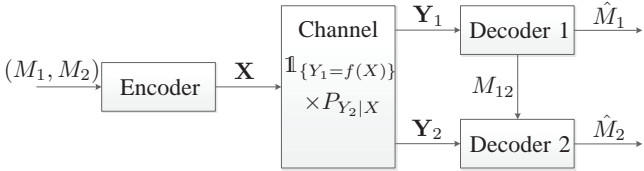}
            \caption{SD-BC with one-sided decoder cooperation.} \label{fig_semi_deterministic_BC}
            \psfragscanoff
        \end{psfrags}
     \end{center}%\vspace{-6mm}
 \end{figure}

%%%%%%%%%%%%%%%%%%%%%%%%%%%%%%%%%%%%%%%%%%%%%%%%%%%%%%%%%%%%%%%%%%%%%%%%%%%%%%%%%%%%%%%%%%%%%%%%%%%%%%%%%%%%%%%%%%%

\par To explore duality, we consider a channel coding problem (Fig. \ref{fig_semi_deterministic_BC}) that we show is \emph{dual} to the WAK problem of interest. By interchanging the roles of the encoders and decoder of the WAK problem, we obtain a semi-deterministic (SD) broadcast channel (BC) where the decoders cooperate via a rate-limited link. This duality naturally extends the well-known duality between point-to-point (PTP) source and channel coding problems. PTP duality has been widely treated in the literature since it was studied by Shannon in 1959 \cite{Shannon_rate_distortion1959} (see \cite{cover_duality2002,pradhan_duality2002,verdu_duality2011} and references therein). Multiuser duality, however, remains obscure, despite the attention it attracted in the last decade \cite{yu_duality2002,wang_duality2004,Successive_Refinemen_Permuter2013,Dikstein_MAC_Action_2015}. We provide principles according to which the two problems can be transformed from one to the other. Moreover, we show that the admissible rate regions of the considered SD-BC and WAK problems are dual. The duality is in the sense that the information measures that define the corner points of both regions coincide, which extends the relation between dual results in the PTP situation.

\par Cooperative communication over noisy channels was extensively treated in the literature since it was introduced by Willems in the context of a multiple-access channel (MAC), in which the encoders are able to hold a conference \cite{Willems83_cooperating}. The Gaussian case was solved by Bross \emph{et al.} in \cite{Wigger_gaussian_cop}, followed by several works involving the compound MAC \cite{Simeone:2009,Wiese_compound_coop2011}. Cooperation between receivers in a broadcast channel (BC) was introduced by Dabora and Servetto \cite{DaboraServetto06BC}. Liang and Veeravalli generalized the work in \cite{DaboraServetto06BC} by examining the problem of a relay-BC (RBC) \cite{LiangVeeravalli07RelayBroadcast}. In both \cite{DaboraServetto06BC} and \cite{LiangVeeravalli07RelayBroadcast}, the capacity region of the physically degraded BC (PD-BC) is characterized. Here we combine cooperation in a SD-BC setting. %The code construction combines superposition coding with the code construction for a degraded relay channel. In this paper we focus on a semi-deterministic BC (SD-BC), in which the deterministic decoder may share information with the probabilistic one via rate-limited noise-free link.%, as illustrated in Fig. \ref{fig_semi_deterministic_BC} and explained below.

\par The SD-BC without cooperation was solved by Gelfand and Pinsker \cite{GP_SemideterministicBC1980}. The coding scheme was based on Marton's scheme for BCs \cite{Marton_BC1979} (see \cite{Lapidoth_senideterministic2012} for a generalization of \cite{GP_SemideterministicBC1980} to the state-dependent case). We derive the capacity region of the SD-BC with cooperation by first deriving an inner bound on the capacity region of the cooperative general BC. The achievable scheme combines rate-splitting with Marton and superposition coding. The cooperation protocol uses binning to increase the transmission rate to the cooperation-aided user. The inner bound is then reduced to the SD-BC case and shown to be tight by providing a converse. The presented converse proof takes a simple and compact form by leveraging telescoping identities \cite{Kramer_telescopic2011}.

\par There is a close relation between the SD-BC with cooperation and a class of SD-RBCs considered in \cite{Liang_Kramer_RBC2007}. We show that a SD-RBC with an orthogonal and deterministic relay is operationally equivalent to the SD-BC with cooperation (see \cite{Kotter_Equivalence2011} for a related work on equivalence between PTP channels in a general network and noiseless bit-pipes with the same capacity). Consequently, the capacity regions of the two problems are the same. However, there are several advantages of our approach. First, we present a capacity achieving coding scheme over a \emph{single} transmission block, while \cite{Liang_Kramer_RBC2007} relies on transmitting many blocks and applying backward decoding. Thus, our scheme avoids the delay introduced by backward decoding. Second, our converse proof is considerably simpler than in \cite{Liang_Kramer_RBC2007}. Finally, considering the SD-BC with a one-sided conferencing link between the decoders gives insight into multiuser channel-source duality \cite{Goldfeld_AK_ISIT2014}.
%
%
%Lapidoth and Wang extended the result in \cite{GP_SemideterministicBC1980} to state-dependent channels where the state sequence is available at the encoder non-causally \cite{Lapidoth_senideterministic2012}. Thus, \cite{Lapidoth_senideterministic2012} extends the work by Gelfand and Pinker on PTP state-dependent channel with non-causal encoder channel state information (CSI) \cite{Gelfand_Pinsker} to the
%SD-BC scenario. The converse proof in \cite{Lapidoth_senideterministic2012} included a new element: the choice of the auxiliary random variable was dependent on the codebook. The converse proof for the semi-deterministic BC with cooperation that is presented in this work generalizes the method used in \cite{Lapidoth_senideterministic2012}.

\par To show the duality between the optimal rate regions of the considered source and channel coding problems, an alternative characterization of the capacity region of the SD-BC is given. The corner points of the alternative region satisfy the correspondence to those of the coordination-capacity region of the WAK problem. The structure of the alternative expression motivates a converse proof technique that generalizes classical techniques. Specifically, our converse uses auxiliary random variables that are not only \emph{chosen as a function of the joint distribution induced by each codebook}, but that are \emph{constructed in a probabilistic manner} (see \cite{Lapidoth_senideterministic2012} for a deterministic codebook-dependent construction of auxiliaries). Allowing a probabilistic construction of the auxiliary random variables introduces additional optimization parameters (i.e., a probability distribution). By optimizing over the probability values, an upper bound on the alternative formulation of the capacity region is tightened to coincide with the achievable region. Probabilistic arguments of a similar nature were previously used in the literature \cite{Dabora_probabilistic2008,Courtade_Weissman_Converse2014,Dikstein_PDBC_Cooperation2014}. The novelty of our approach is the incorporation of such arguments in a converse proof to describe the optimal choice of auxiliaries. Moreover, a closed form formula for the optimal probability values is derived as part of the converse and highlights the dependence of the choice of auxiliaries on the codebook.

\par This paper is organized as follows. In Section \ref{sec_problem definition} we describe the two models of interest - the WAK problem with encoder cooperation and the SD-BC with decoder cooperation. In Section \ref{sec_main_result}, we state capacity results for the WAK and BC models. In Section \ref{sec_duality} we analyse the duality between the two problems and their capacity regions. In Section \ref{sec_RBC} we discuss the relation of the considered SD-BC to a class of SD-RBCs. Section \ref{sec_special cases} presents special cases of the capacity region of the SD-BC, and each case is shown to preserve a dual relation to the corresponding reduced source coding problem. Finally, Section \ref{summary} summarizes the main achievements and insights of this work.
%
%
%Section \ref{sec_bc_converse} gives the converse for the semi-deterministic BC. The methodology used to prove the converse is elaborated.

%%%%%%%%%%%%%%%%%%%%%%%%%%%%%%%%%%%%%%%%%%%%%%%%%%%%%%%%%%%%%%%%%%%%%%%%%%%%%%%%%%%%%%%%%%%%%%%%%%%%%%%%%%%%%%%%%%%
%%%%%%%%%%%%%%%%%%%%%%%%%%%%%%%%%%%%%%%%%%%%%%%%%%%%%%%%%%%%%%%%%%%%%%%%%%%%%%%%%%%%%%%%%%%%%%%%%%%%%%%%%%%%%%%%%%%
%%%%%%%%%%%%%%%%%%%%%%%%%%%%%%%%%%%%%%%%%%%%                         %%%%%%%%%%%%%%%%%%%%%%%%%%%%%%%%%%%%%%%%%%%%%%
%%%%%%%%%%%%%%%%%%%%%%%%%%%%%%%%%%%%%%%%%%%%    Problem Definition   %%%%%%%%%%%%%%%%%%%%%%%%%%%%%%%%%%%%%%%%%%%%%%
%%%%%%%%%%%%%%%%%%%%%%%%%%%%%%%%%%%%%%%%%%%%                         %%%%%%%%%%%%%%%%%%%%%%%%%%%%%%%%%%%%%%%%%%%%%%
%%%%%%%%%%%%%%%%%%%%%%%%%%%%%%%%%%%%%%%%%%%%%%%%%%%%%%%%%%%%%%%%%%%%%%%%%%%%%%%%%%%%%%%%%%%%%%%%%%%%%%%%%%%%%%%%%%%
%%%%%%%%%%%%%%%%%%%%%%%%%%%%%%%%%%%%%%%%%%%%%%%%%%%%%%%%%%%%%%%%%%%%%%%%%%%%%%%%%%%%%%%%%%%%%%%%%%%%%%%%%%%%%%%%%%%

\section{Preliminaries and Problem Definitions}\label{sec_problem definition}

\par We use the following notations. Given two real numbers $a,b$, we denote by $[a\mspace{-3mu}:\mspace{-3mu}b]$ the set of integers $\big\{n\in\mathbb{N}\big| \lceil a\rceil\leq n \leq\lfloor b \rfloor\big\}$. We define $\mathbb{R}_+=\{x\in\mathbb{R}|x\geq 0\}$. Calligraphic letters denote sets, e.g., $\mathcal{X}$, the complement of $\mathcal{X}$ is denoted by $\mathcal{X}^c$, while $|\mathcal{X}|$ stands for its cardinality. $\mathcal{X}^n$ denotes the $n$-fold Cartesian product of $\mathcal{X}$. An element of $\mathcal{X}^n$ is denoted by $x^n=(x_1,x_2,\ldots,x_n)$; whenever the dimension $n$ is clear from the context, vectors (or sequences) are denoted by boldface letters, e.g., $\mathbf{x}$. A substring of $\mathbf{x}\in\mathcal{X}^n$ is denoted by $x_i^j=(x_i,x_{i+1},\ldots,x_j)$, for $1\leq i\leq j \leq n$; when $i=1$, the subscript is omitted. We also define $x^{n\backslash i}=(x_1,\ldots,x_{i-1},x_{i+1},\ldots,x_n)$. Random variables are denoted by uppercase letters, e.g., $X$, with similar conventions for random vectors. The probability of an event $\mathcal{A}$ is denoted by $\mathbb{P}(\mathcal{A})$, while $\mathbb{P}(\mathcal{A}\big|\mathcal{B}\mspace{2mu})$ denotes conditional probability of $\mathcal{A}$ given $\mathcal{B}$. We use $\mathds{1}_\mathcal{A}$ to denote the indicator function of $\mathcal{A}$. The set of all probability mass functions (PMFs) on a finite set $\mathcal{X}$ is denoted by $\mathcal{P}(\mathcal{X})$. PMFs are denoted by the capital letter $P$, with a subscript that identifies the random variable and its possible conditioning. For example, for two jointly distributed random variables $X$ and $Y$, let $P_X$, $P_{X,Y}$ and $P_{X|Y}$ denote, respectively, the PMF of $X$, the joint PMF of $(X,Y)$ and the conditional PMF of $X$ given $Y$. In particular, when $X$ and $Y$ are discrete, $P_{X|Y}$ represents the stochastic matrix whose elements are given by $P_{X|Y}(x|y)=\mathbb{P}\big(X=x|Y=y\big)$. We omit subscripts if the arguments of the PMF are lowercase versions of the random variables.  The expectation of a random variable $X$ is denoted by $\mathbb{E}\big[X\big]$. We use $\mathbb{E}_P$ and $\mathbb{P}_P$ to indicate that an expectation or a probability are taken taken with respect to a PMF $P$ (when the PMF is clear from the context, the subscript is omitted). If the entries of $X^n$ are drawn in an independent and identically distributed (i.i.d.) manner according to $P_X$, then for every $\mathbf{x}\in\mathcal{X}^n$ we have $P_{X^n}(\mathbf{x})=\prod_{i=1}^nP_X(x_i)$ and we write $P_{X^n}(\mathbf{x})=P_X^n(\mathbf{x})$. Similarly, if for every $(\mathbf{x},\mathbf{y})\in\mathcal{X}^n\times\mathcal{Y}^n$ we have $P_{Y^n|X^n}(\mathbf{y}|\mathbf{x})=\prod_{i=1}^nP_{Y|X}(y_i|x_i)$, then we write $P_{Y^n|X^n}(\mathbf{y}|\mathbf{x})=P_{Y|X}^n(\mathbf{y}|\mathbf{x})$. We often use $Q_X^n$ or $Q_{Y|X}^n$ when referring to an i.i.d. sequence of random variables. The conditional product PMF $Q_{Y|X}^n$ given a specific sequence $\mathbf{x}\in\mathcal{X}^n$ is denoted by $Q_{Y|X=\mathbf{x}}^n$.

For every sequence $\mathbf{x}\in\mathcal{X}^n$, the empirical PMF of $\mathbf{x}$ is 
\begin{equation}
\nu_{\mathbf{x}}(a)\triangleq\frac{N(a|\mathbf{x})}{n}
\end{equation}
where $N(a|\mathbf{x})=\sum_{i=1}^n\mathds{1}_{\{x_i=a\}}$.
We use $\mathcal{T}_\epsilon^n(P_X)$ to denote the set of letter-typical sequences of length $n$ with respect to the PMF $P_X$ and the non-negative number $\epsilon$ \cite[Ch. 3]{Massey_Applied}, \cite{Orlitsky_Roche2001}, i.e., we have
\begin{equation}
\mathcal{T}_\epsilon^{n}(P_X)\mspace{-2mu}=\mspace{-2mu}\Big\{\mathbf{x}\in\mathcal{X}^n\Big|\big|\nu_{\mathbf{x}}(a)-P_X(a)\big|\leq\epsilon P_X(a),\mspace{3mu}\forall a\in\mathcal{X}\Big\}.
\end{equation}

\noindent Furthermore, for a PMF $P_{X,Y}$ over $\mathcal{X}\times\mathcal{Y}$ and a fixed sequence $\mathbf{y}\in\mathcal{Y}^n$, we define
\begin{equation}
\mathcal{T}_\epsilon^{n}(P_{X,Y}|\mathbf{y})=\Big\{\mathbf{x}\in\mathcal{X}^n\Big| (\mathbf{x},\mathbf{y})\in\mathcal{T}_\epsilon^n(P_{X,Y})\Big\}.
\end{equation}

%%%%%%%%%%%%%%%%%%%%%%%%%%%%%%%%%%%%%%%%%%%%%%%%%%%%%%%%%%%%%%%%%%%%%%%%%%%%%%%%%%%%%%%%%%%%%%%%%%%%%%%%%%%%%%%%%%%
%%%%%%%%%%%%%%%%%%%%%%%%%%%%       AK Problem with Cooperation - Definiton       %%%%%%%%%%%%%%%%%%%%%%%%%%%%%%%%%%
%%%%%%%%%%%%%%%%%%%%%%%%%%%%%%%%%%%%%%%%%%%%%%%%%%%%%%%%%%%%%%%%%%%%%%%%%%%%%%%%%%%%%%%%%%%%%%%%%%%%%%%%%%%%%%%%%%%

\subsection{The WAK Source Coordination Problem with One-Sided Encoder Cooperation}\label{subsec_ak_problem definition}

\par Consider the source coding problem illustrated in Fig. \ref{fig_coordination_model}. Two source sequences $\mathbf{x}_1\in\mathcal{X}_1^n$ and $\mathbf{x}_2\in\mathcal{X}_2^n$ are available at Encoder 1 and Encoder 2, respectively. The sources are drawn in a pairwise independent and identically distributed (i.i.d.) manner according to the PMF $Q_{X_1,X_2}$ \footnote{We usually use $Q$ to denote a PMF that is fixed as part of the problem's definition, while $P$ is used for PMFs that we optimize over.}. Each encoder communicates with the decoder by sending a message via a noiseless communication link of limited rate. The rate of the link between Encoder $j$ and the decoder is $R_j$ and the corresponding message is $t_j$, where $j=1,2$. Moreover, Encoder 1 can communicate with Encoder 2 over a one-sided communication link of rate $R_{12}$.
%The manner in which the nodes in the setting communicate with each other is by conveying messages that are deterministically chosen from the appropriate alphabets over each of the corresponding links. For example, the message sent from Encoder 1 to the decoder is denoted by $T_1(X_1^n)\in\mathcal{T}_1$, where $\mathcal{T}_1\triangleq\{1,2,\ldots,2^{nR_1}\}$, and it is deterministically chosen as a function of the source sequence $X_1^n$ (the messages $T_{12}(X_1^n)\in\mathcal{T}_{12}$ and $T_2(T_{12},X_2^n)\in\mathcal{T}_2$ are chosen in an analogous manner). The decoder generates an output sequence $y^n\in\mathcal{Y}^n$ as a deterministic function of the received messages. This output sequence is to achieve coordination with the input sequences with respect to a joint distribution in which the random variable $X_1$ is given by applying a deterministic function $f$ on the random variable $Y$. The latter coordination constraint is, in fact, equivalent to a lossless reconstruction requirement of $X_1^n$ at the decoder.

\begin{definition}[Coordination Code]
A $(n,R_{12},R_1,R_2)$ coordination code $\mathcal{L}_n$ for the WAK source coordination problem with one-sided encoder cooperation has:
\begin{enumerate}
\item Three message sets: $\mathcal{T}_{12}=\left[1:2^{nR_{12}}\right]$, $\mathcal{T}_1=\left[1:2^{nR_1}\right]$ and $\mathcal{T}_2=\left[1:2^{nR_2}\right]$.
\item An encoder cooperation function:
\begin{subequations}
\begin{equation}
f_{12}: \mathcal{X}_1^n\to \mathcal{T}_{12}.
\end{equation}

\item Two encoding functions:
\begin{align}
&f_1: \mathcal{X}_1^n\to \mathcal{T}_1\\
&f_2:\mathcal{X}_2^n\times\mathcal{T}_{12}\to \mathcal{T}_2.
\end{align}

\item A decoding function:
\begin{equation}
\phi: \mathcal{T}_1\times\mathcal{T}_2 \to \mathcal{Y}^n.
\end{equation}
\end{subequations}
\end{enumerate}
\end{definition}

%\begin{definition}[Empirical Distribution] The empirical distribution of $(x_1^n,x_2^n,y^n)\in\mathcal{X}_1^n\times\mathcal{X}_2^n\times\mathcal{Y}^n$  is
%\begin{equation}
%\nu_{x_1^n,x_2^n,y^n}(a,b,c)\triangleq\frac{1}{n}\sum_{i=1}^n\mathds{1}_{\big\{(x_{1,i},x_{2,i},y_i)=(a,b,c)\big\}}.
%\end{equation}
%\end{definition}

\begin{definition}[Total Variation]
Let $\mathcal{X}$ be a countable space~\footnote{Countable sample spaces are assumed throughout this work} and let $P,Q\in\mathcal{P}(\mathcal{X})$. The total variation (TV) distance between $P$ and $Q$ is
\begin{equation}
||P-Q||_{TV}=\frac{1}{2}\sum_{a\in\mathcal{X}}\big|P(a)-Q(a)\big|.
\end{equation}
\end{definition}

Let $\mathcal{Q}$ be the set of PMFs defined in \eqref{EQ:PMF_domain} at the bottom of the next page.

\begin{figure*}[!b]
\setcounter{equation}{5}
\hrulefill
\begin{equation}
\mathcal{Q}=\left\{P_{X_1,X_2,Y}\in\mathcal{P}(\mathcal{X}_1\times\mathcal{X}_2\times\mathcal{Y})\mspace{3mu}\vline\begin{array}{lll} \exists f:\mathcal{Y}\to\mathcal{X}_1,\ P_{X_1,X_2,Y}=Q_{X_2}P_{Y|X_2}\mathds{1}_{\{X_1=f(Y)\}},\\
\sum_{y\in\mathcal{Y}}P_{X_1,X_2,Y}(x_1,x_2,y)=Q_{X_1,X_2}(x_1,x_2),\forall (x_1,x_2)\in\mathcal{X}_1\times\mathcal{X}_2\end{array}\right\}.\label{EQ:PMF_domain}
\end{equation}
\hrulefill
\begin{equation}
e_\delta(P_{X_1,X_2,Y},\mathcal{L}_n)\triangleq\mathbb{P}_{\mathcal{L}_n}\bigg(\big|\big|\mspace{2mu}\nu_{\mathbf{X}_1,\mathbf{X}_2,\mathbf{Y}}-P_{X_1,X_2,Y}\big|\big|_{TV}\geq \delta \mspace{2mu}\bigg)=\mspace{-25mu}\sum_{\substack{(\mathbf{x}_1,\mathbf{x}_2,\mathbf{y})\in\mathcal{X}_1^n\times\mathcal{X}_2^n\times\mathcal{Y}^n:\\||\nu_{\mathbf{x}_1,\mathbf{x}_2,\mathbf{y}}-P_{X_1,X_2,Y}||_{TV}\geq\delta}}\mspace{-60mu}Q_{X_1,X_2}^n(\mathbf{x}_1,\mathbf{x}_2)\mathds{1}_{\Big\{\phi\big(f_1(\mathbf{x}_1),f_2(\mathbf{x}_2,f_{12}(\mathbf{x}_1))\big)=\mathbf{y}\Big\}}.\label{EQ:coord_error_prob}
\end{equation}
\hrulefill
\setcounter{equation}{8}
\begin{equation}
e(\mathcal{C}_n)\triangleq\mathbb{P}_{\mathcal{C}_n}\Big((\hat{M}_1,\hat{M}_2)\neq(M_1,M_2)\Big)=2^{-n(R_1+R_2)}\mspace{-35mu}\sum_{(m_1,m_2)\in\mathcal{M}_1\times\mathcal{M}_2}\sum_{\substack{(\mathbf{y}_1,\mathbf{y}_2)\in\mathcal{Y}_1^n\times\mathcal{Y}_2^n:\\\psi_1(\mathbf{y}_1)\neq m_1\ {\scriptsize{\mbox{or}}}\ \psi_2(\mathbf{y}_2,g_{12}(\mathbf{y}_1))\neq m_2}}\mspace{-30mu}Q_{Y_1,Y_2|X}^n\big(\mathbf{y}_1,\mathbf{y}_1\big|g(m_1,m_2)\big).\label{BC_Pe}
\end{equation}
\end{figure*}

\begin{definition}[Coordination Error]
Let $P_{X_1,X_2,Y}\in\mathcal{Q}$ and $\delta>0$. The coordination error $e_\delta(P_{X_1,X_2,Y},\mathcal{L}_n)$ of an $(n,R_{12},R_1,R_2)$ coordination code $\mathcal{L}_n$ with respect to $P_{X_1,X_2,Y}$ is given in \eqref{EQ:coord_error_prob} at the bottom of the page.% \footnote{Being clear from the context, in this work we drop the conditioning on $\mathcal{L}_n$.}.
\end{definition}

\setcounter{equation}{7}

\begin{definition}[Coordination Achievability]
Let $P_{X_1,X_2,Y}\in\mathcal{Q}$. A rate triple $(R_{12},R_1,R_2)$ is $P_{X_1,X_2,Y}$-\emph{achievable} if for every $\epsilon,\delta>0$ there is a sufficiently large $n\in\mathbb{N}$ and a $(n,R_{12},R_1,R_2)$ coordination code $\mathcal{L}_n$ such that $e_\delta(P_{X_1,X_2,Y},\mathcal{L}_n)\leq\epsilon$.
\end{definition}

\begin{definition}[Coordination-Capacity Region]
The \emph{coordination-capacity region} $\mathcal{R}_{\mathrm{WAK}}(P_{X_1,X_2,Y})$ with respect to a PMF $P_{X_1,X_2,Y}\in\mathcal{Q}$ is the closure of the set of $P_{X_1,X_2,Y}$-achievable rate triples $(R_{12},R_1,R_2)$.
\end{definition}

%%%%%%%%%%%%%%%%%%%%%%%%%%%%%%%%%%%%%%%%%%%%%%%%%%%%%%%%%%%%%%%%%%%%%%%%%%%%%%%%%%%%%%%%%%%%%%%%%%%%%%%%%%%%%%%%%%%
%%%%%%%%%%%%%%%%%%%%%%%%%%%%%%%       Semi-Deterministic BC - Definiton       %%%%%%%%%%%%%%%%%%%%%%%%%%%%%%%%%%%%%
%%%%%%%%%%%%%%%%%%%%%%%%%%%%%%%%%%%%%%%%%%%%%%%%%%%%%%%%%%%%%%%%%%%%%%%%%%%%%%%%%%%%%%%%%%%%%%%%%%%%%%%%%%%%%%%%%%%

\subsection{SD-BCs with One-Sided Decoder Cooperation}\label{subsec_bc_problem definition}

\par The SD-BC with cooperation is illustrated in Fig. \ref{fig_semi_deterministic_BC}. The channel has one sender and two receivers. The sender chooses a pair $(m_1,m_2)$ of indices uniformly and independently from the $\left[1:2^{nR_1}\right]\times\left[1:2^{nR_2}\right]$ and maps them to a sequence $\mathbf{x}\in\mathcal{X}^n$, which is the channel input. The sequence $\mathbf{x}$ is transmitted over a BC with transition probability $Q_{Y_1,Y_2|X}=\mathds{1}_{\{Y_1=f(X)\}}Q_{Y_2|X}$. The output sequence $\mathbf{y}_j\in\mathcal{Y}^n_j$, where $j=1,2$, is received by decoder $j$. Decoder $j$ produces an estimate of $m_j$, which is denoted by $\hat{m}_j$. There is a one-sided noiseless cooperation link of rate $R_{12}$ from Decoder 1 to Decoder 2. By conveying a message $m_{12}\in\left[1:2^{nR_{12}}\right]$ over this link, Decoder 1 can share with Decoder 2 information about $\mathbf{y}_1$, $\hat{m}_1$, or both.

\begin{definition}[Code]
A $(n,R_{12},R_1,R_2)$ code $\mathcal{C}_n$ for the SD-BC with one-sided decoder cooperation has:
\begin{enumerate}
\item Three message sets $\mathcal{M}_{12}=\left[1:2^{nR_{12}}\right]$, $\mathcal{M}_1=\left[1:2^{nR_1}\right]$ and $\mathcal{M}_2=\left[1:2^{nR_2}\right]$.
\item An encoding function:
\begin{subequations}\label{def_sdbc_code}
 \begin{equation}
 g: \mathcal{M}_1\times\mathcal{M}_2 \to \mathcal{X}^n.
 \end{equation}

\item A decoder cooperation function:
\begin{equation}
g_{12}: \mathcal{Y}_1^n\to \mathcal{M}_{12}.
\end{equation}

\item Two decoding functions:
\begin{align}
&\psi_1: \mathcal{Y}_1^n\to \mathcal{M}_1\\
&\psi_2: \mathcal{Y}_2^n\times\mathcal{M}_{12}\to \mathcal{M}_2.
\end{align}
\end{subequations}
\end{enumerate}
\end{definition}

\begin{definition}[Error Probability] The average error probability $e(\mathcal{C}_n)$ of an $(n,R_{12},R_1,R_2)$ code $\mathcal{C}_n$ is given in \eqref{BC_Pe} at the bottom of the page. 
%The average error probability for each receiver is defined by
%\begin{equation}
%P_{e,1}(\mathcal{C}_n)=\mathbb{P}\Big(\hat{M}_1\neq M_1\mspace{2mu}\Big|\mspace{2mu}\mathcal{C}_n\Big)\ ;\ %P_{e,2}(\mathcal{C}_n)=\mathbb{P}\Big(\hat{M}_2\neq M_2\mspace{2mu}\Big|\mspace{2mu}\mathcal{C}_n\Big).
%\end{equation}
%\end{subequations}
\end{definition}
\setcounter{equation}{9}

\begin{definition}[Achievability] A rate triple $(R_{12},R_1,R_2)$ is {\it{achievable}} if for any $\epsilon>0$ there is a sufficiently large $n\in\mathbb{N}$ and an $(n,R_{12},R_1,R_2)$ code $\mathcal{C}_n$ such that $e(\mathcal{C}_n)\leq\epsilon$.
%\footnote{As stated in Section \ref{subsec_ak_problem definition}, the conditioning on $\mathcal{C}_n$ will also be dropped subsequently. Accordingly, $P_{e}(\mathcal{C}_n)$ will be denoted by $P_e^{(n)}$.}
\end{definition}

\begin{definition}[Capacity Region]
The capacity region $\mathcal{C}_{\mathrm{BC}}$ of the SD-BC with one-sided encoder cooperation is the closure of the set of achievable rate triples $(R_{12},R_1,R_2)$.
\end{definition}

%%%%%%%%%%%%%%%%%%%%%%%%%%%%%%%%%%%%%%%%%%%%%%%%%%%%%%%%%%%%%%%%%%%%%%%%%%%%%%%%%%%%%%%%%%%%%%%%%%%%%%%%%%%%%%%%%%%
%%%%%%%%%%%%%%%%%%%%%%%%%%%%%%%%%%%%%%%%%%%%%%%%%%%%%%%%%%%%%%%%%%%%%%%%%%%%%%%%%%%%%%%%%%%%%%%%%%%%%%%%%%%%%%%%%%%
%%%%%%%%%%%%%%%%%%%%%%%%%%%%%%%%%%%%%%%%%%%%                         %%%%%%%%%%%%%%%%%%%%%%%%%%%%%%%%%%%%%%%%%%%%%%
%%%%%%%%%%%%%%%%%%%%%%%%%%%%%%%%%%%%%%%%%%%%       Main Results      %%%%%%%%%%%%%%%%%%%%%%%%%%%%%%%%%%%%%%%%%%%%%%
%%%%%%%%%%%%%%%%%%%%%%%%%%%%%%%%%%%%%%%%%%%%                         %%%%%%%%%%%%%%%%%%%%%%%%%%%%%%%%%%%%%%%%%%%%%%
%%%%%%%%%%%%%%%%%%%%%%%%%%%%%%%%%%%%%%%%%%%%%%%%%%%%%%%%%%%%%%%%%%%%%%%%%%%%%%%%%%%%%%%%%%%%%%%%%%%%%%%%%%%%%%%%%%%
%%%%%%%%%%%%%%%%%%%%%%%%%%%%%%%%%%%%%%%%%%%%%%%%%%%%%%%%%%%%%%%%%%%%%%%%%%%%%%%%%%%%%%%%%%%%%%%%%%%%%%%%%%%%%%%%%%%

\section{Main Results}\label{sec_main_result}

\par We state our main results as the coordination-capacity region of the WAK source coordination problem (Section \ref{subsec_ak_problem definition}) and the capacity region of the SD-BC with cooperation (Section \ref{subsec_bc_problem definition}).

\begin{theorem}[WAK Problem Coordination-Capacity]\label{tm_ak_capacity}
The coordination-capacity region $\mathcal{R}_{\mathrm{WAK}}(P_{X_1,X_2,Y})$ of the WAK source coordination problem with one-sided encoder cooperation with respect to a PMF $P_{X_1,X_2,Y}\in\mathcal{Q}$ is the union of rate triples $(R_{12},R_1,R_2)\in\mathbb{R}^3_+$ satisfying:
\begin{subequations}
\begin{align}
R_{12} &\geq I(V;X_1|X_2)\label{region_AK12}\\
R_1 &\geq H(X_1|V,U)\label{region_AK1}\\
R_2 &\geq I(U;X_2|X_1,V)\label{region_AK2}\\
R_1+R_2 &\geq H(X_1|V,U)+I(V,U;X_1,X_2)\label{region_AK1+2}
\end{align}\label{region_AK}
\end{subequations}

\vspace{-4.5mm}
\noindent where the union is over all PMFs $Q_{X_1,X_2}P_{V|X_1}P_{U|X_2,V}P_{Y|X_1,U,V}$ that have $P_{X_1,X_2,Y}$ as a marginal. Moreover, $\mathcal{R}_{\mathrm{WAK}}(P_{X_1,X_2,Y})$ is convex and one may choose $|\mathcal{V}|\leq|\mathcal{X}_1|+3$ and $|\mathcal{U}|\leq|\mathcal{V}|\cdot|\mathcal{X}_2|+3$.
\end{theorem}

See Appendix \ref{appen_proof_tm1} for the proof of Theorem \ref{tm_ak_capacity}.

%%%%%%%%%%%%%%%%%%%%%%%%%%%%%%%%%%%%%%%%%%%%%%%%%%%%%%%%%%%%%%%%%%%%%%%%%%%%%%%%%%%%%%%%%%%%%%%%%%%%%%%%%%%%%%%%%%%
%%%%%%%%%%%%%%%%%%%%%%%%%%%%%%%%%%%%%%%%%%         Fig AK Region      %%%%%%%%%%%%%%%%%%%%%%%%%%%%%%%%%%%%%%%%%%%%%

\begin{figure}[t!]
\begin{center}
\begin{psfrags}
    \psfragscanon
    \psfrag{X}[][][0.9]{$\ \ \ R_1$}
    \psfrag{Y}[][][0.9]{$R_2$}
    \psfrag{Z}[][][0.9]{$0$}
    \psfrag{S}[][][0.9]{$\ \ \ \ I(X_2;U|X_1,V)$}
    \psfrag{K}[][][0.9]{$\ \ \ \ \ \ \ \ +I(X_1;V)$ }
    \psfrag{O}[][][0.9]{$\ \ \ \ \ \ \ \ \ H(X_1|V,U)$}
    \psfrag{T}[][][0.9]{$\ \ \ \ \ \ I(X_2;U|V)$}
    \psfrag{W}[][][0.9]{ }
    \psfrag{P}[][][0.9]{$\ \ \ \ \ \ \ H(X_1)$}
    \psfrag{D}[][][0.9]{$0$}
\includegraphics[scale=0.6]{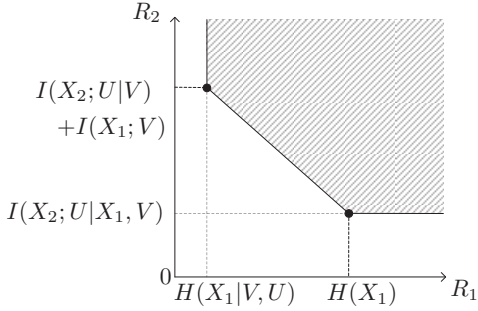}
\caption{Corner points of the coordination-capacity region of the WAK coordination problem with cooperation at the hyperplane where $R_{12}=I(X_1;V|X_2)$.} \label{fig_AK_region}
\psfragscanoff
\end{psfrags}
\end{center}%\vspace{-5mm}
\end{figure}

%%%%%%%%%%%%%%%%%%%%%%%%%%%%%%%%%%%%%%%%%%%%%%%%%%%%%%%%%%%%%%%%%%%%%%%%%%%%%%%%%%%%%%%%%%%%%%%%%%%%%%%%%%%%%%%%%%%

\begin{remark}
For a fixed PMF in Theorem \ref{tm_ak_capacity}, the triples $(R_{12},R_1,R_2)$ at the corner points of $\mathcal{R}_{\mathrm{WAK}}(P_{X_1,X_2,Y})$ are (see Fig.~\ref{fig_AK_region})
\begin{subequations}
\begin{align}
&\big(\mspace{3mu}I(V;X_1|X_2)\mspace{3mu},\mspace{3mu}H(X_1)\mspace{3mu},\mspace{3mu}I(U;X_2|X_1,V)\mspace{3mu}\big)\label{ak_corner1outline}\\
&\big(\mspace{3mu}I(V;X_1|X_2)\mspace{3mu},\mspace{3mu}H(X_1|V,U)\mspace{3mu},\mspace{3mu}I(U;X_2|V)+I(V;X_1)\mspace{3mu}\big).\label{ak_corner2outline}
\end{align}
\end{subequations}
The corner point in (\ref{ak_corner2outline}) is achieved using the coding scheme from \cite{triangular_coord2013} by setting $V=0$ in \cite[Theorem 1]{triangular_coord2013}. However, the rate triple \eqref{ak_corner1outline} does not seem to be achievable for that scheme.
\end{remark}

\begin{remark}
The cardinality bounds on the auxiliary random variables $V$ and $U$ in Theorem \ref{tm_ak_capacity} are established by standard application of the Eggleston-Fenchel-Carath{\'e}odory theorem \cite[Theorem 18]{Eggleston_Convexity1958} twice. The details are omitted.
\end{remark}

The source coordination problem defined in Section \ref{subsec_ak_problem definition} can be transformed into an equivalent rate-distortion problem. This is done by substituting $\mathbf{Y}$, the output of the coordination problem, with the pair $(\hat{\mathbf{X}}_1,\hat{\mathbf{X}}_2)$, where $\hat{\mathbf{X}}_1$ is a lossless reconstruction of the source sequence $\mathbf{X}_1$, while $\hat{\mathbf{X}}_2$ satisfies the distortion constraint
\begin{equation}
\mathbb{E}\left[\sum_{i=1}^nd(X_{2,i},\hat{X}_{2,i})\right]\leq D
\end{equation}
where $d:\ \mathcal{X}_2\times\hat{\mathcal{X}}_2\to \mathbb{R}_+$ is a single-letter distortion measure and $D\in\mathbb{R}_+$ is the distortion constraint. The two models are equivalent in the sense that the rate bounds that describe the optimal rate regions of both problems are the same; the domain over which the union is taken, however, is slightly modified. This gives rise to the following corollary.

\begin{corollary}[WAK Problem Rate-Distortion Region]\label{corollary_ak_rate_dist}
The rate-distortion region $\mathcal{R}_{\mathrm{WAK}}(D)$ for the equivalent rate-distortion problem is the union of rate triples $(R_{12},R_1,R_2)\in\mathbb{R}^3_+$ satisfying \eqref{region_AK}, where the union is over all PMFs $Q_{X_1,X_2}P_{V|X_1}P_{U|X_2,V}$ and the reconstructions $\hat{X}_2$ that are a functions of $(X_1,U,V)$ such that $\mathbb{E}\big[d(X_2,\hat{X}_2)\big]\leq D$.
\end{corollary}

The proof of Corollary \ref{corollary_ak_rate_dist} is similar to that of Theorem \ref{tm_ak_capacity} and is omitted. We next state the capacity region of the SD-BC with cooperation.

\begin{theorem}[SD-BC Capacity Region]\label{tm_bc_capacity}
The capacity region $\mathcal{C}_{\mathrm{BC}}$ of the SD-BC with one-sided encoder cooperation is the union of rate triples $(R_{12},R_1,R_2)\in\mathbb{R}^3_+$ satisfying:
\begin{subequations}
\begin{align}
R_1 &\leq H(Y_1)\label{region_BC1}\\
R_2 &\leq I(V,U;Y_2)+R_{12}\label{region_BC2}\\
R_1+R_2 &\leq H(Y_1|V,U)+I(U;Y_2|V)+I(V;Y_1)\label{region_BC1+2a}\\
R_1+R_2 &\leq H(Y_1|V,U)+I(V,U;Y_2)+R_{12}\label{region_BC1+2b}
\end{align}\label{region_BC}
\end{subequations}

\vspace{-4.5mm}
\noindent where the union is over all PMFs $P_{V,U,Y_1,X}Q_{Y_2|X}$ for which $Y_1=f(X)$. Moreover, $\mathcal{C}_{\mathrm{BC}}$ is convex and one may choose $|\mathcal{V}|\leq|\mathcal{X}|+3$ and $|\mathcal{U}|\leq |\mathcal{X}|$.
\end{theorem}

The proof of Theorem \ref{tm_bc_capacity} is relegated to Appendix \ref{appen_proof_tm_bc}. The achievable scheme combines Marton and superposition coding with rate-splitting and binning. The rather simple converse proof is due to the telescoping identity \cite[Eq. (9) and (11)]{Kramer_telescopic2011}.

\begin{remark}
The derivation of the capacity region in Theorem \ref{tm_bc_capacity} strongly relies on the SD nature of the channel. Since $Y_1=f(X)$, the encoder has full control over the message that is conveyed via the cooperation link. This allows one to design the cooperation protocol at the encoding stage without assuming a particular Markov relation on the coding random variables. Our approach differs from the one taken in \cite{DaboraServetto06BC}, where an inner bound on the capacity region of a BC with two-sided conferencing links between the decoders was derived. In \cite{DaboraServetto06BC}, the decoders cooperate by conveying to each other a compressed versions of their received channel outputs (via a WZ-like coding mechanism). Doing so forced the authors to restrict their coding PMF to satisfy certain Markov relations that must not hold in general. Consequently, the inner bound in \cite{DaboraServetto06BC} is not tight for the SD-BC considered 
here.
\end{remark}

\begin{remark}
The SD-BC with cooperation is strongly related to the SD-RBC that was studied in \cite{Liang_Kramer_RBC2007}. The SD-BC with cooperation is operationally equivalent to a reduced version of the SD-RBC, in which the relay channel is orthogonal and deterministic. Section \ref{sec_RBC} gives a detailed discussion on the relation between the two problems.
\end{remark}

\begin{remark}
The cardinality bounds on the auxiliary random variables in Theorem \ref{tm_bc_capacity} are established using the perturbation method \cite{Perturbation2012} and a standard application of the Eggleston-Fenchel-Carath{\'e}odory theorem. The details are omitted.
\end{remark}

\begin{remark}
The SD-BC with decoder cooperation and the WAK problem with encoder cooperation are duals. A full discussion on the duality between the problems is given in the following section.
\end{remark}

%%%%%%%%%%%%%%%%%%%%%%%%%%%%%%%%%%%%%%%%%%%%%%%%%%%%%%%%%%%%%%%%%%%%%%%%%%%%%%%%%%%%%%%%%%%%%%%%%%%%%%%%%%%%%%%%%%%
%%%%%%%%%%%%%%%%%%%%%%%%%%%%%%%%%%%%%%%%%%%%%%%%%%%%%%%%%%%%%%%%%%%%%%%%%%%%%%%%%%%%%%%%%%%%%%%%%%%%%%%%%%%%%%%%%%%
%%%%%%%%%%%%%%%%%%%%%%%%%%%%%%%%%%                                         %%%%%%%%%%%%%%%%%%%%%%%%%%%%%%%%%%%%%%%%
%%%%%%%%%%%%%%%%%%%%%%%%%%%%%%%%%%       Duality Between the Regions       %%%%%%%%%%%%%%%%%%%%%%%%%%%%%%%%%%%%%%%%
%%%%%%%%%%%%%%%%%%%%%%%%%%%%%%%%%%                                         %%%%%%%%%%%%%%%%%%%%%%%%%%%%%%%%%%%%%%%%
%%%%%%%%%%%%%%%%%%%%%%%%%%%%%%%%%%%%%%%%%%%%%%%%%%%%%%%%%%%%%%%%%%%%%%%%%%%%%%%%%%%%%%%%%%%%%%%%%%%%%%%%%%%%%%%%%%%
%%%%%%%%%%%%%%%%%%%%%%%%%%%%%%%%%%%%%%%%%%%%%%%%%%%%%%%%%%%%%%%%%%%%%%%%%%%%%%%%%%%%%%%%%%%%%%%%%%%%%%%%%%%%%%%%%%%

\begin{figure*}[!b]
\setcounter{equation}{13}
\hrulefill
\begin{equation}
e_\delta(Q_X^\star,\mathcal{C}_n^\star)\triangleq\mathbb{P}_{\mathcal{C}_n^\star}\left(\Big\{(\hat{M}_1,\hat{M}_2)\neq(M_1,M_2)\Big\}\cup \left\{\bigcup_{(m_1,m_2)\in\mathcal{M}_1\times\mathcal{M}_2}\mspace{-20mu}\Big\{ \big|\big|\mspace{2mu}\nu_{g(m_1,m_2),\mathbf{Y}_1,\mathbf{Y}_2}-Q_X^\star Q_{Y_1,Y_2|X}\big|\big|_{TV}\geq \delta \Big\}\right\}\right).\label{EQ:composition_error_prob}
\end{equation}

%&=2^{-n(R_1+R_2)}\mspace{-35mu}\sum_{(m_1,m_2)\in\mathcal{M}_1\times\mathcal{M}_2}\sum_{\substack{(\mathbf{y}_1,\mathbf{y}_2)\in\mathcal{Y}_1^n\times\mathcal{Y}_2^n:\\\psi_1(\mathbf{y}_1)\neq m_1\ {\scriptsize{\mbox{or}}}\ \psi_2(\mathbf{y}_2,g_{12}(\mathbf{y}_1))\neq m_2\\
%{\scriptsize{\mbox{or}}}\ ||\nu_{g(m_1,m_2),\mathbf{y}_1,\mathbf{y}_2}-Q_X^\star Q_{Y_1,Y_2|X}||_{TV}\geq\delta}}\mspace{-30mu}Q_{Y_1,Y_2|X}^n\big(\mathbf{y}_1,\mathbf{y}_1\big|g(m_1,m_2)\big).\numberthis
%\end{align*}
\end{figure*}

%%%%%%%%%%%%%%%%%%%%%%%%%%%%%%%%%%%%%%%%%%%%%%%%%%%%%%%%%%%%%%%%%%%%%%%%%%%%%%%%%%%%%%%%%%%%%%%%%%%%%%%%%%%%%%%%%%
%%%%%%%%%%%%%%%%%%%%%%%%%%%%%%%%%%%     Duality Transformation Principles      %%%%%%%%%%%%%%%%%%%%%%%%%%%%%%%%%%%%
\begin{table*}[!t]
\begin{center}
\caption{Duality transformation principles: the WAK problem with cooperation vs. the SD-BC with cooperation}\label{table_principles}
\begin{tabular}{|c|c|}
\hline
\textbf{WAK Problem with Encoder Cooperation} & \textbf{SD-BC with Decoder Cooperation}\\ \hline\hline
Decoder inputs / Encoder outputs: & Encoder inputs / Decoder outputs: \\
$t_j\in\left[1:2^{nR_j}\right],$ $j=1,2$& $m_j\in\left[1:2^{nR_j}\right],$ $j=1,2$\\ \hline
Encoder inputs / Sources: & Decoder inputs / Channel outputs: \\
$\mathbf{X}_1\ ,\ \mathbf{X}_2$ & $\mathbf{Y}_1\ ,\ \mathbf{Y}_2$ \\ \hline
Decoder output / Coordination sequence: & Encoder output / Channel input: \\
$\mathbf{Y}$ & $\mathbf{X}$ \\ \hline
Encoding functions: & Decoding functions:\\
$f_1:\mathcal{X}_1^n\to \mathcal{T}_1$, & $\psi_1:\mathcal{Y}_1^n\to \mathcal{M}_1$,\\
$f_2:\mathcal{X}_2^n\times\mathcal{T}_{12}\to \mathcal{T}_2$ & $\psi_2:\mathcal{Y}_2^n\times\mathcal{M}_{12}\to \mathcal{M}_2$\\\hline
Encoder cooperation functions: & Decoder cooperation function:\\
$f_{12}:\mathcal{X}_1^n\rightarrow \mathcal{T}_{12}$ & $g_{12}:\mathcal{Y}_1^n\rightarrow \mathcal{M}_{12}$\\ \hline
Decoding functions: & Encoding function: \\
$\phi:\mathcal{T}_1\times\mathcal{T}_2\rightarrow \mathcal{Y}^n$ & $g:\mathcal{M}_1\times\mathcal{M}_2\rightarrow \mathcal{X}^n$\\
%Joint type: & Joint type:\\
%$(x_1^n,x_2^n,y^n)$ & $(x^n,y_1^n,y_2^n)$ \\
%$\in\mathcal{T}_\epsilon^{(n)}(P_YP_{X_2|Y}\mathds{1}_{X_1=f(Y)})$ & $\in\mathcal{T}_\epsilon^{(n)}(P_XP_{Y_2|X}\mathds{1}_{Y_1=f(X)})$ \\
%& where $P_X$ is a fixed input PMF\\
% & where $P_X$ is a fixed input distribution\\
%Joint type: & Joint type:\\
%$(x_1^n,x_2^n,y^n)\in\mathcal{T}_\epsilon^{(n)}(P_YP_{X_2|Y}\mathds{1}_{X_1=f(Y)})$ & %$(x^n,y_1^n,y_2^n)\in\mathcal{T}_\epsilon^{(n)}(P_XP_{Y_2|X}\mathds{1}_{Y_1=f(X)})$ \\
% & where $P_X$ is a fixed input distribution\\
\hline
\end{tabular}\vspace{-3mm}
\end{center}
%\hrulefill
\end{table*}

%%%%%%%%%%%%%%%%%%%%%%%%%%%%%%%%%%%%%%%%%%%%%%%%%%%%%%%%%%%%%%%%%%%%%%%%%%%%%%%%%%%%%%%%%%%%%%%%%%%%%%%%%%%%%%%%%%%

\section{Channel and Source Duality}\label{sec_duality}

\par We examine the WAK coordination problem with encoder cooperation (Fig. \ref{fig_coordination_model}) and the SD-BC with decoder cooperation (Fig. \ref{fig_semi_deterministic_BC}) from a duality perspective. We show that the two problems and their solutions are dual to one another in a manner that naturally extends PTP duality \cite{cover_duality2002,pradhan_duality2002,verdu_duality2011}. In the PTP scenario, two lossy source (or equivalently, source coordination) and channel coding problems are said to be dual if interchanging the roles of the encoder and the decoder in one problem produces the other problem. The solutions of such problems are dual in that they require an optimization of an information measure of the same structure, up to renaming the random variables involved. Solving one problem provides insight into the solution of the other. However, how duality extends to the multiuser case is still obscure. %Therefore, in this subsection, we aim to shed some light on this matter by analyzing the relation between the two models of interest.

%\par In our context and when extending the discussion of duality to the multiuser case,
%The SD-BC with decoder cooperation (Fig. \ref{fig_semi_deterministic_BC}) is dual to the AK coordination problem with encoder cooperation (Fig. \ref{fig_coordination_model}).
\par In the context of multiuser lossy source coding, we favor the framework of source coordination over rate-distortion, since the former provides a natural perspective on the similarities of the two problems. Source coordination inherently accounts for the probabilistic relations among \emph{all} the sequences involved in the problem's definition. However, in a coordination problem, both the input and output (coordination) PMFs are fixed, while in a channel coding problem, the input PMF is optimized. Therefore, for convenience, throughout this section we consider channel codes with codewords of fixed composition, as defined in the following (see also \cite{Successive_Refinemen_Permuter2013}).

\begin{definition}\label{def_bc_code_restrict} \emph{\textbf{(Fixed-Type Codes, Achievability and Capacity)}} An $(n,R_{12},R_1,R_2,Q_X^\star)$ fixed-type code $\mathcal{C}_n^\star$ for the SD-BC with one-sided decoder cooperation consists of three integer sets, an encoding function, a decoder cooperation function,
and two decoding functions as defined in \eqref{def_sdbc_code}.

For any $\delta>0$, the average error probability $e_\delta(Q_X^\star,\mathcal{C}_n^\star)$ of an $(n,R_{12},R_1,R_2,Q_X^\star)$ fixed-type code $\mathcal{C}_n^\star$ is defined in \eqref{EQ:composition_error_prob} at the bottom of the page, where $\hat{M}_1=\psi_1(\mathbf{Y}_1)$ and $\hat{M}_2=\psi_2\big(\mathbf{Y}_2,g_{12}(\mathbf{Y}_1)\big)$.

A rate triple $(R_{12},R_1,R_2)$ is achievable if for any $\epsilon,\delta>0$, there is a sufficiently large $n\in\mathbb{N}$ and a $(n,R_{12},R_1,R_2,Q_X^\star)$ fixed-type code $\mathcal{C}_n^\star$ such that $e_\delta(Q_X^\star,\mathcal{C}_n^\star)\leq\epsilon$. The definition of the capacity region is standard (see, e.g., \cite{CovThom06}).
\end{definition}

\setcounter{equation}{14}

\par Note that for fixed-composition codes \cite{Korner_fixed_comp1979,Pokorny_fixed_comp1985,Liu_fixed_comp1996,chang_fixed_comp2009} and for codes that are drawn in an i.i.d. manner according to $Q^\star_X$, the TV distance in \eqref{EQ:composition_error_prob} is arbitrarily small with high probability. Moreover, the capacity region of the SD-BC with cooperation and a fixed-type code is similar to that stated in Theorem \ref{tm_bc_capacity}. The only difference between the regions is the domain of PMFs over which the union is taken. Specifically, for the BC with a fixed-type code, the union is taken over all PMFs $P_{V,U,Y_1}P_{X|V,U,Y_1}Q_{Y_2|X}$ that have $Q_{X}^\star\mathds{1}_{\{Y_1=f(X)\}}Q_{Y_2|X}$ as a marginal.

\par The WAK and SD-BC problems with cooperation are obtained from each other by interchanging the roles of their encoder(s) and decoder(s) and renaming the random variables involved. A full description of the duality transformation principles is given in Table \ref{table_principles}. The duality is also evident in that the input and output sequences in both problems are jointly typical with respect to a PMF of the same form. Namely, in the source coding problem, the triple $(\mathbf{X}_1,\mathbf{X}_2,\mathbf{Y})$ is coordinated with respect to the PMF
\begin{equation}
Q_{X_2}P_{Y|X_2}\mathds{1}_{\{X_1=f(Y)\}}=P_{Y}\mathds{1}_{\{X_1=f(Y)\}}P_{X_2|Y}.\label{PMF_AK}
\end{equation}
The corresponding triple of sequences $(\mathbf{X},\mathbf{Y}_1,\mathbf{Y}_2)$ in the channel coding problem are jointly typical with high probability with respect to the PMF
\begin{equation}
Q^\star_{X}\mathds{1}_{\{Y_1=f(X)\}}Q_{Y_2|X}.\label{PMF_BC}
\end{equation}
By renaming the random variables according to Table \ref{table_principles}, the two PMFs in \eqref{PMF_AK} and \eqref{PMF_BC} coincide.

%%%%%%%%%%%%%%%%%%%%%%%%%%%%%%%%%%%%%%%%%%%%%%%%%%%%%%%%%%%%%%%%%%%%%%%%%%%%%%%%%%%%%%%%%%%%%%%%%%%%%%%%%%%%%%%%%%%
%%%%%%%%%%%%%%%%%%%%%%%%%%%%%%%%%%%         Fig Region Correspondance       %%%%%%%%%%%%%%%%%%%%%%%%%%%%%%%%%%%%%%%

\begin{figure*}[!t]
\begin{center}
\begin{psfrags}
    \psfragscanon
    \psfrag{X}[][][0.8]{$R_1$}
    \psfrag{Y}[][][0.8]{$\ \ \ R_2$}
    \psfrag{Z}[][][0.8]{$0$}
    \psfrag{S}[][][0.8]{$\ \ \ \ \ \ I(U;X_2|V)$}
    \psfrag{K}[][][0.8]{$\ \ \ \ \ \ \ \ \ \ +I(V;X_1)$}
    \psfrag{O}[][][0.8]{$\ \ \ \ \ \ \ \ \ H(X_1|V,U)$}
    \psfrag{T}[][][0.8]{$\ \ \ \ \ \ \ I(U;X_2|V)$}
    \psfrag{W}[][][0.8]{$\ \ \ \ \ \ \ -I(U;X_1|V)$}
    \psfrag{P}[][][0.8]{$\ \ \ \ \ \ \ H(X_1)$}
    \psfrag{A}[][][0.8]{$R_1$}
    \psfrag{G}[][][0.8]{$\ \ \ R_2$}
    \psfrag{D}[][][0.8]{$0$}
    \psfrag{E}[][][0.8]{$\ \ \ \ \ \ I(U;Y_2|V)$}
    \psfrag{J}[][][0.8]{$\ \ \ \ \ \ \ \ \ \ +I(V;Y_1)$}
    \psfrag{C}[][][0.8]{$\ \ \ \ \ \ \ \ \ H(Y_1|V,U)$}
    \psfrag{F}[][][0.8]{$\ \ \ \ \ \ \ I(U;Y_2|V)$}
    \psfrag{H}[][][0.8]{$\ \ \ \ \ \ \ -I(U;Y_1|V)$}
    \psfrag{B}[][][0.8]{$\ \ \ \ \ \ \ H(Y_1)$}
    \psfrag{U}[][][0.8]{(a)}
    \psfrag{I}[][][0.8]{(b)}
\includegraphics[scale=0.6]{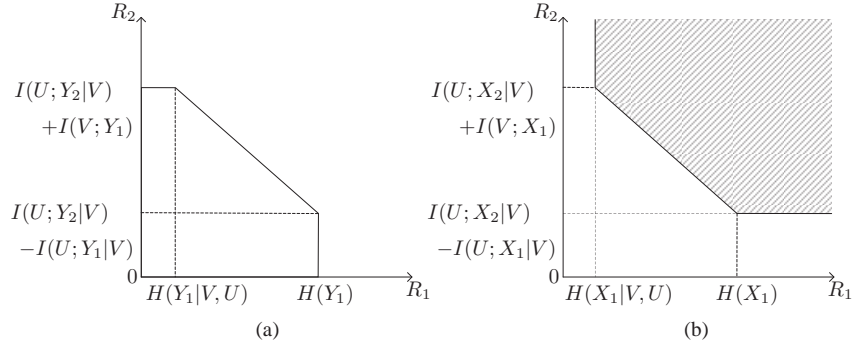}%\vspace{-2mm}
\caption{Corner point correspondence between: (a) the capacity region of the SD-BC with cooperation; (b) the coordination-capacity region of the WAK coordination problem with cooperation. The regions are depicted at the hyperplanes where to $R_{12}=I(V;Y_1)-I(V;Y_2)$ and $R_{12}=I(V;X_1)-I(V;X_2)$, respectively.} \label{fig_regions}
\psfragscanoff
\end{psfrags}
\end{center}\vspace{-3mm}
\end{figure*}

%%%%%%%%%%%%%%%%%%%%%%%%%%%%%%%%%%%%%%%%%%%%%%%%%%%%%%%%%%%%%%%%%%%%%%%%%%%%%%%%%%%%%%%%%%%%%%%%%%%%%%%%%%%%%%%%%%%

\par The duality between the two problems extends beyond the correspondence presented above. The coordination-capacity region of the WAK problem (Theorem \ref{tm_ak_capacity}) and the capacity region of the SD-BC (Theorem \ref{tm_bc_capacity}) are also dual to one another. To see this, the following lemma gives an alternative characterization of the capacity region $\mathcal{C}_{\mathrm{BC}}$.

\begin{lemma}\label{lemma_region_BC_equal} \emph{\textbf{(SD-BC Capacity Alternative Characterization)}}
Let $\mathcal{C}^{(\mathrm{D})}_\mathrm{BC}$ be the region defined by the union of rate triples $(R_{12},R_1,R_2)\in\mathbb{R}^3_+$ satisfying:
\begin{subequations}
\begin{align}
R_{12} &\geq I(V;Y_1)-I(V;Y_2)\label{EQ:region_eq_sdbc12}\\
R_1 &\leq H(Y_1)\label{EQ:region_eq_sdbc1}\\
R_2 &\leq I(V,U;Y_2)+R_{12}\label{EQ:region_eq_sdbc2}\\
R_1+R_2 &\leq H(Y_1|V,U)+I(U;Y_2|V)+I(V;Y_1)\label{EQ:region_eq_sdbc1+2}
\end{align}\label{EQ:region_eq_sdbc}
\end{subequations}

\vspace{-4.5mm}
\noindent where the union is over the domain stated in Theorem \ref{tm_bc_capacity}. Then:
\begin{equation}
\mathcal{C}^{(\mathrm{D})}_\mathrm{BC}=\mathcal{C}_{\mathrm{BC}}.\label{alter_region_equal}
\end{equation}
\end{lemma}

See Appendix \ref{appen_region_euivalence_RBC} for a proof of Lemma \ref{lemma_region_BC_equal} based on bidirectional inclusion arguments.

\begin{remark}
$\mathcal{C}^{(\mathrm{D})}_\mathrm{BC}$ can be established as the capacity region of the SD-BC with cooperation by providing achievability and converse proofs. We refer the reader to \cite{Goldfeld_BC_Cooperation_IEEEI2014} for a full description of the achievability scheme. The proof of the converse is given in Appendix \ref{appen_region_euivalennt_converse}. The converse is established via a novel approach, in which the auxiliaries are not only chosen as a (possibly different) function of the joint distribution induced by each code, but they are also constructed in a probabilistic manner. The need for this probabilistic construction stems from the unique structure of the region $\mathcal{C}^{(\mathrm{D})}_\mathrm{BC}$. Specifically, the lower bound on $R_{12}$ in \eqref{EQ:region_eq_sdbc12} (which is typical to source coding problems where the random source sequences are memoryless) and the fact that $\mathbf{Y}_1$ and $\mathbf{Y}_2$ have memory are the underlying reasons for the usefulness of the approach. Depending on the distribution that stems from the code, a deterministic choice of auxiliaries may result in a $I(V;Y_1)-I(V;Y_2)$ that is too large. By a stochastic choice of the auxiliaries, we circumvent this difficulty and dominate the quantity $I(V;Y_1)-I(V;Y_2)$ to satisfy \eqref{EQ:region_eq_sdbc12}.

\par The converse proof boils down to two key steps. First, we derive an outer bound on the achievable region $\mathcal{C}^{(\mathrm{D})}_\mathrm{BC}$ that is described by three auxiliary random variables $(A,B,C)$. Then, by probabilistically choosing $(V,U)$ from $(A,B,C)$, we show that the outer bound is tight. The second step implies that the outer bound is an alternative formulation of the capacity region. Capacity proofs that rely on alternative descriptions for which the converse is provided have been previously used (see, e.g., \cite{Korner_BC_DegradedMessageSet1977} and \cite{ElGamal_MoreCapable1979}). However, the proof of equivalence typically relies on operational arguments rather than on a probabilistic identification of auxiliaries. Probabilistic arguments of a similar nature to those we present here were also used before \cite{Dabora_probabilistic2008,Courtade_Weissman_Converse2014,Dikstein_PDBC_Cooperation2014}. For instance, in \cite{Dabora_probabilistic2008}, such arguments were used to prove the equivalence between two representations of the compress-and-forward inner bound for the relay channel via time-sharing. Such arguments were also leveraged in \cite{Courtade_Weissman_Converse2014} to characterize the admissible rate-distortion region for the multiterminal source coding problem under logarithmic loss. The novelty of our approach stems from combining these two concepts and essentially using a probabilistic construction to define the auxiliary random variables and establish the tightness of the outer bound. We derive a closed form formula for the optimal probability values, that highlights the dependence of the the auxiliaries on the distribution induced by the code.
\end{remark}

\begin{figure*}[!t]
    \begin{center}
        \begin{psfrags}
            \psfragscanon
            \psfrag{I}[][][1]{$(M_1,M_2)$}
            \psfrag{J}[][][1]{\ \ \ \ \ \ \ \ \ \ Encoder}
            \psfrag{K}[][][1]{\ \ \ \ $X_i$}
            \psfrag{V}[][][1]{\ \ \ \ \ \ \ \ \ \ \ \ \ $Q_{Y_1,Y_2|X,X_1}$}
            \psfrag{S}[][][1]{\ \ \ \ \ \ \ \ \ \ \ \ Channel}
            \psfrag{M}[][][1]{\ \ \ \ \ \ \ \ \ \ \ \ $Y_{1,i}$}
            \psfrag{N}[][][1]{\ \ \ \ \ \ \ \ \ \ \ \ $Y_{2,i}$}
            \psfrag{U}[][][0.67]{\ \ \ \ \ \ \ \ $X_{1,i}\big(Y_1^{1-i}\big)$}
            \psfrag{O}[][][1]{\ \ \ \ \ \ \ \ \ \ Decoder 1}
            \psfrag{P}[][][1]{\ \ \ \ \ \ \ \ \ \ Decoder 2}
            \psfrag{Q}[][][1]{\ \ \ $\hat{M}_1$}
            \psfrag{R}[][][1]{\ \ \ $\hat{M}_2$}
            \psfrag{T}[][][1]{\ \ \ \ Relay}
            \includegraphics[scale = .64]{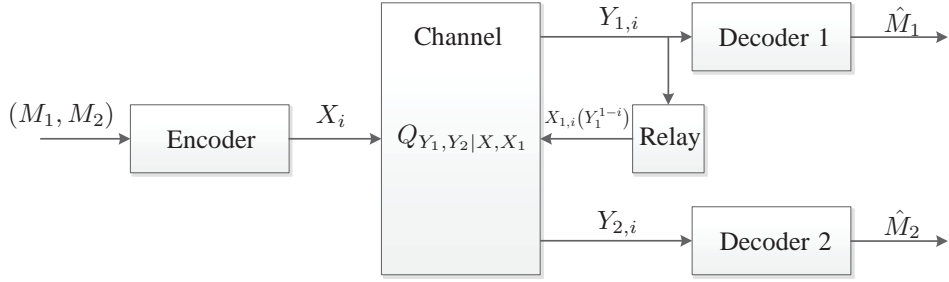}
            \caption{A general RBC.} \label{fig_RBC}
            \psfragscanoff
        \end{psfrags}
     \end{center}%\vspace{-6mm}
 \end{figure*}

%%%%%%%%%%%%%%%%%%%%%%%%%%%%%%%%%%%%%%%%%%%%%%%%%%%%%%%%%%%%%%%%%%%%%%%%%%%%%%%%%%%%%%%%%%%%%%%%%%%%%%%%%%%%%%%%%%%

\par The duality between $\mathcal{R}_{\mathrm{WAK}}(P_{X_1,X_2,Y})$ in \eqref{region_AK} and $\mathcal{C}^{(\mathrm{D})}_\mathrm{BC}$ in \eqref{EQ:region_eq_sdbc} is expressed as a correspondence between the information measures at their corner points. The values of $(R_{12},R_1,R_2)$ at the corner points of the coordination-capacity region of the WAK problem are
\begin{subequations}
\begin{align}
&\mspace{-20mu}\big(\mspace{3mu}I(V;X_1|X_2)\mspace{3mu},\mspace{3mu}H(X_1)\mspace{3mu},\mspace{3mu}I(U;X_2|X_1,V)\mspace{3mu}\big)\label{ak_corner1}\\
&\mspace{-20mu}\big(\mspace{3mu}I(V;X_1|X_2)\mspace{3mu},\mspace{3mu}H(X_1|V,U)\mspace{3mu},\mspace{3mu}I(U;X_2|V)+I(V;X_1)\mspace{3mu}\big)\label{ak_corner2}
\end{align}\label{ak_corner}
\end{subequations}

\vspace{-4.5mm}
\noindent while the triple $(R_{12},R_1,R_2)$ at the corner points of capacity region of the SD-BC with cooperation are
\begin{align*}
&\big(\mspace{3mu}I(V;Y_1)\mspace{-3mu}-\mspace{-3mu}I(V;Y_2)\mspace{3mu},\mspace{3mu}H(Y_1)\mspace{3mu},\mspace{3mu}I(U;Y_2|V)\mspace{-3mu}-\mspace{-3mu}I(U;Y_1|V)\mspace{3mu}\big)\\
&\big(\mspace{2mu}I(V;Y_1)\mspace{-3mu}-\mspace{-3mu}I(V;Y_2)\mspace{3mu},\mspace{3mu}H(Y_1|V,U)\mspace{3mu},\mspace{3mu}I(U;Y_2|V)\mspace{-3mu}+\mspace{-3mu}I(V;Y_1)\mspace{2mu}\big).\numberthis\label{bc_corner}
\end{align*}
We show that \eqref{ak_corner} and \eqref{bc_corner} correspond by first rewriting the value of $R_{12}$ in \eqref{ak_corner} as
%\vspace{-1.7mm}
\begin{equation}
R_{12}=I(V;X_1|X_2)\stackrel{(a)}=I(V;X_1)-I(V;X_2)\label{R12_ak_corner_mod}
\end{equation}
where (a) is due to the Markov relation $V-X_1-X_2$. Moreover, the value of $R_2$ in \eqref{ak_corner1} is rewritten as
\begin{equation}
R_2=I(U;X_2|X_1,V)\stackrel{(a)}=I(U;X_2|V)-I(U;X_1|V)\label{R2_ak_corner_mod}
\end{equation}
where (a) is since $U-(X_2,V)-X_1$ forms a Markov chain. By substituting \eqref{R12_ak_corner_mod}-\eqref{R2_ak_corner_mod} into \eqref{ak_corner} and renaming the random variables according to Table \ref{table_principles}, the corner points of both regions coincide (see Fig. \ref{fig_regions}).
%\vspace{-1mm}

\par Chronologically, upon observing the duality between the two problem settings, we solved the WAK problem first. Then, based on past experience (cf., e.g., \cite{Successive_Refinemen_Permuter2013} and \cite{Dikstein_MAC_Action_2015}), our focus turned to the dual SD-BC with cooperation. Since the capacity region is defined by the corner points of a union of polytopos, the structure of the capacity region for the SD-BC was evident. Thus, duality was key in obtaining the results of this work. We note that the relation between our result for the SD-BC with cooperation and the SD-RBC (that is discussed in the following section) was observed only at a later stage.

%%%%%%%%%%%%%%%%%%%%%%%%%%%%%%%%%%%%%%%%%%%%%%%%%%%%%%%%%%%%%%%%%%%%%%%%%%%%%%%%%%%%%%%%%%%%%%%%%%%%%%%%%%%%%%%%%%%
%%%%%%%%%%%%%%%%%%%%%%%%%%%%%%%%%%%%%%%%%%%%%%%%%%%%%%%%%%%%%%%%%%%%%%%%%%%%%%%%%%%%%%%%%%%%%%%%%%%%%%%%%%%%%%%%%%%
%%%%%%%%%%%%%%%%%%%%%%%%                                                           %%%%%%%%%%%%%%%%%%%%%%%%%%%%%%%%
%%%%%%%%%%%%%%%%%%%%%%%%          RELATION TO THE SEMI-DETERMINISTIC RBC           %%%%%%%%%%%%%%%%%%%%%%%%%%%%%%%%
%%%%%%%%%%%%%%%%%%%%%%%%                                                           %%%%%%%%%%%%%%%%%%%%%%%%%%%%%%%%
%%%%%%%%%%%%%%%%%%%%%%%%%%%%%%%%%%%%%%%%%%%%%%%%%%%%%%%%%%%%%%%%%%%%%%%%%%%%%%%%%%%%%%%%%%%%%%%%%%%%%%%%%%%%%%%%%%%
%%%%%%%%%%%%%%%%%%%%%%%%%%%%%%%%%%%%%%%%%%%%%%%%%%%%%%%%%%%%%%%%%%%%%%%%%%%%%%%%%%%%%%%%%%%%%%%%%%%%%%%%%%%%%%%%%%%

\section{Relation to the SD-RBC}\label{sec_RBC}

\par The SD-BC with cooperation is strongly related to the SD-RBC that was studied in \cite{Liang_Kramer_RBC2007}. A general RBC is illustrated in Fig. \ref{fig_RBC} (for the full definition see \cite[Section II]{Liang_Kramer_RBC2007}). The RBC is SD if the PMF $Q_{Y_1|X,X_1}$ only takes on the values 0 or 1. To see the correspondence between the SD-RBC and the BC of interest, let $Y_2=(Y_{21},Y_{22})$ and let the channel transition PMF factorize as
\begin{equation}
Q_{Y_1,Y_{21},Y_{22}|X,X_1}=Q_{Y_{21}|X}\mathds{1}_{\{Y_1=f(X)\}}Q_{Y_{22}|X_1}. \label{RBC_factorization}
\end{equation}
\eqref{RBC_factorization} implies that the channel from the encoder to the decoders is orthogonal to the one between the decoders. Suppose the relay channel is deterministic with capacity $R_{12}$ and let $Y_{22}=f_R(X_1)$. The SD-RBC obtained under these assumptions is referred to as the $R_{12}$-reduced SD-RBC and its capacity region is denoted by $\mathcal{C}_{\mathrm{RBC}}(R_{12})$. As stated in the following lemma, the $R_{12}$-reduced SD-RBC is operationally equivalent to the SD-BC with cooperation. By operational equivalence, we mean that for every achievable rate tuple in one problem, there exists a code (that achieves these rates) that can be transformed into a code (with the same rates) for the other problem. The transformation mechanism treats the code for each model as a black-box and is described as part of the proof of Lemma \ref{RBC_lemma_equivalence} given in Appendix \ref{appen_RBC_lemma_equivalence}.

\begin{lemma}[Operational Equivalence]\label{RBC_lemma_equivalence}
For every $(R_1,R_2)\in\mathcal{C}_{RBC}(R_{12})$, there is an $(n,R_1,R_2)$ code $\mathcal{C}_n^{(\mathrm{RBC})}(R_{12})$ for the $R_{12}$-reduced SD-RBC that can be transformed into an $(n,R_{12},R_1,R_2)$ code $\mathcal{C}_n^{(\mathrm{BC})}$ for the SD-BC with cooperation, and vice versa. Namely, for every $(R_{12},R_1,R_2)\in\mathcal{C}_{\mathrm{BC}}$, there is a $(n,R_{12},R_1,R_2)$ code $\mathcal{C}_n^{(\mathrm{BC})}$ for the SD-BC with cooperation that can be transformed into an $(n,R_1,R_2)$ code $\mathcal{C}_n^{(\mathrm{RBC})}(R_{12})$ for the $R_{12}$-reduced SD-RBC.
\end{lemma}

\par Lemma \ref{RBC_lemma_equivalence} implies that the capacity regions of the SD-BC with cooperation and the $R_{12}$-reduced SD-RBC coincide. Using the result of \cite[Theorem 8]{Liang_Kramer_RBC2007}, the capacity region $\mathcal{C}_{RBC}(R_{12})$ of the $R_{12}$-reduced SD-RBC is the union of rate pairs $(R_1,R_2)\in\mathbb{R}^2_+$ satisfying:
\begin{align*}
R_1&\leq H(Y_1|X_1)\\
R_2&\leq I(V,U,X_1;Y_{21})+H(Y_{22}|Y_{21})\\
R_1\mspace{-3mu}+\mspace{-3mu}R_2&\leq H(Y_1|V,U,X_1)\mspace{-3mu}+\mspace{-3mu}I(U;Y_{21}|V,X_1)\mspace{-3mu}+\mspace{-3mu}I(V;Y_1|X_1)\\
R_1\mspace{-3mu}+\mspace{-3mu}R_2&\leq H(Y_1|V,U,X_1)\mspace{-3mu}+\mspace{-3mu}I(V,U,X_1;Y_{21})\mspace{-3mu}+\mspace{-3mu}H(Y_{22}|Y_{21})\numberthis\label{region_RBC_reduced}
\end{align*}
where the union is over all PMFs $P_{V,U,X,X_1}Q_{Y_{21}|X}\mathds{1}_{\{Y_1=f(X)\}}\mathds{1}_{\{Y_{22}=f_R(X_1)\}}$. In Appendix \ref{appen_region_RBC} we simplify the region in \eqref{region_RBC_reduced} and show that it coincides with the capacity region of the SD-BC with cooperation from Theorem \ref{tm_bc_capacity}.
%\begin{subequations}
%\begin{align}
%R_1&\leq H(Y_1)\label{region_RBC_reduced2_R1}\\
%R_2&\leq I(V,U;Y_{21})+R_{12}\label{region_RBC_reduced2_R2}\\
%R_1+R_2&\leq H(Y_1|V,U)+I(U;Y_{21}|V)+I(V;Y_1)\label{region_RBC_reduced2_R1+21}\\
%R_1+R_2&\leq H(Y_1|V,U)+I(V,U;Y_{21})+R_{12}\label{region_RBC_reduced2_R1+22}
%\end{align}\label{region_RBC_reduced2}
%\end{subequations}
%
%\vspace{-7mm}

%\noindent where the union is over all joint distributions $P_{V,U,X}P_{Y_{21}|X}\mathds{1}_{\{Y_1=f(X)\}}$. In Appendix \ref{appen_region_euivalence_RBC} we establish that the region characterized by (\ref{region_RBC_reduced2}) coincides with the capacity region of the SD-BC with cooperation stated in Theorem \ref{tm_bc_capacity}.
%\ \\
\par The advantage of the approach taken in this work compared to that in \cite{Liang_Kramer_RBC2007} is threefold. First, we achieve capacity over a \emph{single} transmission block, while the scheme in \cite{Liang_Kramer_RBC2007} (which, as a consequence of Lemma \ref{RBC_lemma_equivalence}, can also be used for the SD-BC with cooperation) transmits a large number of blocks and applies backward decoding. The substantial delay introduced by a backward decoding process implies the superiority of our scheme for practical uses. The reduction of the multi-block coding scheme in \cite{Liang_Kramer_RBC2007} to our single-block scheme is consistent with the results in \cite{Cover_Kim_Relay2008}. The authors of \cite{Cover_Kim_Relay2008} showed that for the primitive relay channel (i.e., a relay channel with a noiseless link from relay to the receiver), the decode-and-forward and compress-and-forward multi-block coding schemes can be applied with only a single transmission block.
The second advantage of our approach is the simple and concise converse proof that follows using telescoping identities \cite[Eq. (9) and (11)]{Kramer_telescopic2011}. Finally, focusing on the SD-BC with cooperation (rather than the SD-RBC) highlights the duality with the cooperative WAK source coordination problem (as discussed in Section \ref{sec_duality}), and gives insight into the relations between multiuser channel and source coding problems.

%Consequently, the two coding schemes induce different characterizations of the capacity region. The difference is expressed in replacing the lower bound on $R_{12}$ in (\ref{region_BC12}) with the additional bound on the sum rate stated in (\ref{region_RBC_reduced2_R1+22}). Due to the particular structure of the region in (\ref{region_BC}), which involves the lower bound on the cooperation rate $R_{12}$, our approach for the converse proof is essential to establish optimality

%%%%%%%%%%%%%%%%%%%%%%%%%%%%%%%%%%%%%%%%%%%%%%%%%%%%%%%%%%%%%%%%%%%%%%%%%%%%%%%%%%%%%%%%%%%%%%%%%%%%%%%%%%%%%%%%%%%
%%%%%%%%%%%%%%%%%%%%%%%%%%%%%%%%%%%%%%%%%%%%%%%%%%%%%%%%%%%%%%%%%%%%%%%%%%%%%%%%%%%%%%%%%%%%%%%%%%%%%%%%%%%%%%%%%%%
%%%%%%%%%%%%%%%%%%%%%%%%                                                           %%%%%%%%%%%%%%%%%%%%%%%%%%%%%%%%
%%%%%%%%%%%%%%%%%%%%%%%%      Special Cases of the Capacity Region of the BC       %%%%%%%%%%%%%%%%%%%%%%%%%%%%%%%%
%%%%%%%%%%%%%%%%%%%%%%%%                                                           %%%%%%%%%%%%%%%%%%%%%%%%%%%%%%%%
%%%%%%%%%%%%%%%%%%%%%%%%%%%%%%%%%%%%%%%%%%%%%%%%%%%%%%%%%%%%%%%%%%%%%%%%%%%%%%%%%%%%%%%%%%%%%%%%%%%%%%%%%%%%%%%%%%%
%%%%%%%%%%%%%%%%%%%%%%%%%%%%%%%%%%%%%%%%%%%%%%%%%%%%%%%%%%%%%%%%%%%%%%%%%%%%%%%%%%%%%%%%%%%%%%%%%%%%%%%%%%%%%%%%%%%

\section{Special Cases}\label{sec_special cases}

\par We consider special cases of the capacity region of the SD-BC with decoder cooperation and show that the dual relation discussed in Section \ref{sec_duality} is preserved for each special case.

\subsection{Deterministic BCs with Decoder Cooperation}

\begin{corollary}[Deterministic BC Capacity Region]
The capacity region of a deterministic BC (DBC) is the union of rate triples $(R_{12},R_1,R_2)\in\mathbb{R}^3_+$ satisfying:
\begin{align*}
R_1&\leq H(Y_1)\\
R_2&\leq H(Y_2)+R_{12}\\
R_1+R_2&\leq H(Y_1,Y_2)\numberthis\label{region_det_BC}
\end{align*}
\noindent where the union is over all input PMFs $P_X$.
\end{corollary}

\begin{IEEEproof}
Achievability follows from Theorem \ref{tm_bc_capacity} by taking $V=0$ and $U=Y_2$. A converse follows by the Cut-Set bound.\end{IEEEproof}

\par The DBC is dual to the SW source coding problem with one-sided encoder cooperation (see \cite{Draper_Coop_SW2004} and \cite{Nguyen_Coop_DBC2005}). The SW setting is obtained from the WAK coordination problem by also adding a lossless reproduction requirement to the second source. A proper choice of the auxiliary random variables, $\mathcal{R}_{\mathrm{WAK}}(P_{X_1,X_2,Y})$ reduces to the optimal rate region for the SW problem, which is the set of rate triples $(R_{12},R_1,R_2)\in\mathbb{R}_+^3$ satisfying:
\begin{align*}
R_1&\geq H(X_1|X_2)-R_{12}\\
R_2&\geq H(X_2|X_1)\\
R_1+R_2&\geq H(X_1,X_2)\numberthis\label{region_SW}
\end{align*}
(see Appendix \ref{appen_SW_region_corr} for the derivation of (\ref{region_SW})). Examining the regions from \eqref{region_det_BC} and \eqref{region_SW}, reveals the correspondence between their corner points.

\subsection{PD-BCs with Decoder Cooperation}

\begin{corollary}[PD-BC Capacity Region]\label{cor_PDBC}
The capacity region $\mathcal{C}_{\mathrm{PD}}$ for the PD-BC with $Y_1=X$ coincides with the results in \cite{DaboraServetto06BC} and \cite{Dikstein_PDBC_Cooperation2014} and is the union of rate triples $(R_{12},R_1,R_2)\in\mathbb{R}^3_+$ satisfying:
\begin{subequations}
\begin{align}
R_1 &\leq H(X|U)\label{dikstein1}\\
R_2 &\leq I(U;Y_2)+R_{12}\label{dikstein2}\\
R_1+R_2 &\leq H(X)\label{dikstein1+2}
\end{align}\label{region_dikstein_special}
\end{subequations}

\vspace{-4.5mm}
\noindent where the union is over all PMFs $P_{U,X}Q_{Y_2|X}$.
\end{corollary}

\begin{IEEEproof}
The capacity region of the PD-BC was originally derived in \cite{DaboraServetto06BC} where it was described as the union of rate triples $(R_{12},R_1,R_2)\in\mathbb{R}^3_+$ satisfying:
\begin{align*}
R_1 &\leq I(X;Y_1|U)\\
R_2 &\leq I(U;Y_2)+R_{12}\\
R_2 &\leq I(U;Y_1)\numberthis\label{region_dabora}
\end{align*}
\noindent where the union is over all PMFs $P_{U,X}Q_{Y_1|X}Q_{Y_2|Y_1}$.

\par An equivalent characterization of region in (\ref{region_dabora}) was later given in \cite{Dikstein_PDBC_Cooperation2014} as the union over the domain stated above of rate triples $(R_{12},R_1,R_2)\in\mathbb{R}^3_+$ satisfying:
\begin{align*}
R_1 &\leq I(X;Y_1|U)\\
R_2 &\leq I(U;Y_2)+R_{12}\\
R_1+R_2 &\leq I(X;Y_1).\numberthis\label{region_dikstein}
\end{align*}

Since a SD-BC in which $Y_1=X$ is also PD, substituting $Y_1=X$ into \eqref{region_dikstein} yields the region from Corollary \ref{cor_PDBC}. By substituting $Y_1=X$, setting $U=0$, and relabeling $V$ as $U$ in the capacity of the SD-BC with cooperation stated in Theorem \ref{tm_bc_capacity}, we obtain an achievable region given by the union over the domain stated in Corollary \ref{cor_PDBC} of rate triples $(R_{12},R_1,R_2)\in\mathbb{R}^3_+$ satisfying:
\begin{subequations}
\begin{align}
R_2 &\leq I(U;Y_2)+R_{12}\label{general_case1}\\
R_1+R_2 &\leq H(X).\label{general_case2}
\end{align}\label{region_BC_special}
\end{subequations}

\vspace{-4.5mm}
\noindent Denote the region in \eqref{region_BC_special} by $\mathcal{R}_{\mathrm{SD}}$. Since $\mathcal{R}_{\mathrm{SD}}$ is an achievable region, clearly $\mathcal{R}_{\mathrm{SD}}\subseteq\mathcal{C}_{\mathrm{PD}}$. On the other hand, the opposite inclusion $\mathcal{C}_{\mathrm{PD}}\subseteq\mathcal{R}_{\mathrm{SD}}$ also holds, because the rate bound \eqref{dikstein1} does not appear in $\mathcal{R}_{\mathrm{SD}}$, while \eqref{dikstein2}-\eqref{dikstein1+2} and the domain over which the union is taken are preserved.\end{IEEEproof}

\par The dual source coding problem for the PD-BC with cooperation where $Y_1=X$ is a model in which the output sequence is a lossless reproduction of $\mathbf{X}_1$. The latter setting is a special case of the WAK problem with cooperation, that is obtained by taking $f$ (the coordination function) to be the identity function. The corresponding coordination-capacity region is given by (\ref{region_AK}) (with a slight modification of the domain over which the union is taken). However, an equivalent coordination-capacity region that is characterized by a single auxiliary random variable has yet to be derived. Since the capacity region of the PD-BC with cooperation where $Y_1=X$ is described using a single auxiliary (as in (\ref{region_dikstein_special})), the lack of such a characterization for the region of the dual problem makes the comparison problematic. Nonetheless, recalling that the capacity region of the considered PD-BC is also given by (\ref{region_BC}) while substituting $Y_1=X$ emphasizes that the duality holds.

\section{Summary and Concluding Remarks}\label{summary}

\par We considered the WAK empirical coordination problem with one-sided encoder cooperation and derived its coordination-capacity region. The capacity-achieving coding scheme combined WZ coding, binning and superposition coding. Furthermore, a SD-BC in which the decoders can cooperate via a one-sided rate-limited link was considered and its capacity region was found. Achievability was established by deriving an inner bound on the capacity region of a general BC that was shown to be tight for the SD scenario. The coding strategy that achieved the inner bound combined rate-splitting, Marton and superposition coding, and binning (used for the cooperation protocol). The converse for the SD case leveraged telescoping identities that resulted in a concise and a simple proof. The relation between the SD-BC with cooperation and the SD-RBC was examined. The two problems were shown to be operationally equivalent under proper assumptions and the correspondence between their capacity regions was established.

\par The cooperative WAK and SD-BC problems were inspected from a channel-source duality perspective. Transformation principles between the two settings that naturally extend duality relations between PTP models were presented. It was shown that the duality between the WAK and the SD-BC problems induces a duality between their capacities that is expressed in a correspondence between the corner points of the two regions. To this end, the capacity region of the SD-BC was restated as an alternative expression. The converse was based on a novel approach where the construction of the auxiliary random variables is probabilistic and depends on the distribution induced by the code. The probabilistic construction introduced additional optimization parameters (the probability values) that were used to tighten the outer bound to coincide with the alternative achievable region. To conclude the discussion, several special cases of the BC setting and their corresponding capacity regions were inspected.

%We finally note that extensions of the result for the SD-BC to the state-dependent case (i.e., combining the setting in \cite{Lapidoth_senideterministic2012} and the one considered in this paper) and to the scenario of a cognitive semi-deterministic interference channel are currently being investigated. Furthermore, a SD-BC with a one-sided cooperation link in the opposite direction (i.e., from Decoder 2 to Decoder 1) and its corresponding dual source coding problem are also being considered.

%%%%%%%%%%%%%%%%%%%%%%%%%%%%%%%%%%%%%%%%%%%%%%%%%%%%%%%%%%%%%%%%%%%%%%%%%%%%%%%%%%%%%%%%%%%%%%%%%%%%%%%%%%%%%%%%%%%
%%%%%%%%%%%%%%%%%%%%%%%%%%%%%%%%%%%%%%%%%%%%%%%%%%%%%%%%%%%%%%%%%%%%%%%%%%%%%%%%%%%%%%%%%%%%%%%%%%%%%%%%%%%%%%%%%%%
%%%%%%%%%%%%%%%%%%%%%%%%%%%%%%%%%%%%                                         %%%%%%%%%%%%%%%%%%%%%%%%%%%%%%%%%%%%%%
%%%%%%%%%%%%%%%%%%%%%%%%%%%%%%%%%%%%                APPENDICES               %%%%%%%%%%%%%%%%%%%%%%%%%%%%%%%%%%%%%%
%%%%%%%%%%%%%%%%%%%%%%%%%%%%%%%%%%%%                                         %%%%%%%%%%%%%%%%%%%%%%%%%%%%%%%%%%%%%%
%%%%%%%%%%%%%%%%%%%%%%%%%%%%%%%%%%%%%%%%%%%%%%%%%%%%%%%%%%%%%%%%%%%%%%%%%%%%%%%%%%%%%%%%%%%%%%%%%%%%%%%%%%%%%%%%%%%
%%%%%%%%%%%%%%%%%%%%%%%%%%%%%%%%%%%%%%%%%%%%%%%%%%%%%%%%%%%%%%%%%%%%%%%%%%%%%%%%%%%%%%%%%%%%%%%%%%%%%%%%%%%%%%%%%%%

\appendices

%%%%%%%%%%%%%%%%%%%%%%%%%%%%%%%%%%%%%%%%%%%%%%%%%%%%%%%%%%%%%%%%%%%%%%%%%%%%%%%%%%%%%%%%%%%%%%%%%%%%%%%%%%%%%%%%%%%
%%%%%%%%%%%%%%%%%%%%%%%%%%%%%%%%%%%%%%%%%%%%%%%%%%%%%%%%%%%%%%%%%%%%%%%%%%%%%%%%%%%%%%%%%%%%%%%%%%%%%%%%%%%%%%%%%%%
%%%%%%%%%%%%%%%                        APPENDIX A - Proof of Theorem 1                    %%%%%%%%%%%%%%%%%%%%%%%%%
%%%%%%%%%%%%%%%%%%%%%%%%%%%%%%%%%%%%%%%%%%%%%%%%%%%%%%%%%%%%%%%%%%%%%%%%%%%%%%%%%%%%%%%%%%%%%%%%%%%%%%%%%%%%%%%%%%%
%%%%%%%%%%%%%%%%%%%%%%%%%%%%%%%%%%%%%%%%%%%%%%%%%%%%%%%%%%%%%%%%%%%%%%%%%%%%%%%%%%%%%%%%%%%%%%%%%%%%%%%%%%%%%%%%%%%

\section{Proof of Theorem \ref{tm_ak_capacity}}\label{appen_proof_tm1}

%%%%%%%%%%%%%%%%%%%%%%%%%%%%%%%%%%%%%%%%%%%%%%%%%%%%%%%%%%%%%%%%%%%%%%%%%%%%%%%%%%%%%%%%%%%%%%%%%%%%%%%%%%%%%%%%%%%
%%%%%%%%%%%%%%%                              Achievability                                %%%%%%%%%%%%%%%%%%%%%%%%%
%%%%%%%%%%%%%%%%%%%%%%%%%%%%%%%%%%%%%%%%%%%%%%%%%%%%%%%%%%%%%%%%%%%%%%%%%%%%%%%%%%%%%%%%%%%%%%%%%%%%%%%%%%%%%%%%%%%

\subsection{Achievability}

For any $P_{X_1,X_2,Y}\in\mathcal{Q}$, the direct proof is based on a coding scheme that achieves the corner points of $\mathcal{R}_{\mathrm{WAK}}(P_{X_1,X_2,Y})$. The corner points are stated in \eqref{ak_corner1}-\eqref{ak_corner2} and illustrated in Fig \ref{fig_AK_region}. Fix a PMF $P_{X_1,X_2,Y}\in\mathcal{Q}$, $\epsilon,\delta>0$ and a PMF $P_{X_1,X_2,V,U,Y}=Q_{X_1,X_2}P_{V|X_1}P_{U|X_2,V}P_{Y|X_1,U,V}$ that has $P_{X_1,X_2,Y}$ as a marginal. Recall that $P_{X_1,X_2,Y}$ factors as $Q_{X_2}P_{Y|X_2}\mathds{1}_{\{X_1=f(Y)\}}$ and that it has the source PMF $Q_{X_1,X_2}$ as a marginal. 

The error probability analysis of the subsequently described coding scheme follows by standard random coding arguments. Namely, we evaluate the expected error probability over the ensemble of codebooks and use the union bound to account for each error event separately. Being standard, the details are omitted and only the consequent rate bounds required for reliability are stated.

\par \textbf{Codebook Generation:} A codebook $\mathcal{C}_{V}$ that comprises $2^{nR_V}$ codewords $\mathbf{v}(i)$, where $i\in[1:2^{nR_v}]$, each generated according to $P_V^n$. The codebook $\mathcal{C}_V$ is randomly partitioned into $2^{nR_{12}}$ bins indexed by $t_{12}\in[1:2^{nR_{12}}]$ and denoted by $\mathcal{B}_V(t_{12})$. For every $i\in[1:2^{nR_v}]$ a codebook $\mathcal{C}_U(i)$ is generated. Each codebook $\mathcal{C}_U(i)$ is assembled of $2^{nR_U}$ codewords $\mathbf{u}(i,j)$, $j\in[1:2^{nR_u}]$, generated according to $P^n_{U|V=\mathbf{v}(i)}$. Each $\mathcal{C}_U(i)$ codebook is randomly partitioned into $2^{nR_2'}$ bins $\mathcal{B}_U(i,t_2')$, where $t_2\in[1:2^{nR_2'}]$. Moreover, the set $\mathcal{T}_\epsilon^n(Q_{X_1})$ is partitioned into $2^{nR'_1}$ bins $\mathcal{B}_{X_1}(t_1')$, where $t_1'\in[1:2^{nR_1'}]$. To achieve \eqref{ak_corner1}, consider the following scheme:

\par \textbf{Encoding at Encoder 1:} Upon receiving $\mathbf{x}_1$, Encoder 1 searches a pair of indices $(i,t_1')\in[1:2^{nR_V}]\times[1:2^{nR_1'}]$ such that $\big(\mathbf{x}_1,\mathbf{v}(i)\big)\in\mathcal{T}_\epsilon^n(P_{X_1,V})$ and $\mathbf{x}_1\in\mathcal{B}_{X_1}(t_1')$. A concatenation of $i$ and $t_1'$ is conveyed to the decoder. The bin index of $\mathbf{v}(i)$, i.e., the index $t_{12}\in[1:2^{nR_{12}}]$ such that $\mathbf{v}(i)\in\mathcal{B}_V(t_{12})$, is conveyed to Encoder 2 via the cooperation link. Taking
\begin{equation}
R_V>I(V;X_1)\label{ak_coop_cp1_rb1}
\end{equation}
ensures that such a codeword $\mathbf{v}(i)$ is found with high probability.

\par \textbf{Decoding at Encoder 2:} Given the source sequence $\mathbf{x}_2$ and the bin index $t_{12}$, Encoder 2 searches for an index $\hat{i}\in[1:2^{nR_V}]$ such that $\mathbf{v}(\hat{i})\in\mathcal{B}_V(t_{12})$ and $\big(\mathbf{x}_2,\mathbf{v}(\hat{i})\big)\in\mathcal{T}_\epsilon^n(P_{X_2,V})$. Reliable decoding follows by taking
\begin{equation}
R_V-R_{12}<I(V;X_2).\label{ak_coop_cp1_rb2}
\end{equation}

\textbf{Encoding at Encoder 2:} After decoding $\mathbf{v}(\hat{i})$, Encoder 2 searches for an index $j\in[1:2^{nR_U}]$, such that $\mathbf{u}(\hat{i},j)\in\mathcal{C}_U(\hat{i})$ and $\big(\mathbf{x}_2,\mathbf{v}(\hat{i}),\mathbf{u}(\hat{i},j)\big)\in\mathcal{T}_\epsilon^n(P_{X_2,V,U})$. The bin number of the chosen $\mathbf{u}(\hat{i},j)$, that is, the index $t'_2\in[1:2^{nR'_2}]$ such that $\mathbf{u}(\hat{i},j)\in\mathcal{B}_U\big(\hat{i},t_2'\big)$, is conveyed to the decoder. If
\begin{equation}
R_U>I(U;X_2|V)\label{ak_coop_cp1_rb3}
\end{equation}
then a codeword $\mathbf{u}(\hat{i},j)$ as needed is found with high probability.

\par \textbf{Decoding and Output Generation:} Upon receiving $(i,t_1')$ from Encoder 1 and $t_2'$ from Encoder 2, the decoder first identifies the codeword $\mathbf{v}(i)\in\mathcal{C}_V$ that is associated with $i$. Then it searches the bin $\mathcal{B}_{X_1}(t_1')$ for a sequence $\hat{\mathbf{x}}_1$ such that $\big(\mathbf{v}(i),\hat{\mathbf{x}}_1\big)\in\mathcal{T}_\epsilon^n(P_{X_1,V})$. A reliable lossless reconstruction of $\mathbf{x}_1$ follows provided that
\begin{equation}
R_1'>H(X_1|V).\label{ak_coop_cp1_rb4}
\end{equation}
Given $\big(\mathbf{v}(i),\hat{\mathbf{x}}_1\big)$, the decoder searches for an index $\hat{j}\in[1:2^{nR_u}]$, such that $\mathbf{u}(i,\hat{j})\in\mathcal{B}_U\big(i,t_2'\big)$ and $\big(\hat{\mathbf{x}}_1,\mathbf{v}(i),\mathbf{u}(i,\hat{j})\big)\in\mathcal{T}_\epsilon^n(P_{X_1,V,U})$. To ensure error-free decoding, we take
\begin{equation}
R_U-R_2'<I(U;X_1|V).\label{ak_coop_cp1_rb5}
\end{equation}
Finally, an output sequence $\mathbf{y}$ is generated according to $P^n_{Y|X_1=\hat{\mathbf{x}}_1,U=\mathbf{u}(i,\hat{j}),V=\mathbf{v}(i)}$. The structure of the joint PMF implies that the output sequence admits the desired coordination constraint.
\par By taking $(R_1,R_2)=(R_1'+R_V,R_2')$ and applying the Fourier-Motzkin elimination (FME) on \eqref{ak_coop_cp1_rb1}-\eqref{ak_coop_cp1_rb5}, we obtain the rate bounds
\begin{align*}
R_{12}&>I(V;X_1)-I(V;X_2)=I(V;X_1|X_2)\\
R_1&> H(X_1|V)+I(V;X_1)=H(X_1)\\
R_2&> I(U;X_2|V)-I(U;X_1|V)=I(U;X_2|X_1,V)\numberthis
\end{align*}
which imply that \eqref{ak_corner1} is achievable.

\begin{figure*}[!b]
\setcounter{equation}{38}
\hrulefill
\begin{equation}
P_{\mathbf{X}_1,\mathbf{X}_2,T_{12},T_1,T_2,\mathbf{Y}}(\mathbf{x}_1,\mathbf{x}_2,t_{12},t_1,t_2,\mathbf{y})=Q_{X_1,X_2}^n(\mathbf{x}_1,\mathbf{x}_2)\mathds{1}_{\big\{t_{12}=f_{12}(\mathbf{x}_1)\big\}\cap\big\{t_1=f_1(\mathbf{x}_1)\big\}\cap\big\{t_2=f_2(\mathbf{x}_2,t_{12})\big\}\cap\big\{\mathbf{y}=\phi(t_1,t_2)\big\}}.\label{EQ:induced_coord_PMF}
\end{equation}
\end{figure*}

\setcounter{equation}{36}
To establish the achievability of \eqref{ak_corner2} requires no binning of the codebooks $\mathcal{C}_U(i)$, where $i\in[1:2^{nR_v}]$.

\par \textbf{Encoding at Encoder 1:} Given $\mathbf{x}_1$, Encoder 1 finds $\mathbf{v}(i)\in\mathcal{C}_V$ in a similar manner and conveys its bin index $t_{12}$ to Encoder 2. Moreover, it conveys the bin index of the received $\mathbf{x}_1$, say $t_1'$, to the decoder. Again, by having \eqref{ak_coop_cp1_rb1}, such a codeword $\mathbf{v}(i)$ is found with high probability.

\par \textbf{Decoding at Encoder 2:} Performed in a similar manner as before. We again take \eqref{ak_coop_cp1_rb2} to ensure reliable decoding of $\mathbf{v}(i)$. As before, the decoded codeword is denoted by $\mathbf{v}(\hat{i})$.

\textbf{Encoding at Encoder 2:} Encoder 2 finds a codeword $\mathbf{u}(\hat{i},j)\in\mathcal{C}_U(\hat{i})$ in a manner similar to that presented in the previous scheme. Now, however, it sends to the decoder a concatenation of $\hat{i}$ and $j$. This decoding process has a vanishing probability of error if \eqref{ak_coop_cp1_rb3} holds.

\textbf{Decoding and Output Generation:} Upon receiving $t_1'$ and $(\hat{i},j)$ from Encoder 1 and 2, respectively, the decoder first finds the $\mathbf{v}(\hat{i})\in\mathcal{C}_V$ that is associated with $\hat{i}$ and the $\mathbf{u}(\hat{i},j)\in\mathcal{C}_U(\hat{i})$ that is associated with $(\hat{i},j)$. Given $\big(\mathbf{v}(\hat{i}),\mathbf{u}(\hat{i},j)\big)$, it searches the bin $\mathcal{B}_{X_1}(t_1')$ for a sequence $\hat{\mathbf{x}}_1$ such that $\big(\hat{\mathbf{x}}_1,\mathbf{v}(\hat{i}),\mathbf{u}(\hat{i},j)\big)\in\mathcal{T}_\epsilon^n(P_{X_1,V,U})$. A reliable lossless reconstruction of $\mathbf{x}_1$ is ensured provided
\begin{equation}
R_1'>H(X_1|V,U).\label{ak_coop_cp1_rb6}
\end{equation}
Finally, an output sequence $\mathbf{y}$ is generated in the same manner as in the coding scheme for \eqref{ak_corner1}.
\par Taking $(R_1,R_2)=(R_1',R_V+R_U)$ and applying FME on \eqref{ak_coop_cp1_rb1}-\eqref{ak_coop_cp1_rb3} and \eqref{ak_coop_cp1_rb6} yields the following bounds:
\begin{align*}
& R_{12}>I(V;X_1)-I(V;X_2)=I(V;X_1|X_2)\\
& R_1> H(X_1|V,U)\\
& R_2> I(V;X_1)+I(U;X_2|V).\numberthis
\end{align*}
This concludes the proof of achievability for \eqref{ak_corner2}.

\subsection{Converse}

We show that given an achievable rate triple $(R_{12},R_1,R_2)$, there exists a PMF $P_{X_1,X_2,V,U,Y}=Q_{X_1,X_2}P_{V|X_1}P_{U|X_2,V}P_{Y|X_1,U,V}$ that has $Q_{X_2}P_{Y|X_2}\mathds{1}_{\{X_1=f(Y)\}}$ as a marginal, such that the inequalities in \eqref{region_BC} are satisfied. Fix an achievable tuple $(R_{12},R_1,R_2)$ and $\delta,\epsilon>0$, and let $\mathcal{L}_n$ be the corresponding coordination code for some sufficiently large $n\in\mathbb{N}$. The joint distribution on $\mathcal{X}_1^n\times\mathcal{X}_2^n\times\mathcal{T}_{12}\times\mathcal{T}_1\times\mathcal{T}_2\times\mathcal{Y}^n$ induced by $\mathcal{L}_n$ is given in \eqref{EQ:induced_coord_PMF} at the bottom of the page. All subsequent multi-letter information measures are calculated with respect to $P_{\mathbf{X}_1,\mathbf{X}_2,T_{12},T_1,T_2,\mathbf{Y}}$ or its marginals.

\setcounter{equation}{39}
Since $(R_{12},R_1,R_2)$ is achievable, $X_1^n$ can be reconstructed at the decoder with a small probability of error. By Fano's inequality we have
\begin{align}
H(X_1^n|T_1,T_2)\leq (1+\epsilon nR)\triangleq n\epsilon_n\label{ak_Fano}
\end{align}
where $\epsilon_n=\frac{1}{n}+\epsilon R$. 

\par Next, by the structure of the single-letter PMF $P_{X_1,X_2,V,U,Y}$, we rewrite the mutual information measure in \eqref{region_AK2} as
\begin{align*}
R_2 &\geq I(U;X_2|X_1,V) \stackrel{(a)}=I(V;X_2|X_1)+I(U;X_2|X_1,V)\\
&=I(V,U;X_2|X_1)\numberthis\label{ak_coop_alt_rb1}.
\end{align*}
where (a) is because $V-X_1-X_2$ forms a Markov chain.

For the lower bound on $R_{12}$, consider
\begin{align*}
  nR_{12}&\geq H(T_{12})\\
         &\stackrel{(a)}\geq I(T_{12};X_1^n|X_2^n)\\
         &= \sum_{i=1}^n I(T_{12};X_{1,i}|X_{1,i+1}^n,X_2^{n\backslash i},X_{2,i})\\
         &\stackrel{(b)}= \sum_{i=1}^n I(T_{12},X_{1,i+1}^n,X_2^{n\backslash i};X_{1,i}|X_{2,i})\\
         &\geq \sum_{i=1}^n I(T_{12},X_{1,i+1}^n,X_2^{i-1};X_{1,i}|X_{2,i})\\
         &\stackrel{(c)}= \sum_{i=1}^n I(V_i;X_{1,i}|X_{2,i})\numberthis\label{ak_coop_rb12}
\end{align*}
where (a) is because $T_{12}$ is determined by $X_1^n$ and since conditioning cannot increase entropy, (b) is since $(X_1^n,X_2^n)$ are pairwise i.i.d., and (c) defines $V_i\triangleq(T_{12},X_{1,i+1}^n,X_2^{i-1})$, for every $i\in[1:n]$.

\par Next, for $R_1$ we have
\begin{align*}
  nR_1&\geq H(T_1)\\
      &\geq H(T_1|T_{12},T_2)\numberthis\\
      &\stackrel{(a)}= I(T_1;X_1^n|T_{12},T_2)\\
      &= H(X_1^n|T_{12},T_2)-H(X_1^n|T_{12},T_1,T_2)\\
      &\stackrel{(b)} \geq \sum_{i=1}^nH(X_{1,i}|T_{12},T_2,X_{1,i+1}^n)-n\epsilon_n\\
      &\geq \sum_{i=1}^nH(X_{1,i}|T_{12},T_2,X_{1,i+1}^n,X_2^{i-1})-n\epsilon_n\\
      &\stackrel{(c)}= \sum_{i=1}^nH(X_{1,i}|V_i,U_i)-n\epsilon_n  \numberthis \label{ak_coop_rb1}
\end{align*}
where (a) is because $T_1$ is determined by $X_1^n$, (b) uses \eqref{ak_Fano} and the mutual information chain rule, while in (c) we define $U_i\triangleq T_2$, for every $i\in[1:n]$, and use the definition of $V_i$.

\par To bound $R_2$ consider
\begin{align*}
  nR_2&\geq H(T_2)\\
      &\geq H(T_2|X_1^n)\\
      &\geq I(T_2;X_2^n|X_1^n)\\
      &= \sum_{i=1}^n I(T_2;X_{2,i}|X_1^{n\backslash i},X_2^{i-1},X_{1,i})\\
      &\stackrel{(a)}= \sum_{i=1}^n I(T_2,X_1^{n\backslash i},X_2^{i-1};X_{2,i}|X_{1,i})\\
      &\stackrel{(b)}= \sum_{i=1}^n I(T_2,T_{12},X_1^{n\backslash i},X_2^{i-1};X_{2,i}|X_{1,i})\\
      &\stackrel{(c)}\geq \sum_{i=1}^n I(V_i,U_i;X_{2,i}|X_{1,i})     \numberthis \label{ak_coop_rb2}
\end{align*}
where (a) is because $(X_1^n,X_2^n)$ are pairwise i.i.d., (b) is because $T_{12}$ is determined by $X_1^n$, while (c) follows since conditioning cannot increase entropy and from the definitions of $V_i$ and $U_i$.

\par For the sum of rates, we have
%\begin{equation}
  %n(R_1+R_2)\geq H(T_1,T_2)=H(T_2)+H(T_1|T_2).\label{ak_coop_sum}
%\end{equation}
%For the first term in \eqref{ak_coop_sum}, we have
\begin{align*}
&n(R_1+R_2)\\
          &\geq H(T_1,T_2)\\
          &\stackrel{(a)}= I(T_1,T_2;X_1^n,X_2^n)\\    
          &\stackrel{(b)}= H(X_1^n)-H(X_1^n|T_1,T_2)+I(T_{12},T_2;X_2^n|X_1^n)\\    
          &\stackrel{(c)}\geq \sum_{i=1}^n\Big[H(X_{1,i})+I(T_{12},T_2,X_1^{n\backslash i},X_2^{i-1};X_{2,i}|X_{1,i})\Big]\mspace{-3mu}-\mspace{-3mu}n\epsilon_n\\
          &\geq \sum_{i=1}^n\Big[H(X_{1,i})\mspace{-3mu}+\mspace{-3mu}I(T_{12},T_2,X_{1,i+1}^n,X_2^{i-1};X_{2,i}|X_{1,i})\Big]\mspace{-3mu}-\mspace{-3mu}n\epsilon_n\\
          &\stackrel{(d)}= \sum_{i=1}^n\Big[H(X_{1,i}|V_i,U_i)+I(V_i,U_i;X_{1,i},X_{2,i})\Big]-n\epsilon_n\numberthis\label{ak_coop_rb1+2}
\end{align*}
where:\\
(a) is because $(T_1,T_2)$ are determined by $(X_1^n,X_2^n)$;\\
(b) is since $X_1^n$ defines $(T_{12},T_1)$;\\
(c) uses \eqref{ak_Fano}, the mutual information chain rule and the pairwise i.i.d. nature of $(X_1^n,X_2^n)$;\\
(b) uses the mutual information chain rule and the definition of $(V_i,U_i)$.

\par The upper bounds in \eqref{ak_coop_rb12}, \eqref{ak_coop_rb1}, \eqref{ak_coop_rb2} and \eqref{ak_coop_rb1+2} are rewritten by introducing a time-sharing random variable $T$ that is independent of $(X_1^n,X_2^n,T_{12},T_1,T_2,Y^n)$ and is uniformly distributed over $[1:n]$. The rate bound on $R_{12}$ is rewritten as
\begin{align}
R_{12}&\geq \frac{1}{n}\sum_{t=1}^n I(V_t;X_{1,t}|X_{2,t},T=t)\\
      &=\sum_{t=1}^n \mathbb{P}\big(T=t\big)I(V_t;X_{1,t}|X_{t,q},T=t)\\
      &=I(V_T;X_{1,T}|X_{2,T},T)\\
      &\stackrel{(a)}=I(V_T,T;X_{1,T}|X_{2,T})\label{ak_coop_rb12_Q}
\end{align}
where (a) follows because $T$ is independent of the pair $(X_{1,T},X_{2,T})$ (see property 1 in \cite[Section IIV-B]{Cuff_Permuter_Cover_Coordination}).
By rewriting \eqref{ak_coop_rb1}, \eqref{ak_coop_rb2} and \eqref{ak_coop_rb1+2} in an analogous manner, the region obtained is convex. This follows from the presence of the time-sharing random variable $T$ in the conditioning of all the mutual information and entropy terms.

Next, define $X_1\triangleq X_{1,T}$, $X_2\triangleq X_{2,T}$, $V\triangleq(V_T,T)$, $U\triangleq U_T$ and $Y\triangleq Y_T$. Notice that $(X_1,X_2)\sim Q_{X_1,X_2}$ and then use the time-mixing property from \cite[Section IIV-B, Property 2]{Cuff_Permuter_Cover_Coordination} to get
\begin{align*}
R_{12} &\geq I(V;X_1|X_2)\\
R_1 &\geq H(X_1|V,U)-\epsilon_n\\
R_2 &\geq I(V,U;X_2|X_1)\\
R_1+R_2 &\geq H(X_1|V,U)+I(V,U;X_1,X_2)-\epsilon_n.\numberthis\label{single_letter_r1+r2}
\end{align*}

To complete the converse, the following Markov relations must be shown to hold.
\begin{subequations}
\begin{align}
&V-X_1-X_2\label{ak_markov1}\\
&U-(X_2,V)-X_1\label{ak_markov2}\\
&Y-(X_1,U,V)-X_2.\label{ak_markov3}
\end{align}\label{ak_markov}
\end{subequations}
We prove that the Markov relations in \eqref{ak_markov} hold for every $t\in[1:n]$. Upon doing so, showing that the relations hold in their single-letter (as stated in \eqref{ak_markov}) is straightforward.

\begin{figure*}[!b]
\setcounter{equation}{64}
\hrulefill
\begin{equation}
\Big(\mathbf{v}(m_{10},m_{20}),\mathbf{u}_1(m_{10},m_{20},m_1,i_1),\mathbf{u}_2(m_{10},m_{20},m_{22},i_2)\Big)\in\mathcal{T}_\epsilon^n(P_{V,U_1,U_2}).\label{bc_achiev_encoder_typicality}
\end{equation}
\end{figure*}

\setcounter{equation}{57}

\par For \eqref{ak_markov1}, recall that $V_t=(T_{12},X_{1,t+1}^n,X_2^{t-1})$, for every $t\in[1:n]$, and consider
\begin{align*}
    0 &\leq I(T_{12},X_{1,t+1}^n,X_2^{t-1};X_{2,t}|X_{1,t})\\
      &\stackrel{(a)}\leq I(X_1^{n\backslash t},X_2^{t-1};X_{2,t}|X_{1,t})\stackrel{(b)}=0
\end{align*}
where (a) is because conditioning cannot increase entropy and sicen $T_{12}$ is determined by $X_1^n$, while (b) uses the pairwise i.i.d. nature of $(X_1^n,X_2^n)$. Thus \eqref{ak_markov1} holds.

\par To establish \eqref{ak_markov2}, we use Lemma 1 in \cite{Kaspi85}. Since $U_t=T_2$, for every $t\in[1:n]$, we have
\begin{align*}
    0 &\leq I(T_2;X_{1,t}|X_{2,t},T_{12},X_{1,t+1}^n,X_2^{t-1})\\
      &\leq I(T_2;X_{1,t},X_1^{t-1}|X_{2,t},T_{12},X_{1,t+1}^n,X_2^{t-1})\numberthis\label{ak_coop_markov21}.
\end{align*}
Set
\begin{align*}
    &A_1=X_{1,t+1}^n\ ,\quad A_2=(X_{1,t},X_1^{t-1}),\\
    &B_1=X_{2,t+1}^n\ ,\quad B_2=(X_{2,t},X_2^{t-1}).
\end{align*}
Accordingly, \eqref{ak_coop_markov21} is rewritten as
\begin{equation}
0 \leq I(T_2;A_2|T_{12},A_1,B_2).\label{ak_coop_markov22}
\end{equation}
Noting that $(A_1,A_2)$ and $(T_{12},B_1,B_2)$ determine $T_{12}$ and $T_2$, respectively, and that $P_{A_1,A_2,B_1,B_2}=P_{A_1,B_1}P_{A_2,B_2}$. The result of \cite[Lemma 1, Conclusion 2]{Kaspi85} thus implies
\begin{equation}
    0 \leq I(T_2;A_2|T_{12},A_1,B_2)=0 \label{ak_coop_markov23}
\end{equation}
which establishes \eqref{ak_markov2}.

\par For \eqref{ak_markov3} note the the structure of the joint PMF from \eqref{EQ:induced_coord_PMF} implies that for any $i\in[1:n]$, the marginal distribution of $(X_1^n,X_2^n,T_2,T_{12},Y^n)$ factors as:
\begin{align*}
P(&x_1^n,x_2^n,t_2,t_{12},y^n)\\&=P(x_2^{i-1})P(x_{1,i},x_{2,i})P(x_{1,i+1}^n,x_{2,i+1}^n)\\&\times\mathds{1}_{\big\{t_2=f_2(x_2^n,t_{12})\big\}}P(x_1^{i-1},t_{12},y^n|x_2^{i-1},x_{1,i}^n,t_2).\numberthis
\end{align*}
Consequently, by further marginalizing over $X_1^{i-1}$, we get
\begin{align*}
&P(x_{1,i}^n,x_2^n,t_2,t_{12},y^n)\\&=P(x_2^{i-1})P(x_{1,i},x_{2,i})P(x_{1,i+1}^n,x_{2,i+1}^n)\mathds{1}_{\big\{t_2=f_2(x_2^n,t_{12})\big\}}\\&\times P(t_{12}|x_2^{i-1},x_{1,i}^n,t_2)P(y^n|x_2^{i-1},x_{1,i},x_{1,i+1}^n,t_2,t_{12}).\numberthis
\end{align*}
The structure of the conditional distribution of $Y^n$ given $(X_{1,i}^n,X_2^n,T_2,T_{12})$ implies that 
\begin{equation}
    Y^n-(T_2,T_{12},X_{1,i+1}^n,X_2^{i-1},X_{1,i})-(X_{2,i},X_{2,i+1}^n)
\end{equation}
forms a Markov chain, and in particular we have
\begin{equation}
    Y_i-(T_2,T_{12},X_{1,i+1}^n,X_2^{i-1},X_{1,i})-X_{2,i}
\end{equation}
for every $i\in[1:n]$. Taking $\delta,\epsilon\to 0$ and $n\to\infty$ concludes the converse.

\section{Proof of Theorem \ref{tm_bc_capacity}}\label{appen_proof_tm_bc}

%%%%%%%%%%%%%%%%%%%%%%%%%%%%%%%%%%%%%%%%%%%%%%%%%%%%%%%%%%%%%%%%%%%%%%%%%%%%%%%%%%%%%%%%%%%%%%%%%%%%%%%%%%%%%%%%%%%
%%%%%%%%%%%%%%%                              Achievability                                %%%%%%%%%%%%%%%%%%%%%%%%%
%%%%%%%%%%%%%%%%%%%%%%%%%%%%%%%%%%%%%%%%%%%%%%%%%%%%%%%%%%%%%%%%%%%%%%%%%%%%%%%%%%%%%%%%%%%%%%%%%%%%%%%%%%%%%%%%%%%

\subsection{Achievability}\label{SUBAPPEN:achievability}

To establish achievability, we show that for any fixed $\epsilon>0$, a PMF
\begin{equation}
P_{V,U,Y_1}P_{X|V,U,Y_1}Q_{Y_2|X}
\end{equation}
for which $Y_1=f(X)$, and a rate triple $(R_{12},R_1,R_2)$ that satisfies \eqref{region_BC},
there is a sufficiently large $n\in\mathbb{N}$ and a corresponding $(n,R_{12},R_1,R_2)$ code $\mathcal{C}_n$, such that $e(\mathcal{C}_n)\leq \epsilon$. We first derive an achievable region for a general BC with a one-sided conferencing link between the decoders with a channel transition matrix $Q_{Y_1,Y_2|X}$. The region is described using three auxiliaries (rather than two). Then, by a proper choice of the auxiliaries, we achieve $\mathcal{C}_{\mathrm{BC}}$. Fix a PMF
\begin{equation}
P_{V,U_1,U_2,X,Y_1,Y_2}=P_{V,U_1,U_2,X}Q_{Y_1,Y_2|X}\label{input_joint_dist}
\end{equation}
and an $\epsilon>0$, and consider the following coding scheme.

%%%%%%%%%%%%%%%%%%%%%%%%%%%%%%%            Codebook Generation           %%%%%%%%%%%%%%%%%%%%%%%%%%%%%%%%%%%%%%%%%%%
%%%%%%%%%%%%%%%%%%%%%%%%%%%%%%%%%%%%%%%%%%%%%%%%%%%%%%%%%%%%%%%%%%%%%%%%%%%%%%%%%%%%%%%%%%%%%%%%%%%%%%%%%%%%%%%%%%%%

\textbf{Codebook Generation:} Split each message $m_j$, $j=1,2$, into two sub-messages denoted by $(m_{j0},m_{jj})$. The pair $m_0\triangleq(m_{10},m_{20})$ is referred to as a \emph{public message} while $m_{jj}$ serve as \emph{private message} $j$. The rates associated with $m_{j0}$ and $m_{jj}$, $j=1,2$, are denoted by $R_{j0}$ and $R_{jj}$, while the corresponding alphabets are $\mathcal{M}_{j0}$ and $\mathcal{M}_{jj}$, respectively. Accordingly, we have
\begin{equation}
R_j=R_{j0}+R_{jj},\ \ j=1,2.\label{bc_achiev_partial_rates}
 \end{equation}
We also denote $R_0\triangleq R_{10}+R_{20}$ and $\mathcal{M}_0\triangleq\hat{\hat{m}}_0$. The random variables $M_0$ and $M_{jj}$, for $j=1,2$, are associated with the public message and private message $j$, respectively. Furthermore, $M_0$, $M_{11}$ and $M_{22}$ are independent and uniform over $\mathcal{M}_0$, $\mathcal{M}_{11}$ and $\mathcal{M}_{22}$, respectively.

%-----------------------------          Public Message Codebook          -------------------------------------------
%-------------------------------------------------------------------------------------------------------------------

Partition $\mathcal{M}_0$ into $2^{nR_{12}}$ equal-sized bins $\mathcal{B}(m_{12})$, where $m_{12}\in\mathcal{M}_{12}$. Generate a public message codebook, denoted by $\mathcal{C}_V$, that comprises $2^{nR_0}$ $v$-codewords $\mathbf{v}(m_0)$, $(m_0)\in\mathcal{M}_0$, each drawn according to $P_V^n$ independent of all the other $v$-codewords. 

%-----------------------------          Private Message 1 Codebook       -------------------------------------------
%-------------------------------------------------------------------------------------------------------------------

For each $\mathbf{v}(m_0)\in\mathcal{C}_V$, generate two codebooks $C_{U_j}(m_0)$, $j=1,2$, each comprises $2^{n(R_{jj}+R_j')}$ codewords $\mathbf{u}_j$ that are independently drawn according to $P^n_{U_j|V=\mathbf{v}(m_0)}$. The $u_j$-codewords in $C_{U_j}(m_0)$ are labeled as $\mathbf{u}_j(m_0,m_{jj},i_j)$, where $(m_{jj},i_j)\in\mathcal{M}_{jj}\times\mathcal{I}_j$ and $\mathcal{I}_j=[1:2^{nR'_j}]$. Based on this labeling, the codebook $C_{U_j}(m_0)$ has a $u_j$-bin associated with every $m_{jj}\in\mathcal{M}_{jj}$, each containing $2^{nR_j'}$ $u_j$-codewords.

%
%%-----------------------------          Private Message 2 Codebook       -------------------------------------------
%%-------------------------------------------------------------------------------------------------------------------
%
%For each $\mathbf{v}(m_0)\in\mathcal{C}_V$ also generate a codebook $C_{U_2}(m_0)$ that comprises $2^{nR_{22}}$ $u_2$-codewords, each associated with a private message $m_{22}\in\mathcal{M}_{22}$. Each $u_2$-codeword in $C_{U_2}(m_{20})$ is drawn according to $Q_{U_2|V}^n\big(\mathbf{u}_2\big|\mathbf{v}(m_{20})\big)$ independently of all the other $u_2$-codewords.
%
%
%\par The structure of the codebook, while overlooking the binning of $\mathcal{C}_V$ for simplicity, is illustrated in Fig. \ref{fig_codebook}.
%%%%%%%%%%%%%%%%%%%%%%%%%%%%%%%%%%%%%%%%%%%%%%%%%%%%%%%%%%%%%%%%%%%%%%%%%%%%%%%%%%%%%%%%%%%%%%%%%%%%%%%%%%%%%%%%%%%
%%%%%%%%%%%%%%%%%%%%%%%%%%%%%%%%%%%%%                 Encoding                %%%%%%%%%%%%%%%%%%%%%%%%%%%%%%%%%%%%%
%%%%%%%%%%%%%%%%%%%%%%%%%%%%%%%%%%%%%%%%%%%%%%%%%%%%%%%%%%%%%%%%%%%%%%%%%%%%%%%%%%%%%%%%%%%%%%%%%%%%%%%%%%%%%%%%%%%

\par\textbf{Encoding:} To transmit the message pair $(m_1,m_2)=\big((m_{10},m_{11}),(m_{20},m_{22})\big)$, the encoder searches for a pair $(i_1,i_2)\in\mathcal{I}_1\times\mathcal{I}_2$ that satisfies \eqref{bc_achiev_encoder_typicality} on the bottom of the page, where $\mathbf{v}(m_0)\in\mathcal{C}_V$ and $\mathbf{u}_j(m_0,m_j,i_j)\in\mathcal{C}_{U_j}(m_0)$, for $j=1,2$. If the set of appropriate index pairs contains more than one element, the encoder chooses the component-wise minimal pair; if the set is empty, the encoder sets $(i_1,i_2)=(1,1)$. The channel input sequence $\mathbf{x}$ is then randomly generated according to $P^n_{X|V,U_1,U_2}$ and is transmitted over the channel.

%%%%%%%%%%%%%%%%%%%%%%%%%%%%%%%%%%%%%             Decoding            %%%%%%%%%%%%%%%%%%%%%%%%%%%%%%%%%%%%%%%%%%%%%%
%%%%%%%%%%%%%%%%%%%%%%%%%%%%%%%%%%%%%%%%%%%%%%%%%%%%%%%%%%%%%%%%%%%%%%%%%%%%%%%%%%%%%%%%%%%%%%%%%%%%%%%%%%%%%%%%%%%%

\begin{figure*}[!b]
\setcounter{equation}{69}
\hrulefill
\begin{align*}
P_{M_1,M_2,\mathbf{X},\mathbf{Y}_1,\mathbf{Y}_2,M_{12},\hat{M}_1,\hat{M}_2}(m_1,m_2,\mathbf{x},\mathbf{y}_1,\mathbf{y}_2,m_{12},\hat{m}_1,\hat{m}_2)=&2^{-n(R_1+R_2)}\mathds{1}_{\big\{\mathbf{x}=g(m_1,m_2)\big\}\cap\left\{\bigcap_{i=1}^n\big(y_{1,i}=f(x_i)\big)\right\}}Q_{Y_2|X}^n(\mathbf{y}_2|\mathbf{x})\\&\mspace{25mu}\times\mathds{1}_{\big\{m_{12}=g_{12}(\mathbf{y}_1)\big\}\cap\big\{\hat{m}_1=\psi_1(\mathbf{y}_1)\big\}\cap\big\{\hat{m}_2=\psi_2(\mathbf{y}_2,m_{12})\big\}}.\numberthis\label{EQ:induced_bc_PMF}
\end{align*}
\end{figure*}

\setcounter{equation}{65}
%------------------------------------             Decoder 1              -------------------------------------------
%-------------------------------------------------------------------------------------------------------------------

\par\textbf{Decoding and Cooperation:} \underline{Decoder 1:} Searches for a unique pair $(\hat{m}_0,\hat{m}_{11})\in\mathcal{M}_0\times\mathcal{M}_{11}$ for which there is an index $\hat{i}_1\in\mathcal{I}_1$, such that
\begin{equation}
\Big(\mathbf{v}(\hat{m}_0),\mathbf{u}_1(\hat{m}_0,\hat{m}_{11},\hat{i}_1),\mathbf{y}_1\Big)\in\mathcal{T}_\epsilon^n(P_{V,U_1,Y_1})\label{bc_achiev_decoder1} \end{equation}
where $\mathbf{v}(\hat{m}_0)\in\mathcal{C}_V$ and $\mathbf{u}_1(\hat{m}_0,\hat{m}_{11},\hat{i}_1)\in\mathcal{C}_{U_1}(\hat{m}_0)$. If such a unique triple is found, then $\hat{m}_1=\big(\hat{m}_{10},\hat{m}_{11}\big)$ is declared as the decoded message; otherwise, an error is declared.

%----------------------------------             Cooperation              -------------------------------------------
%-------------------------------------------------------------------------------------------------------------------

\noindent\underline{Cooperation:} Given $(\hat{m}_0,\hat{m}_{11},\hat{i}_1)$, Decoder 1 conveys the bin number of $\hat{m}_0$ to Decoder 2 via the cooperation link. Namely, Decoder 1 shares with Decoder 2 the index $\hat{m}_{12}\in\mathcal{M}_{12}$, such that $\hat{m}_0\in\mathcal{B}(\hat{m}_{12})$.

%------------------------------------             Decoder 2              -------------------------------------------
%-------------------------------------------------------------------------------------------------------------------

\noindent\underline{Decoder 2:} Upon receiving $\hat{m}_{12}$ from Decoder 1 and $\mathbf{y}_2$ from the channel, Decoder 2 searches for a unique pair $(\hat{\hat{m}}_0,\hat{\hat{m}}_{22})\in\mathcal{M}_0\times\mathcal{M}_{22}$ for which there is an $\hat{\hat{i}}_2\in\mathcal{I}_2$, such that
\begin{equation}
\Big(\mathbf{v}(\hat{\hat{m}}_0),\mathbf{u}_2(\hat{\hat{m}}_0,\hat{\hat{m}}_{22},\hat{\hat{i}}_2),\mathbf{y}_2\Big)\in\mathcal{T}_\epsilon^n(P_{V,U_2,Y_2})\label{bc_achiev_decoder2}
\end{equation}
where $\hat{\hat{m}}_0\in\mathcal{B}(\hat{m}_{12})$, $\mathbf(\hat{\hat{m}}_0)\in\mathcal{C}_V$ and $\mathbf{u}_2(\hat{\hat{m}}_0,\hat{\hat{m}}_{22},\hat{\hat{i}}_2)\in\mathcal{C}_{U_2}(\hat{\hat{m}}_0)$. If such a unique triple is found, then $\hat{\hat{m}}_2\triangleq(\hat{\hat{m}}_{20},\hat{\hat{m}}_{22})$ is declared as the decoded message; otherwise, an error is declared.

\par By standard error probability analysis (see Appendix \ref{appen_analysis}) and existence arguments, an $(n,R_{12},R_1,R_2)$ code $\mathcal{C}_n$ that achieves reliability is extracted provided that
\begin{align*}
R'_1+R'_2 &> I(U_1;U_2|V)\\
R_{11}+R'_1 &< I(U_1;Y_1|V)\\
R_{20}+R_1+R'_1 &< I(V,U_1;Y_1)\\
R_{22}+R'_2 &< I(U_2;Y_2|V)\\
R_{10}+R_2+R'_2-R_{12}&<I(V,U_2;Y_2).\numberthis\label{EQ:bc_achiev_rb}
\end{align*}
Applying FME on \eqref{EQ:bc_achiev_rb} while using \eqref{bc_achiev_partial_rates} yields the rate bounds
\begin{align*}
R_1 &< I(V,U_1;Y_1)\\
R_2 &< I(V,U_2;Y_2)\mspace{-2mu}+\mspace{-2mu}R_{12}\\
R_1+R_2 &< I(V,U_1;Y_1)\mspace{-2mu}+\mspace{-2mu}I(U_2;Y_2|V)\mspace{-2mu}-\mspace{-2mu}I(U_1;U_2|V)\\
R_1+R_2 &< I(U_1;Y_1|V)\mspace{-2mu}+\mspace{-2mu}I(V,U_2;Y_2)\mspace{-2mu}-\mspace{-2mu}I(U_1;U_2|V)\mspace{-2mu}+\mspace{-2mu}R_{12}.\numberthis\label{EQ:region_inner}
\end{align*}
By setting $U_1=Y_1$ and $U_2=U$, the bounds in \eqref{EQ:region_inner} reduce to \eqref{region_BC}. Note that this choice of auxiliaries is valid as they satisfy the Markov relations stated in Theorem \ref{tm_bc_capacity}. This shows that $\mathcal{C}_{\mathrm{BC}}$ is achievable.

\begin{remark}
The cooperation protocol described in the proof is reminiscent of the WZ coding technique. The cooperation link is used to convey a \emph{bin} of the common message codeword $\mathbf{v}$ (rather than the codeword itself) from 1st decoder to the 2nd. As part of the joint typicality decoding rule in \eqref{bc_achiev_decoder2}, the channel input $\mathbf{y}_2$ is used as correlated side information to isolate the actual $v$-codeword from the bin. This correlation is induced from the channel transition probability and the underlying Markov relations (with respect to the PMF in \eqref{input_joint_dist}). %The resemblance to the WZ scheme is also reflected in the rate bound in (\ref{rb5}).
\end{remark}
%%%%%%%%%%%%%%%%%%%%%%%%%%%%%%%%%%%%%%%%%%%%%%%%%%%%%%%%%%%%%%%%%%%%%%%%%%%%%%%%%%%%%%%%%%%%%%%%%%%%%%%%%%%%%%%%%%%
%%%%%%%%%%%%%%%                                Converse                                   %%%%%%%%%%%%%%%%%%%%%%%%%
%%%%%%%%%%%%%%%%%%%%%%%%%%%%%%%%%%%%%%%%%%%%%%%%%%%%%%%%%%%%%%%%%%%%%%%%%%%%%%%%%%%%%%%%%%%%%%%%%%%%%%%%%%%%%%%%%%%

\subsection{Converse}\label{proof_bc_converse}

\par We show that if a rate triple $(R_{12},R_1,R_2)$ is achievable, then there exists a PMF $P_{V,U,Y_1,X}Q_{Y_2|X}$ for which $Y_1=f(X)$, such that the inequalities in \eqref{region_BC} are satisfied. Fix an achievable tuple $(R_{12},R_1,R_2)$ and an $\epsilon>0$, and let $\mathcal{C}_n$ be the corresponding $(n,R_{12},R_1,R_2)$ code for some sufficiently large $n\in\mathbb{N}$. The joint distribution on $\mathcal{M}_1\times\mathcal{M}_2\times\mathcal{X}^n\times\mathcal{Y}_1^n\times\mathcal{Y}_2^n\times\mathcal{M}_{12}\times\mathcal{M}_1\times\mathcal{M}_2$ induced by $\mathcal{C}_n$ is given in \eqref{EQ:induced_bc_PMF} at the bottom of the page. All subsequent multi-letter information measures are calculated with respect to the PMF from \eqref{EQ:induced_bc_PMF} or its marginals.
\setcounter{equation}{70}

Since $e(\mathcal{C}_n)\leq\epsilon$, Fano's inequality gives
\begin{subequations}
\begin{align}
H(M_1|Y_1^n)&\leq 1+\epsilon nR_1\triangleq n\epsilon_n^{(1)}\label{bc_converse_Fano1}\\
H(M_2|M_{12},Y_2^n)&\leq 1+\epsilon nR_2\triangleq n\epsilon_n^{(2)}\label{bc_converse_Fano2}
\end{align}
where $\epsilon_n^{(j)}\triangleq \frac{1}{n}+\epsilon R_j$, for $j=1,2$. Define
\begin{equation}
\epsilon_n=\max\big\{\epsilon_n^{(1)},\epsilon_n^{(2)}\big\}.
\end{equation}\label{bc_converse_Fano}
\end{subequations}

It follows that
\begin{align*}
  nR_1&=H(M_1)\\
      &\stackrel{(a)}\leq I(M_1;Y_1^n)+n\epsilon_n\\
      &\stackrel{(b)}=I(X^n;Y_1^n)+n\epsilon_n\\
      &\stackrel{(c)}\leq\sum_{i=1}^nH(Y_{1,i})+n\epsilon_n\numberthis \label{bc_converse_r1UB}
\end{align*}
where (a) uses \eqref{bc_converse_Fano}, (b) is by the Markov chain $M_1-X^n-Y_1^n$ and the Data Processing Inequality, while (c) follows because $Y_1^n$ is a function of $X^n$ and since conditioning cannot increase entropy.

%\par For the upper bound on $R_1$, consider
%\begin{align*}
%  nR_1&=H(M_1)\numberthis \label{bc_converse_r1UB1}\\
%      &=H(M_1|M_2)\\
%      &\stackrel{(a)}\leq I(M_1;Y_1^n|M_2)+n\epsilon_n\\
%      &\stackrel{(b)}=H(Y_1^n|M_2)-H(Y_1^n|M_1,M_2,X^n)+n\epsilon_n\numberthis \label{bc_converse_r1UB2}\\
%      &\stackrel{(c)}=\sum_{i=1}^nH(Y_{1,i}|M_2,Y_1^{i-1})+n\epsilon_n\\
%      &\stackrel{(d)}=\sum_{i=1}^nH(Y_{1,i}|B_i,C_i)+n\epsilon_n
%\end{align*}
%where:\\
%  (a) follows from (\ref{bc_converse_Fano1}) and (\ref{bc_converse_Fano});\\
%  (b) follows because $X^n$ is a deterministic function of $(M_1,M_2)$;\\
%  (c) follows because $Y_1^n$ is a deterministic function of $X^n$;\\
%  (d) follows by defining $B_i\triangleq M_2$ and from the definition of $C_i$.

\par To bound $R_2$ consider
\begin{align*}
  nR_2&= H(M_2)\numberthis \label{bc_converse_r2UB1}\\
      &\stackrel{(a)}\leq I(M_2;M_{12},Y_2^n)+n\epsilon_n\\
      &= I(M_2;Y_2^n|M_{12})+I(M_2;M_{12})+n\epsilon_n\numberthis \label{bc_converse_r2UB2}\\
      &\stackrel{(b)}\leq \sum_{i=1}^n I(M_2;Y_{2,i}|M_{12},Y_{2,i+1}^n)+nR_{12}+n\epsilon_n\\
      &\stackrel{(c)}\leq \sum_{i=1}^n I(V_i,U_i;Y_{2,i})+nR_{12}+n\epsilon_n\numberthis\label{bc_converse_r2UB}
\end{align*}
where (a) uses \eqref{bc_converse_Fano}, (b) is because a uniform distribution maximizes entropy, while (c) defines $V_i\triangleq (M_{12},Y_1^{i-1},Y_{2,i+1}^n)$ and $U_i\triangleq M_2$, for every $i\in[1:n]$.

\begin{figure*}[!b]
\setcounter{equation}{85}
\hrulefill
\begin{align*}
\mathcal{E}= \bigcap_{(\hat{i}_1,\hat{i}_2)\in\mathcal{I}_1\times\mathcal{I}_2}\Big\{\big(\mathbf{V}(1),\mathbf{U}_1(1,1,\hat{i}_1),\mathbf{U}_2(1,1,\hat{i}_2)\big)\notin\mathcal{T}_\epsilon^n(P_{V,U_1,U_2})\Big\}\numberthis\label{EQ:analysis_event0}.
\end{align*}
\end{figure*}

\setcounter{equation}{75}

\par For the sum of rates, we first write
\begin{equation}
  n(R_1+R_2)= H(M_1,M_2)=H(M_2)+H(M_1|M_2).\label{bc_converse_UBsum}
\end{equation}
By the independence of $M_1$ and $M_2$ and by \eqref{bc_converse_Fano}, we have
\begin{equation}
H(M_1|M_2)\leq H(Y_1^n|M_2)+n\epsilon_n.\label{EQ:bc_converse_r1UB3}
\end{equation}
Moreover, we bound $H(M_2)$ as
\begin{align*}
H(M_2)&\stackrel{(a)}\leq I(M_2;Y_2^n|M_{12})+I(M_2;M_{12})+n\epsilon_n\\
      &\begin{multlined}[b][.85\columnwidth]\stackrel{(b)}= \sum_{i=1}^n \Big[ I(M_2;Y_{2,i}^n|M_{12},Y_1^{i-1})\\-I(M_2;Y_{2,i+1}^n|M_{12},Y_1^i)\Big]+I(M_2;M_{12})+n\epsilon_n\end{multlined}\\
      &\begin{multlined}[b][.85\columnwidth]\stackrel{(c)}= \sum_{i=1}^n \Big[ I(M_2;Y_{2,i+1}^n|M_{12},Y_1^{i-1})+I(U_i;Y_{2,i}|V_i)\\-I(M_2;Y_{1,i},Y_{2,i+1}^n|M_{12},Y_1^{i-1})\\+I(M_2;Y_{1,i}|M_{12},Y_1^{i-1})\Big]+I(M_2;M_{12})+n\epsilon_n
      \end{multlined}\\
      &\begin{multlined}[b][.85\columnwidth]= \sum_{i=1}^n \Big[I(U_i;Y_{2,i}|V_i)\\-I(M_2;Y_{1,i}|M_{12},Y_1^{i-1},Y_{2,i+1}^n)\Big]\\+I(M_2;M_{12},Y_1^n)+n\epsilon_n
      \end{multlined}\\
      &\begin{multlined}[b][.85\columnwidth]\stackrel{(d)}= \sum_{i=1}^n \Big[ I(U_i;Y_{2,i}|V_i)-I(U_i;Y_{1,i}|V_i)\Big]\\+I(M_2;Y_1^n)+n\epsilon_n
      \end{multlined}\numberthis\label{bc_converse_r2UB3}
\end{align*}
where:\\
  (a) is by repeating steps \eqref{bc_converse_r2UB1}-\eqref{bc_converse_r2UB2} in the upper bounding of $R_2$;\\
  (b) uses a telescoping identity \cite[Eq. (9) and (11)]{Kramer_telescopic2011};\\
  (c) uses the definitions of $V_i$ and $U_i$;\\
  (d) again uses the definition of $V_i$ and $U_i$ (second term) and the Markov relation $M_{12}-Y_1^n-M_2$ (third term).

\par Inserting \eqref{EQ:bc_converse_r1UB3} and \eqref{bc_converse_r2UB3} into \eqref{bc_converse_UBsum} results in
\begin{align*}
&n(R_1+R_2)\\
&\leq \sum_{i=1}^n \Big[I(U_i;Y_{2,i}|V_i)\mspace{-2mu}-\mspace{-2mu}I(U_i;Y_{1,i}|V_i)\Big]\mspace{-2mu}+\mspace{-2mu}H(Y_1^n)\mspace{-2mu}+\mspace{-2mu}2n\epsilon_n\numberthis\label{bc_converse_sumUB_temp}\\
&\leq \sum_{i=1}^n \Big[H(Y_{1,i}|V_i,U_i)\mspace{-2mu}+\mspace{-2mu}I(U_i;Y_{2,i}|V_i)\mspace{-2mu}+\mspace{-2mu}I(V_i;Y_{1,i})\Big]\mspace{-2mu}+\mspace{-2mu}2n\epsilon_n.\numberthis\label{bc_converse_sumUB}
\end{align*}

Finally, note that
\begin{align*}
&H(Y_1^n)-\sum_{i=1}^nH(Y_{1,i}|V_i)\\
&\stackrel{(a)}=\sum_{i=1}^nI(Y_{2,i+1}^n;Y_{1,i}|M_{12},Y_1^{i-1})+I(M_{12};Y_1^n)\\
&\stackrel{(b)}\leq\sum_{i=1}^nI(Y_1^{i-1};Y_{2,i}|M_{12},Y_{2,i+1}^n)+H(M_{12})\\
&\stackrel{(c)}\leq\sum_{i=1}^nI(V_i;Y_{2,i})+nR_{12}\numberthis\label{EQ:bc_converse_sumUB_temp}
\end{align*}
where (a) is the mutual information chain rule and the definition of $V_i$, (b) is the Csisz{\'a}r sum identity, and (c) is because conditioning cannot increase entropy and since a uniform distribution maximizes it.

By plugging \eqref{EQ:bc_converse_sumUB_temp} into \eqref{bc_converse_sumUB_temp}, we obtain
\begin{align*}
&n(R_1+R_2)\\
&\leq \sum_{i=1}^n \Big[H(Y_{1,i}|V_i,U_i)+I(V_i,U_i;Y_{2,i})\Big]+nR_{12}+2n\epsilon_n.\numberthis\label{bc_converse_sumUB2}
\end{align*}

\par The upper bounds in \eqref{bc_converse_r1UB}, \eqref{bc_converse_r2UB}, \eqref{bc_converse_sumUB} and \eqref{bc_converse_sumUB2} can be rewritten by introducing a time-sharing random variable $T$ that is independent of $(M_1,M_2,X^n,Y_1^n,Y_2^n,M_{12})$ and is uniformly distributed over $[1:n]$. For instance, the bound in \eqref{bc_converse_r2UB} is rewritten as
\begin{align*}
  R_2&\leq \frac{1}{n}\sum_{t=1}^{n}I(V_t,U_t;Y_{2,t})+R_{12}+\epsilon_n\\
        &=\sum_{t=1}^{n}\mathbb{P}\big(T=t\big)I(V_t,U_t;Y_{2,t}|T=t)+R_{12}+\epsilon_n\\
        &=I(V_T,U_T;Y_{2,T}|T)+R_{12}+\epsilon_n\\
        &\leq I(T,V_T,U_T;Y_{2,T})+R_{12}+\epsilon_n.\numberthis\label{bc_converse_r12UBQ}
\end{align*}
By rewriting the rate bounds \eqref{bc_converse_r1UB}, \eqref{bc_converse_sumUB} and \eqref{bc_converse_sumUB2} in a similar manner, the region obtained is convex. Next, let $Y_1\triangleq Y_{1,T},\ Y_2\triangleq Y_{2,T},\ V\triangleq (V_T,T)$ and $U\triangleq U_T$. We have
\begin{align*}
R_1&\leq H(Y_1)+\epsilon_n\\
R_2&\leq I(V,U;Y_2)+R_{12}+\epsilon_n\\
R_1+R_2&\leq H(Y_1|V,U)+I(U;Y_2|V)+I(V;Y_1)+2\epsilon_n\\
R_1+R_2&\leq H(Y_1|V,U)+I(V,U;Y_2)+R_{12}+2\epsilon_n.\numberthis\label{bc_converse_UBnoQ}
\end{align*}

To complete the proof we need to show that the PMF of $(V,U,X,Y_1,Y_2)$ factors as $P_{V,U,Y_1,X}Q_{Y_2|X}$, which boils down to the Markov relation
\begin{equation}
(V,U,Y_1)-X-Y_2.\label{Markov_sdbc}
\end{equation}
The proof of \eqref{Markov_sdbc} is given in Appendix \ref{APPEN:Markov_Proof}. Taking $\epsilon\to 0$ and $n\to\infty$ establishes the converse.

%
%
%\begin{remark}
%As shown in the direct part of the proof, the cooperation scheme employed by Decoder 1 is based on a WZ-like coding. The choice of $V_i=(M_{12},Y_1^{i-1},Y_{2,i+1}^n)$ is consistent with the auxiliary random variable in the converse proof for the classical WZ setting.
%\end{remark}
%\par Note that the auxiliary random variables $(V,Q)$ admit the Markov relation
%\begin{equation}
%(Y_1,Y_2)-X-(A,B,C)\label{bc_converse_Markov}
%\end{equation}
%due to the deterministic and the memoryless property of the channel and since the channel is without feedback. By taking the limit as $P_e^{(n)}\rightarrow 0$, we obtain the rate bounds (\ref{bc_converse_UBfinal}). Thus, we obtain that $\mathcal{C}\subseteq\mathcal{R}_2$.

%%%%%%%%%%%%%%%%%%%%%%%%%%%%%%%%%%%%%%%%%%%%%%%%%%%%%%%%%%%%%%%%%%%%%%%%%%%%%%%%%%%%%%%%%%%%%%%%%%%%%%%%%%%%%%%%%%%
%%%%%%%%%%%%%%%%%%%%%%%%%             APPENDIX C - ERROR PROB. ANALYSIS           %%%%%%%%%%%%%%%%%%%%%%%%%%%%%%%%%
%%%%%%%%%%%%%%%%%%%%%%%%%%%%%%%%%%%%%%%%%%%%%%%%%%%%%%%%%%%%%%%%%%%%%%%%%%%%%%%%%%%%%%%%%%%%%%%%%%%%%%%%%%%%%%%%%%%

\section{Error Probability Analysis for Theorem \ref{tm_bc_capacity}}\label{appen_analysis}

\begin{figure*}[!b]
\setcounter{equation}{88}
\hrulefill
\begin{align*}
\mathbb{E}e(C_n)&\leq\mathbb{P}\big(\mathcal{E}\big)+\Big(1-\mathbb{P}\big(\mathcal{E}\big)\Big)\Vast(\sum_{j=1}^2\vasti[\underbrace{\mathbb{P}\Big(\mathcal{D}_j^c(1,1,i_j)\Big|\mspace{2mu}\mathcal{E}^c\Big)}_{P_j^{[1]}}+\underbrace{\mathbb{P}\left(\bigcup_{\tilde{i}_j,\tilde{m}_{jj}\neq 1}\mathcal{D}_j(1,\tilde{m}_{jj},\tilde{i}_j)\vasti|\mspace{2mu}\mathcal{E}^c\right)}_{P_j^{[2]}}\vasti]\\
   &+\underbrace{\mathbb{P}\left(\bigcup_{\tilde{m}_0\neq 1}\mathcal{D}_1(\tilde{m}_0,1,i_1)\vasti|\mspace{2mu}\mathcal{E}^c\right)}_{P_1^{[3]}}+\underbrace{\mathbb{P}\left(\bigcup_{\substack{\tilde{i}_1,\tilde{m}_0\neq 1,\\ \tilde{m}_{11}\neq 1}}\mathcal{D}_1(\tilde{m}_0,\tilde{m}_{11},\tilde{i})\vast|\mspace{2mu}\mathcal{E}^c\right)}_{P_1^{[4]}}+\underbrace{\mathbb{P}\left(\bigcup_{\substack{\tilde{m}_0\neq 1:\\\tilde{m}_0\in\mathcal{B}(m_{12})}}\mathcal{D}_2(\tilde{m}_0,1,i_2)\vast|\mspace{2mu}\mathcal{E}^c\right)}_{P_2^{[3]}}\\&+\underbrace{\mathbb{P}\left(\bigcup_{\substack{\tilde{i}_2,\tilde{m}_0\neq 1,\tilde{m}_{22}\neq 1:\\ \tilde{m}_0\in\mathcal{B}(m_{12})}}\mathcal{D}_2(\tilde{m}_0,\tilde{m}_{22},\tilde{i}_2)\vast|\mspace{2mu}\mathcal{E}^c\right)}_{P_2^{[4]}}\Vast).\numberthis\label{EQ:error_prob_UB}
\end{align*}
\end{figure*}

%%%%%%%%%%%%%%%%%%%%%%%%%%%%%%%%%%%%%             Analysis            %%%%%%%%%%%%%%%%%%%%%%%%%%%%%%%%%%%%%%%%%%%%%%
%%%%%%%%%%%%%%%%%%%%%%%%%%%%%%%%%%%%%%%%%%%%%%%%%%%%%%%%%%%%%%%%%%%%%%%%%%%%%%%%%%%%%%%%%%%%%%%%%%%%%%%%%%%%%%%%%%%%
Recall that $(M_0,M_{11},M_{22})$ is a triple of random variables that represents the transmitted messages. Since the analysis considers the expected error probability over the ensemble of codebooks, by the symmetry of the codebook construction we may assume that $(M_0,M_{11},M_{22})=\mathbf{1}\triangleq(1,1,1)$. With some abuse of notation, we denote by $i_j$ the index chosen by the encoder from the $u_j$-bin that is associated with the transmitted messages (recall that for a fixed codebook $i_1$ and $i_2$ are deterministically defined by the transmitted messages).

\setcounter{equation}{86}
\textbf{Encoding Error:} An encoding error occurs if the $v$-,$u_1$- and $u_2$-codewords chosen by the encoder are not jointly typical. This is described by the event stated in \eqref{EQ:analysis_event0} at the bottom of the page, where $$\big(\mathbf{V}(1),\mathbf{U}_1(1,1,\hat{i}_1),\mathbf{U}_2(1,1,\hat{i}_2)\big)\sim P_V^nP_{U_1|V}^nP_{U_2|V}^n$$ and $(i_1,i_2)$ are chosen according to the encoding rule from Subsection B in Appendix \ref{appen_proof_tm_bc}. Namely, an encoding error occurs if there is no pair of indices $(\hat{i}_1,\hat{i}_2)\in\mathcal{I}_1\times\mathcal{I}_2$ that satisfies \eqref{bc_achiev_encoder_typicality}. By the Multivariate Covering Lemma \cite[Lemma 8.2]{ElGammalKim10LectureNotes}, $\mathbb{P}\big(\mathcal{E}\big)\to 0$ as $n\to\infty$ if we have
\begin{equation}
R'_1+R'_2>I(U_1;U_2|V).\label{auxiliary_sumrate}
\end{equation}

\textbf{Decoding Errors:} To account for decoding errors, for any $(m_0,m_{jj},\hat{i}_j)\in\mathcal{M}_0\times\mathcal{M}_{jj}\times\mathcal{I}_j$ and $j=1,2$, define the following event
\begin{align*}
&\mathcal{D}_j(m_0,m_{jj},\hat{i}_j)\\&=\Big\{\big(\mathbf{V}(m_0),\mathbf{U}_j(m_0,m_{jj},\hat{i}_j),\mathbf{Y}_j\big)\in\mathcal{T}_\epsilon^n(P_{V,U_j,Y_j})\Big\}\numberthis\label{EQ:analysis_events}
\end{align*}
where $\big(\mathbf{V}(m_0),\mathbf{U}_j(m_0,m_{jj},\hat{i}_j)\big)\sim P_V^nP_{U_j|V}^n$ and $\mathbf{Y}_j$ is distributed according to the channel transition probability conditioned on the input sequence that corresponds to $(m_0,m_{11},m_{22})=\mathbf{1}$ and $(i_1,i_2)$.

%Let $(i_1,i_2)\in\mathcal{I}_1\times\mathcal{I}_2$ denote the pair of indices that were originally chosen by the encoder and $(I_1,I_2)$ denote the corresponding random variables. Define
%\begin{equation}
%\mathcal{K}=\Big\{(M_1,M_2,I_1,I_2)=(m_1,m_2,i_1,i_2)\Big\}.
%\end{equation}
Let $C_n$ be a random variable that represents a random codebook that adheres to the scheme from Appendix \ref{appen_proof_tm_bc}. By the union bound, the average error probability over the ensemble of codebooks is bounded as shown in \eqref{EQ:error_prob_UB} at the bottom of the page. Note that $\big\{P_j^{[k]}\big\}_{k=1}^4$ correspond to decoding errors by Decoder $j$, where $j=1,2$. We proceed with the following steps:

\begin{enumerate}
%------------------------------------------------------------------------------------------------------------------------------------------
%------------------------------------------------------------- 1 --------------------------------------------------------------------------
%------------------------------------------------------------------------------------------------------------------------------------------
\item $P_j^{[1]}$, for $j=1,2$, vanishes to 0 as $n\rightarrow\infty$ by the law of large numbers.\\

%------------------------------------------------------------------------------------------------------------------------------------------
%------------------------------------------------------------- 2 --------------------------------------------------------------------------
%------------------------------------------------------------------------------------------------------------------------------------------
\item
To upper bound $P_j^{[2]}$, $j=1,2$, consider:
\begin{align*}
P_j^{[2]}&\stackrel{(a)}\leq\sum_{\tilde{i}_j,\tilde{m}_{jj}\neq 1}2^{-n\big(I(U_j;Y_j|V)-\delta_j^{[2]}(\epsilon)\big)}\\
         &\leq2^{n(R_{jj}+R'_j)}2^{-n\big(I(U_j;Y_j|V)-\delta_j^{[2]}(\epsilon)\big)}\\
         &=2^{n\big(R_{jj}+R'_j-I(U_j;Y_j|V)+\delta_j^{[2]}(\epsilon)\big)}
\end{align*}
where (a) follows since for every $\tilde{m}_{jj}\neq 1$ and $\tilde{i}_j\in\mathcal{I}_j$, $\mathbf{U}_j(1,\tilde{m}_{jj},\tilde{i}_j)$ is independent of $\mathbf{Y}_j$ while both of them are drawn conditioned on $\mathbf{V}(1)$. Moreover, $\delta_j^{[2]}(\epsilon)\to 0$ as $\epsilon\to 0$. Hence, to ensure that $P_j^{[2]}$ vanishes as $n\to\infty$, we take:
\begin{equation}
     R_{jj}+R'_j<I(U_j;Y_j|V)-\delta_j^{[2]}(\epsilon),\ j=1,2.\label{EQ:r1_RB}
\end{equation}

%------------------------------------------------------------------------------------------------------------------------------------------
%------------------------------------------------------------- 3 --------------------------------------------------------------------------
%------------------------------------------------------------------------------------------------------------------------------------------

\item
For $P_1^{[4]}$, we have:
\begin{align*}
P_1^{[4]}&\stackrel{(a)}\leq\sum_{\substack{\tilde{i}_1,\tilde{m}_0\neq 1,\\ \tilde{m}_{11}\neq 1}}2^{-n\big(I(V,U_1;Y_1)-\delta_1^{[4]}(\epsilon)\big)}\\
         &\stackrel{(b)}\leq2^{n(R_{20}+R_1+R'_1)}\cdot2^{-n\big(I(V,U_1;Y_1)-\delta_1^{[4]}(\epsilon)\big)}\\
         &=2^{n\big(R_{20}+R_1+R'_1-I(V,U_1;Y_1)+\delta_1^{[4]}(\epsilon)\big)}
\end{align*}
where (a) follows since for every $(\tilde{m}_0,\tilde{m}_{11})\neq \mathbf{1}$ and $\tilde{i}_1\in\mathcal{I}_1$, $\mathbf{V}(\tilde{m}_0)$ and $\mathbf{U}_1(\tilde{m}_0,\tilde{m}_{11},\tilde{i}_1)$ are drawn together by independent of $\mathbf{Y}_1$, while (b) uses $R_0=R_{10}+R_{20}$. Again, $\delta_1^{[4]}(\epsilon)\to 0$ as $\epsilon\to 0$, and therefore, we have that $P_1^{[4]}\to 0$ as $n\to\infty$ if
\begin{equation}
     R_{20}+R_1+R'_1<I(V,U_1;Y_1)-\delta_1^{[4]}(\epsilon).\label{EQ:r2_RB}
\end{equation}

%------------------------------------------------------------------------------------------------------------------------------------------
%------------------------------------------------------------- 4 --------------------------------------------------------------------------
%------------------------------------------------------------------------------------------------------------------------------------------

\item
By repeating similar arguments as before while keeping in mind that the search space of $m_0$ at Decoder 2 is of size $2^{n(R_0-R_{12})}$ (as a consequence of the binning of $\mathcal{M}_0$ and the cooperation protocol), we have that $P_2^{[4]}$ decays with $n$ provided that
\begin{equation}
R_{10}+R_2+R'_2-R_{12}<I(V,U_2;Y_2)-\delta_2^{[4]}(\epsilon)\label{EQ:r4_RB}
\end{equation}
where $\delta_2^{[4]}(\epsilon)\to 0$ as $\epsilon\to 0$.

%------------------------------------------------------------------------------------------------------------------------------------------
%------------------------------------------------------------- 5 --------------------------------------------------------------------------
%------------------------------------------------------------------------------------------------------------------------------------------

\item
By repeating similar steps to upper bound $P_1^{[3]}$, the obtained rate bound is redundant. This is since for every $\tilde{m}_0\neq 1$ and $\tilde{i}_1\in\mathcal{I}_1$, the sequences $\mathbf{V}(\tilde{m}_0)$ and $\mathbf{U}_1(\tilde{m}_0,1,\tilde{i}_1)$ are independent of $\mathbf{Y}_1$. Hence, to ensure that $P_1^{[3]}$ vanishes to 0 as $n\to\infty$, we take
\begin{equation}
R_{10}+R_{20}<I(V,U_1;Y_1)-\delta_1^{[3]}(\epsilon)\label{EQ:analysis_RB3}
\end{equation}
where $\delta_1^{[3]}(\epsilon)\to 0$ as $\epsilon\to 0$. But the right-hand side (RHS) of \eqref{EQ:analysis_RB3} coincides with the RHS of \eqref{EQ:r2_RB}, while the left-hand side (LHS) is with respect to $R_{10}+R_{20}$ only. Clearly, \eqref{EQ:r2_RB} is the dominating constraint. In a similar manner one finds that the rate bound that ensures that $P_2^{[3]}$ can be made arbitrarily small with $n$ is redundant (due to \eqref{EQ:r4_RB}).
\end{enumerate}

\par Summarizing the above results, we get that the RHS of \eqref{EQ:error_prob_UB} decays as the blocklength $n\to\infty$ if the conditions in \eqref{EQ:bc_achiev_rb} are met. By standard existence arguments, a vanishing expected average error probability (over the ensemble of codes) ensures that there exists a reliable $(n,R_{12},R_1,R_2)$ code $\mathcal{C}_n$ for all rate triples that satisfy \eqref{EQ:bc_achiev_rb}.

\section{Proof of Lemma \ref{lemma_region_BC_equal}}\label{appen_region_euivalence_RBC}

\par To show $\mathcal{C}^{(\mathrm{D})}_\mathrm{BC}\subseteq\mathcal{C}_{\mathrm{BC}}$, let $(R_{12},R_1,R_2)\in\mathcal{C}^{(\mathrm{D})}_\mathrm{BC}$ be a rate triple achieved by $(V,U,X)$. Setting $V^\star=V$ and $U^\star=U$, implies that the same rate triple $(R_{12},R_1,R_2)$ is contained in $\mathcal{C}_{\mathrm{BC}}$, as it is achieved by $(V^\star,U^\star,X)$ (since substituting \eqref{EQ:region_eq_sdbc12} into \eqref{EQ:region_eq_sdbc1+2} yields \eqref{region_BC1+2b}).

\par To see that $\mathcal{C}_{\mathrm{BC}}\subseteq\mathcal{C}_{\mathrm{BC}}^{(\mathrm{D})}$, let $(R_{12},R_1,R_2)\in\mathcal{C}_{\mathrm{BC}}$ be a rate triple achieved by $(V,U,X)$. Further assume that
\begin{equation}
R_{12}<I(V;Y_1)-I(V;Y_2)
\end{equation}
(otherwise, all four inequalities in (\ref{EQ:region_eq_sdbc}) clearly hold). Accordingly, there is a real number $\gamma>0$, such that
\begin{equation}
R_{12}=I(V;Y_1)-I(V;Y_2)-\gamma.\label{appen_RBC2_R12_equal}
\end{equation}
%The definition of $V^\star$ follows similar lines to these presented in Section \ref{sec_bc_converse}.
Define $V^\star\triangleq(\Theta,\widetilde{V})$, where $\Theta\sim\mbox{Ber}\left(\lambda\right)$, $\lambda\in[0,1]$, is a binary random variable independent of $(V,U,X)$ that takes values in $\mathcal{O}=\{\theta_1,\theta_2\}$, and
\begin{eqnarray}
\widetilde{V}=\begin{cases}
V\ ,& \Theta=\theta_1\\
\emptyset\ ,& \Theta=\theta_2
\end{cases}.\label{appen_RBC2_Vstar}
\end{eqnarray}
Furthermore, set
\begin{equation}
\lambda=\frac{I(V;Y_1)-I(V;Y_2)-\gamma}{I(V;Y_1)-I(V;Y_2)}\label{appen_RBC2_lambda}
\end{equation}
and $U^\star=(V,U)$.

\par With respect to this choice of $(V^\star,U^\star)$, consider
\begin{align*}
I(V^\star;Y_1)-I(V^\star;Y_2)&=\lambda\Big[I(V;Y_1)-I(V;Y_2)\Big]\\
                             &\stackrel{(a)}=I(V;Y_1)-I(V;Y_2)-\gamma\\
                             &\stackrel{(b)}=R_{12}\numberthis\label{appen_RBC2_R12_holds}
\end{align*}
where (a) uses the choice of $\lambda$ in \eqref{appen_RBC2_lambda} and (b) follows from \eqref{appen_RBC2_R12_equal}. Thus, \eqref{EQ:region_eq_sdbc12} holds.

\par Next, by the definition of $U^\star$ and because \eqref{region_BC1}-\eqref{region_BC2} are valid, we obtain \eqref{EQ:region_eq_sdbc1}-\eqref{EQ:region_eq_sdbc2}. It remains to be shown that \eqref{EQ:region_eq_sdbc1+2} holds. Consider the following:
\begin{align*}
&H(Y_1|V^\star,U^\star)+I(U^\star;Y_2|V^\star)+I(V^\star;Y_1)\\
&=H(Y_1|V^\star,U^\star)+I(V^\star,U^\star;Y_2)+I(V^\star;Y_1)-I(V^\star;Y_2)\\
&\stackrel{(a)}=H(Y_1|V,U)+I(V,U;Y_2)+R_{12}\\
&\stackrel{(b)}\geq R_1+R_2\numberthis\label{appen_RBC2_R1+2_holds}
\end{align*}
where (a) uses the definition of $U^\star$ and \eqref{appen_RBC2_R12_holds}, while (b) is by \eqref{region_BC1+2b}. Consequently \eqref{EQ:region_eq_sdbc1+2} is valid and the inclusion $\mathcal{C}_{\mathrm{BC}}\subseteq\mathcal{C}^{(\mathrm{D})}_\mathrm{BC}$ follows.

%%%%%%%%%%%%%%%%%%%%%%%%%%%%%%%%%%%%%%%%%%%%%%%%%%%%%%%%%%%%%%%%%%%%%%%%%%%%%%%%%%%%%%%%%%%%%%%%%%%%%%%%%%%%%%%%%%
%%%%%%%%%%%%%%%%%%%               APPENDIX E - Explicit Converse for Lemma 9           %%%%%%%%%%%%%%%%%%%%%%%%%%%%%
%%%%%%%%%%%%%%%%%%%%%%%%%%%%%%%%%%%%%%%%%%%%%%%%%%%%%%%%%%%%%%%%%%%%%%%%%%%%%%%%%%%%%%%%%%%%%%%%%%%%%%%%%%%%%%%%%%

\section{Explicit Converse for Lemma \ref{lemma_region_BC_equal}}\label{appen_region_euivalennt_converse}

\par The converse for Theorem \ref{tm_bc_capacity} is established using a novel approach that generalizes the classical technique used for converse proofs. Our approach relies on two key properties. First, the construction of the auxiliary random variables depends on the distribution induced by the code. Second, the auxiliaries are constructed in a probabilistic manner.

%\par Initially, tight bounds on the transmission rates of an optimal code are derived. The only inequalities in these rate bounds originate from applying Fano's theorem. This derivation may result in defining a number of auxiliary random variables that exceeds the number that was used in the proof of achievability. Nonetheless, once this outer bound on the capacity region is established, it is shown to be contained inside the achievable region that was derived in the direct part. To do this, we consider an arbitrary rate tuple that is inside the outer bound and is achieved by a \emph{given} input and auxiliary distribution. It is then shown that there exists an input and auxiliary distribution of the structure that corresponded to the inner bound so that the same rate tuple is also contained in it. In general, this proof of existence relies on the specific structure of the given joint distribution and the construction of the corresponding auxiliaries is probabilistic. By leveraging the above technique, the converse proof for the SD-BC with cooperation, which would be cumbersome to solve using the standard approach, is established.

\par We show that if a rate triple $(R_{12},R_1,R_2)$ is achievable, then there is a PMF $P_{V,U,Y_1,X}Q_{Y_2|X}$ for which $Y_1=f(X)$, such that the inequalities in \eqref{EQ:region_eq_sdbc} are satisfied. To do so, we first state an upper bound on $\mathcal{C}^{(\mathrm{D})}_\mathrm{BC}$ and then establish its inclusion in $\mathcal{C}^{(\mathrm{D})}_\mathrm{BC}$. The upper bound is stated in the following lemma.

\begin{lemma}[Upper Bound on the Capacity Region]\label{alt_lemma_upper_bound}
Let $\mathcal{R}_\mathrm{O}$ be the region defined by the union of rate triples $(R_{12},R_1,R_2)\in\mathbb{R}^3_+$ satisfying:
\begin{subequations}
\begin{align}
R_{12}&\geq I(A;Y_1|C)-I(C;Y_2|A)\label{alt_bc_converse_finalUB12}\\
R_1&\leq H(Y_1|B,C)\label{alt_bc_converse_finalUB1}\\
R_2&\leq I(B;Y_2|A)+R_{12}\label{alt_bc_converse_finalUB2}\\
R_1+R_2&\leq H(Y_1|A,B,C)+I(B;Y_2|A,C)+I(A;Y_1|C)\label{alt_bc_converse_finalUB1+2}
\end{align}\label{alt_bc_converse_UBfinal}
\end{subequations}

\vspace{-2.5mm}
\noindent where the union is over all PMFs $P_{A,B,C,Y_1,X}Q_{Y_2|X}$ for which $Y_1=f(X)$. The following inclusion holds:
\begin{equation}
\mathcal{C}^{(\mathrm{D})}_\mathrm{BC}\subseteq\mathcal{R}_\mathrm{O}.\label{alt_bc_converse_inclusion1}
\end{equation}
\end{lemma}

\begin{IEEEproof}
By similar arguments to those given in Subsection B of Appendix \ref{appen_proof_tm_bc}, since $(R_{12},R_1,R_2)$ is achievable and by Fano's inequality, we have
\begin{subequations}
\begin{align}
H(M_1|Y_1^n)&\leq n\epsilon_n\label{alt_bc_converse_Fano1}\\
H(M_2|M_{12},Y_2^n)&\leq n\epsilon_n\label{alt_bc_converse_Fano2}
\end{align}
\end{subequations}
\noindent where $\epsilon_n$ is defined as in \eqref{bc_converse_Fano}. It follows that
\begin{align*}
nR_{12}&\geq H(M_{12})\\
       &\stackrel{(a)}= I(M_{12};Y_1^n)\\
       &\stackrel{(b)}=\sum_{i=1}^n \Big[I(M_{12},Y_{2,i+1}^n;Y_1^i)-I(M_{12},Y_{2,i}^n;Y_1^{i-1})\Big]\\
       &\begin{multlined}[b][.85\columnwidth]=\sum_{i=1}^n \Big[I(M_{12},Y_{2,i+1}^n;Y_{1,i}|Y_1^{i-1})\\-I(Y_1^{i-1};Y_{2,i}|M_{12},Y_{2,i+1}^n)\Big]\end{multlined}\\
       &\stackrel{(c)}=\sum_{i=1}^n \Big[I(A_i;Y_{1,i}|C_i)-I(C_i;Y_{2,i}|A_i)\Big]\numberthis\label{alt_bc_converse_r12UB}
\end{align*}
where (a) is because $M_{12}$ is defined by $Y_1^n$, (b) is a telescoping identity, while (d) is by defining $A_i\triangleq(M_{12},Y_{2,i+1}^n)$ and $C_i\triangleq Y_1^{i-1}$, for every $i\in[1:n]$.

\par For the upper bound on $R_1$, consider
\begin{align*}
  nR_1&=H(M_1)\\
      &=H(M_1|M_2)\\
      &\stackrel{(a)}\leq I(M_1;Y_1^n|M_2)+n\epsilon_n\\
      &\stackrel{(b)}=H(Y_1^n|M_2)-H(Y_1^n|M_1,M_2,X^n)+n\epsilon_n\\
      &\stackrel{(c)}=\sum_{i=1}^nH(Y_{1,i}|M_2,Y_1^{i-1})+n\epsilon_n\\
      &\stackrel{(d)}=\sum_{i=1}^nH(Y_{1,i}|B_i,C_i)+n\epsilon_n\numberthis\label{alt_bc_converse_r1UB}
\end{align*}
where (a) uses \eqref{alt_bc_converse_Fano1}, (b) is since $X^n$ is a function of $(M_1,M_2)$, (c) is because $Y_1^n$ is determined by $X^n$, while (d) defines $B_i\triangleq M_2$, for every $i\in[1:n]$, and uses the definition of $C_i$.

\par To bound $R_2$ we have
\begin{align*}
  nR_2&\stackrel{(a)}\leq I(M_2;Y_2^n|M_{12})+I(M_2;M_{12})+n\epsilon_n\\
      &\leq \sum_{i=1}^n I(M_2;Y_{2,i}|M_{12},Y_{2,i+1}^n)+H(M_{12})+n\epsilon_n\\
      &\stackrel{(b)}\leq \sum_{i=1}^n I(B_i;Y_{2,i}|A_i)+nR_{12}+n\epsilon_n\numberthis\label{alt_bc_converse_r2UB}
\end{align*}
where (a) is by repeating steps \eqref{bc_converse_r2UB1}-\eqref{bc_converse_r2UB2} in Appendix \ref{appen_proof_tm_bc}, while (b) is by the definition of $(A_i,B_i)$ and because a uniform distribution maximizes entropy.

\par Finally, for the sum of rates, we begin from step \eqref{bc_converse_sumUB_temp} in Appendix \ref{appen_proof_tm_bc} and note that the auxiliaries in Appendix \ref{appen_proof_tm_bc} can be rewritten in terms of $(A_i,B_i,C_i)$ as $V_i=(A_i,C_i)$ and $U_i=B_i$. We thus have
\begin{align*}
&n(R_1+R_2)\\&\begin{multlined}[b][.85\columnwidth]\leq \sum_{i=1}^n \Big[I(B_i;Y_{2,i}|A_i,C_i)-I(B_i;Y_{1,i}|A_i,C_i)\Big]\\+H(Y_1^n)+2n\epsilon_n\end{multlined}\\
          &\begin{multlined}[b][.85\columnwidth]\stackrel{(a)}= \sum_{i=1}^n \Big[H(Y_{1,i}|A_i,B_i,C_i)+I(B_i;Y_{2,i}|A_i,C_i)\\+I(A_i;Y_{1,i}|C_i)\Big]+2n\epsilon_n\end{multlined}\numberthis\label{alt_bc_converse_sumUB}
\end{align*}
where (a) is from the mutual information chain rule and the definition of $(A_i,B_i,C_i)$.

%can be rewritten by introducing a time-sharing random variable $Q$ that is uniformly distributed over the set $[\{1,2,...,n\}$. The upper bound in (\ref{bc_converse_r12UB}) can then be rewritten as
%\begin{align*}
%  R_{12}&\geq \frac{1}{n}\sum_{i=1}^{n}\Big[I(A_i;Y_{1,i}|C_i)-I(C_i;Y_{2,i}|A_i)\Big]\\
%        &=\sum_{i=1}^{n}\mathbb{P}\big[Q=i\big]\Big[I(A_Q;Y_{1,Q}|C_Q,Q=i)-I(C_Q;Y_{2,Q}|A_Q,Q=i)\Big]\\
%        &=I(A_Q;Y_{1,Q}|C_Q,Q)-I(C_Q;Y_{2,Q}|A_Q,Q).\numberthis\label{bc_converse_r12UBQ}
%\end{align*}
%By rewriting the rate bounds (\ref{alt_bc_converse_r1UB}), (\ref{alt_bc_converse_r2UB}) and (\ref{alt_bc_converse_sumUB}) in the same manner, the region obtained is convex. This follows from the presence of the time sharing random variable $Q$ in the conditioning of all the mutual information terms.

%\par Next, let $Y_1\triangleq Y_{1,Q},\ Y_2\triangleq Y_{2,Q},\ A\triangleq (A_Q,Q),\ B\triangleq B_Q$ and $C\triangleq (C_Q,Q)$. We have

\par By standard time-sharing arguments, we rewrite the bounds in \eqref{alt_bc_converse_r12UB}-\eqref{alt_bc_converse_sumUB} as
\begin{align*}
R_{12}&\geq I(A;Y_1|C)-I(C;Y_2|A)\\
R_1&\leq H(Y_1|B,C)+\epsilon_n\\
R_2&\leq I(B;Y_2|A)+R_{12}+\epsilon_n\\
R_1+R_2&\leq H(Y_1|A,B,C)+I(B;Y_2|A,C)\\&\mspace{170mu}+I(A;Y_1|C)+2\epsilon_n\numberthis\label{alt_bc_converse_UBnoQ}
\end{align*}
which are the bounds from \eqref{alt_bc_converse_UBfinal} with small added terms such as $\epsilon_n$. Taking $\epsilon\to 0$ and $n\to\infty$, these terms approach 0. The proof is completed by showing that the Markov relations stated in Lemma \ref{alt_lemma_upper_bound} hold. This follows by arguments similar to those presented in Appendix \ref{appen_proof_tm_bc}.
\end{IEEEproof}

\par Based on Lemma \ref{alt_lemma_upper_bound}, the inclusion relation stated in the following lemma completes the proof of the converse.
\begin{lemma}[Tightness of Upper Bound]\label{alt_lemma_inclusion}
The following inclusion holds:
\begin{equation}
\mathcal{R}_\mathrm{O}\subseteq\mathcal{C}^{(\mathrm{D})}_{\mathrm{BC}}.\label{alt_bc_converse_inclusion2}
\end{equation}
\end{lemma}

\begin{IEEEproof}
Let $(R_{12},R_1,R_2)\in\mathcal{R}_\mathrm{O}$ be achieved by a given tuple of random variables $(A,B,C,X)$. We show that there exists a pair of random variables $(V,U)$, such that $(R_{12},R_1,R_2)\in\mathcal{C}^{(\mathrm{D})}_{\mathrm{BC}}$ and is achieved by $(V,U,X)$. We define $(V,U)$ as follows. Let $\Theta\sim\mbox{Ber}\left(\lambda\right)$, $\lambda\in[0,1]$, be a binary random variable independent of $(A,B,C,X)$ that takes values in $\mathcal{O}=\{\theta_1,\theta_2\}$. Define the random variable
\begin{eqnarray}
\widetilde{V}=\begin{cases}
(A,C)\ ,& \Theta=\theta_1\\
\emptyset\ ,& \Theta=\theta_2
\end{cases}.\label{bc_converse_V}
\end{eqnarray}
Set $V\triangleq(\Theta,\widetilde{V})$ and
\begin{equation}
U=(A,B,C)\label{bc_converse_U}
\end{equation}
and note that $(V,U)$ preserves the Markov structure
\begin{equation}
(Y_1,Y_2)-X-(U,V)\label{bc_converse_UVMarkov}
\end{equation}
since, as stated in Lemma \ref{alt_lemma_upper_bound}, $(Y_1,Y_2)-X-(A,B,C)$ forms a Markov chain.

First, consider the case when
\begin{equation}
I(A,C;Y_1)-I(A,C;Y_2)\leq 0\label{bc_converse_case1}.
\end{equation}
By setting $\lambda=1$ we have
\begin{equation}
I(V;Y_1)-I(V;Y_2)\leq 0\stackrel{(a)}\leq R_{12}\numberthis\label{bc_converse_vy1-vy2_first}
\end{equation}
where (a) is since $R_{12}\geq0$, which establishes \eqref{EQ:region_eq_sdbc12}. \eqref{EQ:region_eq_sdbc1} holds since $H(Y_1|B,C)\leq H(Y_1)$.

\par For \eqref{EQ:region_eq_sdbc2}, note that the definition of $(V,U)$ in \eqref{bc_converse_V}-\eqref{bc_converse_U} implies that
\begin{equation}
(A,B,C,X,Y_1,Y_2)-U-V\label{bc_converse_UVMarkov2}
\end{equation}
forms a Markov chain. Consequently, we obtain
\begin{equation}
I(V,U;Y_2)=I(A,B,C;Y_2)\label{bc_converse_Iuvy2eq}
\end{equation}
which yields
\begin{equation}
I(V,U;Y_2)+R_{12}\geq I(B;Y_2|A)+R_{12}\stackrel{(a)}\geq R_2\label{bc_converse_Iuvy2}
\end{equation}
where (a) uses \eqref{alt_bc_converse_finalUB2}. This shows that \eqref{EQ:region_eq_sdbc2} also holds.

\par For the sum rate, we rewrite \eqref{EQ:region_eq_sdbc1+2} as
\begin{align*}
&H(Y_1|V,U)+I(U;Y_2|V)+I(V;Y_1)\\&=H(Y_1|V,U)+I(V,U;Y_2)+I(V;Y_1)-I(V;Y_2)\numberthis\label{bc_converse_sumUV_alt}
\end{align*}
and obtain an explicit expression for each of the information measures in the RHS of \eqref{bc_converse_sumUV_alt} in terms of $(A,B,C,X)$. Based on similar arguments to those presented before, we have
\begin{equation}
H(Y_1|V,U)= H(Y_1|A,B,C)\label{bc_converse_y1|uv}
\end{equation}
while the other two information measures in \eqref{bc_converse_sumUV_alt} were previously evaluated in \eqref{bc_converse_vy1-vy2_first} and \eqref{bc_converse_Iuvy2eq}. Inserting \eqref{bc_converse_vy1-vy2_first}, \eqref{bc_converse_Iuvy2eq} and \eqref{bc_converse_y1|uv} into \eqref{bc_converse_sumUV_alt} results in
\begin{align*}
H(Y_1|V,&U)+I(U;Y_2|V)+I(V;Y_1)\\
&\stackrel{(a)}\geq H(Y_1|A,B,C)+I(B;Y_2|A,C)+I(A;Y_1|C)\\&\stackrel{(b)}\geq R_1+R_2\numberthis\label{bc_converse_1+2final}
\end{align*}
where (a) is because $\lambda=1$ and the mutual information chain rule, while (b) uses \eqref{alt_bc_converse_finalUB1+2}. This satisfies \eqref{EQ:region_eq_sdbc1+2}.

\par To conclude the proof it is left to consider the case where
\begin{equation}
I(A,C;Y_1)-I(A,C;Y_2) > 0\label{bc_converse_case2}.
\end{equation}
This time set
\begin{equation}
\lambda=\min\left\{1,\bigg(\frac{I(A;Y_1|C)-I(A,C;Y_2)+I(A;Y_2)}{I(A;Y_1|C)-I(A,C;Y_2)+I(C;Y_1)}\bigg)^+\right\}\label{bc_converse_lambda}
\end{equation}
where $(x)^+=\max\big\{0,x\big\}$, and consider the following.
\begin{align*}
I(V;&Y_1)-I(V;Y_2)\\&=\lambda\Big[I(A,C;Y_1)-I(A,C;Y_2)\Big]\numberthis\label{bc_converse_vy1-vy2_first2}\\
                 &=\lambda\Big[I(A;Y_1|C)-I(A,C;Y_2)+I(C;Y_1)\Big]\numberthis\label{bc_converse_cases2}\\   %&\stackrel{(a)}=I(A;Y_1|C)-I(A,C;Y_2)+I(A;Y_2)\\
                 &\stackrel{(a)}\leq I(A;Y_1|C)-I(C;Y_2|A)\stackrel{(b)}\leq R_{12}\numberthis\label{bc_converse_vy1-vy22}
\end{align*}
where (b) relies \eqref{alt_bc_converse_finalUB12}, while step (a) is justified as follows. If $\lambda=1$ we have
\begin{equation}
I(A;Y_2)\geq I(C;Y_1).\label{bc_converse_case21}
\end{equation}
Using \eqref{bc_converse_case21}, we rewrite \eqref{bc_converse_cases2} as
\begin{align*}
&\lambda\Big[I(A;Y_1|C)-I(A,C;Y_2)+I(C;Y_1)\Big]\\&\stackrel{(a)}=I(A;Y_1|C)-I(C;Y_2|A)+I(C;Y_1)-I(A;Y_2)\\
                 &\stackrel{(b)}\leq I(A;Y_1|C)-I(C;Y_2|A)
\end{align*}
where (a) follows because $\lambda=1$ and by the mutual information chain rule, while (b) is by \eqref{bc_converse_case21}. On the other hand, if
\begin{equation}
\lambda=\frac{I(A;Y_1|C)-I(A,C;Y_2)+I(A;Y_2)}{I(A;Y_1|C)-I(A,C;Y_2)+I(C;Y_1)}\label{bc_converse_case2_lambda}
\end{equation}
then
\begin{equation}
I(A;Y_2)< I(C;Y_1)\label{bc_converse_case22}
\end{equation}
and we rewrite \eqref{bc_converse_cases2} as
\begin{align*}
\lambda\Big[I(A;Y_1|C)&-I(A,C;Y_2)+I(C;Y_1)\Big]\\&\stackrel{(a)}=I(A;Y_1|C)-I(A,C;Y_2)+I(A;Y_2)\\
                 &=I(A;Y_1|C)-I(C;Y_2|A)
\end{align*}
where (a) uses \eqref{bc_converse_case2_lambda}. The case $\lambda=0$ is trivial, and we omit the derivation of (a) in \eqref{bc_converse_vy1-vy22}. We conclude that \eqref{EQ:region_eq_sdbc12} is satisfied. \eqref{EQ:region_eq_sdbc1}-\eqref{EQ:region_eq_sdbc2} follow by the same arguments presented above, while for \eqref{EQ:region_eq_sdbc1+2} we have
\begin{align*}
H(Y_1&|V,U)+I(U;Y_2|V)+I(V;Y_1)\\&\begin{multlined}[b][.87\columnwidth]\stackrel{(a)}=H(Y_1|A,B,C)+I(A,B,C;Y_2)\\+\lambda\Big[I(A,C;Y_1)-I(A,C;Y_2)\Big]\end{multlined}\\
&\stackrel{(b)}\geq H(Y_1|A,B,C)+I(B;Y_2|A,C)+I(A;Y_1|C)\\
&\stackrel{(c)}\geq R_1+R_2\numberthis\label{bc_converse_1+2final2}
\end{align*}
where (a) is by \eqref{bc_converse_UVMarkov2} and \eqref{bc_converse_sumUV_alt}, (c) uses \eqref{alt_bc_converse_finalUB1+2}, while the derivation of (b) relies on evaluating the terms of interest for the three possible values of $\lambda$. First, by \eqref{bc_converse_case2}, $\lambda=0$ if and only if
\begin{equation}
I(A;Y_1|C)\leq I(C;Y_2|A)\label{bc_converse_case_lambda0}
\end{equation}
which implies
\begin{align*}
&H(Y_1|A,B,C)+I(A,B,C;Y_2)\\&\mspace{200mu}+\lambda\Big[I(A,C;Y_1)-I(A,C;Y_2)\Big]\\
&=H(Y_1|A,B,C)+I(B;Y_2|A,C)+I(A,C;Y_2)\\
&\geq H(Y_1|A,B,C)+I(B;Y_2|A,C)+I(C;Y_2|A)\\
&\geq H(Y_1|A,B,C)+I(B;Y_2|A,C)+I(A;Y_2|C).\numberthis\label{bc_converse_1+2final_case0}
\end{align*}
If $\lambda=1$, by the mutual information chain rule we have
\begin{align*}
&H(Y_1|A,B,C)+I(A,B,C;Y_2)\\&\mspace{200mu}+\lambda\Big[I(A,C;Y_1)-I(A,C;Y_2)\Big]\\
&\geq H(Y_1|A,B,C)+I(B;Y_2|A,C)+I(A;Y_1|C).\numberthis\label{bc_converse_1+2final_case1}
\end{align*}
Finally, if $\lambda$ is as in \eqref{bc_converse_case2_lambda}, we obtain
\begin{align*}
&H(Y_1|A,B,C)+I(A,B,C;Y_2)\\&\mspace{200mu}+\lambda\Big[I(A,C;Y_1)-I(A,C;Y_2)\Big]\\
&=H(Y_1|A,B,C)+I(A;Y_2)+I(B;Y_2|A,C)+I(A;Y_1|C)\\
&\geq H(Y_1|A,B,C)+I(B;Y_2|A,C)+I(A;Y_1|C).\numberthis\label{bc_converse_1+2final_case2}
\end{align*}
We find that \eqref{EQ:region_eq_sdbc1+2} is also satisfied, thus concluding that \eqref{EQ:region_eq_sdbc} holds for the choice of $(V,U)$ and $\lambda$ stated in \eqref{bc_converse_V}-\eqref{bc_converse_U} and \eqref{bc_converse_lambda}, respectively. This implies that $\mathcal{R}_\mathrm{O}\subseteq \mathcal{C}^{(\mathrm{D})}_\mathrm{BC}$.
\end{IEEEproof}

Lemma \ref{alt_lemma_inclusion} completes the converse and characterizes the region in \eqref{EQ:region_eq_sdbc} as the capacity region of the SD-BC with cooperation.

\begin{remark}
The definition of $V$ in \eqref{bc_converse_V} is probabilistic and thought $\lambda$ depends on the joint distribution of $(A,B,C,X)$ that is induced by the code. \end{remark}

%%%%%%%%%%%%%%%%%%%%%%%%%%%%%%%%%%%%%%%%%%%%%%%%%%%%%%%%%%%%%%%%%%%%%%%%%%%%%%%%%%%%%%%%%%%%%%%%%%%%%%%%%%%%%%%%%%%
%%%%%%%%%%%%%%%%%%%%%%%         APPENDIX F - RBC Capacity Region Reduction        %%%%%%%%%%%%%%%%%%%%%%%%%%%%%%%%%
%%%%%%%%%%%%%%%%%%%%%%%%%%%%%%%%%%%%%%%%%%%%%%%%%%%%%%%%%%%%%%%%%%%%%%%%%%%%%%%%%%%%%%%%%%%%%%%%%%%%%%%%%%%%%%%%%%%

\section{Derivation of the Region in (\ref{region_BC}) from (\ref{region_RBC_reduced})}\label{appen_region_RBC}

Denote the region in \eqref{region_RBC_reduced} by $\mathcal{R}$. Note that $\mathcal{C}_{\mathrm{BC}}$ is achievable from $\mathcal{R}$ by taking $X_1$ to be independent of $(V,U,X)$ and applying a coding scheme where the transmission rate via the relay channel is $R_{12}$. This implies that $\mathcal{C}_{\mathrm{BC}}\subseteq\mathcal{R}$.

\par To see that $\mathcal{R}\subseteq\mathcal{C}_{\mathrm{BC}}$ recall that the proof of Theorem 8 in \cite{Liang_Kramer_RBC2007} relies on Theorem 4 in that same work, which characterized an upper bound on the capacity region of a general RBC. In the proof of Theorem 4 (see \cite[Appendix II]{Liang_Kramer_RBC2007}), the auxiliary random variables $V_i$ and $U_i$ are defined as
\begin{equation}
V_i\triangleq(M_0,Y_1^{i-1},Y_{2,i+1}^n)\ \ ;\ \ U_i\triangleq(M_2,Y_1^{i-1},Y_{2,i+1}^n).
\end{equation}
$M_0$ is a common message that was also considered in \cite{Liang_Kramer_RBC2007}. Since $X_{1,i}$ is a function of $Y_1^{i-1}$, it is also a function of $V_i$ (and\textbackslash or $U_i$) for every $i\in[1:n]$. In particular, this implies that $X_1$ is a function of $V$. Consequently, the information measures defining  $\mathcal{R}$ are then upper bounded as follows. For $R_1$ we have
\begin{equation}
R_1\leq H(Y_1|X_1)\leq H(Y_1).\label{appen_RBC1_R1}
\end{equation}
For the $R_2$ consider
\begin{align*}
R_2&\leq I(V,U,X_1;Y_{21})+H(Y_{22}|Y_{21})\\
   &\stackrel{(a)}\leq I(V,U;Y_{21})+H(Y_{22})\\
   &\stackrel{(b)}\leq I(V,U;Y_{21})+R_{12}\numberthis\label{appen_RBC1_R2}
\end{align*}
where (a) is because $X_1$ is defined by $V$ and since conditioning cannot increase entropy, while (b) follows because the relay channel is deterministic with capacity $R_{12}$.

\par For the first bound on the sum of rates, we have
\begin{align*}
R_1&+R_2\\&\leq H(Y_1|V,U,X_1)+I(U;Y_{21}|V,X_1)+I(V;Y_1|X_1)\\
       &\stackrel{(a)}\leq H(Y_1|V,U)+I(U;Y_{21}|V)+I(V;Y_1). \numberthis\label{appen_RBC1_R1+21}
\end{align*}
Here (a) is justified similarly to step (a) in \eqref{appen_RBC1_R2}.

Finally, the second bound on $R_1+R_2$ is upper bounded as
\begin{align*}
R_1+R_2&\leq H(Y_1|V,U,X_1)\mspace{-2mu}+\mspace{-2mu}I(V,U,X_1;Y_{21})\mspace{-2mu}+\mspace{-2mu}H(Y_{22}|Y_{21})\\
       &\stackrel{(a)}\leq H(Y_1|V,U)+I(V,U;Y_{21})+H(Y_{22})\\
       &\stackrel{(b)}\leq H(Y_1|V,U)+I(V,U;Y_{21})+R_{12}.\numberthis\label{appen_RBC1_R1+22}
\end{align*}
Again, (a) and (b) follow by the same arguments as (a) and (b) in \eqref{appen_RBC1_R2}.

\par To complete the proof, it remains to be shown that taking the union only over PMFs in which $X_1$ is independent of $(V,U,X)$ exhausts the entire region. This follows since the rate bounds in \eqref{appen_RBC1_R1}-\eqref{appen_RBC1_R1+22} do not involve $X_1$ nor $Y_{22}$. Relabeling $Y_{21}$ as $Y_2$ shows that  $\mathcal{R}\subseteq\mathcal{C}_{\mathrm{BC}}$ and completes the proof.

%%%%%%%%%%%%%%%%%%%%%%%%%%%%%%%%%%%%%%%%%%%%%%%%%%%%%%%%%%%%%%%%%%%%%%%%%%%%%%%%%%%%%%%%%%%%%%%%%%%%%%%%%%%%%%%%%%%
%%%%%%%%%%%%%%%%%%%%%%%%%%         APPENDIX G - SW Region Correspondance          %%%%%%%%%%%%%%%%%%%%%%%%%%%%%%%%%
%%%%%%%%%%%%%%%%%%%%%%%%%%%%%%%%%%%%%%%%%%%%%%%%%%%%%%%%%%%%%%%%%%%%%%%%%%%%%%%%%%%%%%%%%%%%%%%%%%%%%%%%%%%%%%%%%%%

\section{Proof of Lemma \ref{RBC_lemma_equivalence}}\label{appen_RBC_lemma_equivalence}

Fix $(R_1,R_2)\in\mathcal{C}_{\mathrm{RBC}}(R_{12})$ and let $\big\{\mathcal{C}_n^{(\mathrm{RBC})}(R_{12})\big\}_{n\in\mathbb{N}}$ be the sequence of $(n,R_1,R_2)$ codes for the $R_{12}$-reduced SD-RBC that adhere to the coding scheme described in \cite[Appendix I]{Liang_Kramer_RBC2007}. Accordingly, $e\big(\mathcal{C}^{(\mathrm{RBC})}_n(R_{12})\big)\to 0$ as $n\to\infty$ and the induced codewords, channel inputs, and channel outputs are jointly-typical with high probability \footnote{$e(\mathcal{C}_n)$ stands for the error probability of the code $\mathcal{C}_n$ defined analogously to \eqref{BC_Pe}}. Since the channel from Decoder 1 to Decoder 2 is deterministic, there are approximately $2^{nH(Y_{22})}$ different possible relay channel outputs $\mathbf{y}_{22}$. Recall that the capacity of the orthogonal and deterministic relay of the $R_{12}$-reduced SD-RBC is exactly $R_{12}$, i.e., $H(Y_{22})=R_{12}$. For every sequence $\mathbf{y}_{22}\in\mathcal{T}_\epsilon^n(Q_{Y_{22}})$ (here $\epsilon>0$ corresponds to the margin between the region achieved by the $n$th code in the sequence and $(R_1,R_2)$), define the following subset of $\mathbf{x}_1$ codewords:
 \begin{equation}
 \mathcal{V}(\mathbf{y}_{22})\mspace{-3mu}=\mspace{-3mu}\Big\{\mathbf{x}_1\in\mathcal{C}^{(\mathrm{RBC})}_n(R_{12})\Big|f_R(x_{1,i})=y_{22,i},\forall i\in[1:n]\Big\}.
 \end{equation}
Consider a SD-BC with cooperation and associate a cooperation message $m_{12}$, where $m_{12}\in\mathcal{M}_{12}$, with every set $\mathcal{V}(\mathbf{y}_{22})$. To use $\mathcal{C}^{(\mathrm{RBC})}_n$ for the SD-BC with cooperation, Decoder 1 waits for the $n$-symbol transmission to end and then shares with Decoder 2 the message $m_{12}$ associated with a set $\mathcal{V}(\mathbf{y}_{22})$ that contains the intended $\mathbf{x}_1$ codeword (i.e., such that $\mathbf{x}_1\in\mathcal{V}(\mathbf{y}_{22})$). Given $m_{12}$, Decoder 2 recovers the sequence $\mathbf{y}_{22}$ and proceeds with the decoding process of the $R_{12}$-reduced SD-RBC coding scheme. This results in a sequence of $(n,R_{12},R_1,R_2)$ codes $\big\{\mathcal{C}_n^{(\mathrm{BC})}\big\}_{n\in\mathbb{N}}$ for the SD-BC with cooperation.
\par Next, fix $(R_{12},R_1,R_2)\in\mathcal{C}_{\mathrm{BC}}$ and let $\big\{\mathcal{C}^{(\mathrm{BC})}_n\big\}_{n\in\mathbb{N}}$ be the sequence of $(n,R_{12},R_1,R_2)$ codes for the SD-BC with cooperation described in Appendix \ref{appen_proof_tm_bc}. Consider an $R_{12}$-reduced SD-RBC and map each cooperation message $m_{12}\in\mathcal{M}_{12}$ to a codeword $\mathbf{x}_1(m_{12})$. Since the capacity of the channel between the decoders is $R_{12}$, there is a sufficient number of different codewords $\mathbf{x}_1$ (i.e., sufficient to cover the space of cooperation messages $\mathcal{M}_{12}=\left[1:2^{nR_{12}}\right]$) that can be conveyed via this channel. To use $\mathcal{C}^{(\mathrm{BC})}_n$ for the $R_{12}$-reduced SD-RBC, transmit $B$ blocks, each of length $n$, and denote the messages transmitted by $(m_1^{(b)},m_2^{(b)})\in\mathcal{M}_1\times\mathcal{M}_2$, where $b\in[1:B]$. In the subsequent coding scheme, the transmission of the 1st block is disregarded, while during every block $b\geq 2$, the messages $(m_1^{(b-1)},m_2^{(b-1)})$ are reliably transmitted over the channel. Accordingly, the scheme forfeits the decoding of the messages $(m_1^{(B)},m_2^{(B)})$, which implies that the average rate pair $\left(\frac{B-1}{B}R_1,\frac{B-1}{B}R_2\right)$, over $B$ blocks, is achievable. By taking $B\to\infty$, the transmission rates approach $(R_1,R_2)$.

\par The coding scheme for the $R_{12}$-reduced SD-RBC during block $b\geq 2$ is as follows. First, note that the channel output $\mathbf{y}_1^{(b-1)}$ at Decoder 1 during the previous block is known at the relay at the beginning of block $b$. Thus, during block $b$, the encoder transmits the codeword $\mathbf{x}$ that corresponds to the message pair $(m_1^{(b-1)},m_2^{(b-1)})$, while the relay transmits the codeword $\mathbf{x}_1\Big(m_{12}^{(b-1)}\big(\mathbf{y}_1^{(b-1)}\big)\Big)$. At the end of transmission block $b$, Decoder 2 uses the induced relay output $\mathbf{y}_{22}^{(b)}$ to reliably decode $m_{12}^{(b-1)}\big(\mathbf{y}_1^{(b-1)}\big)$. Both decoders then proceed with the decoding process for the SD-BC with cooperation to decode the messages $(m_1^{(b-1)},m_2^{(b-1)})$. By taking $n$ to infinity, this coding scheme achieves $\left(\frac{B-1}{B}R_1,\frac{B-1}{B}R_2\right)$, over $B$ blocks, for the $R_{12}$-reduced SD-RBC.

%%%%%%%%%%%%%%%%%%%%%%%%%%%%%%%%%%%%%%%%%%%%%%%%%%%%%%%%%%%%%%%%%%%%%%%%%%%%%%%%%%%%%%%%%%%%%%%%%%%%%%%%%%%%%%%%%%%
%%%%%%%%%%%%%%%%%%%%%%%%%%         APPENDIX G - SW Region Correspondance          %%%%%%%%%%%%%%%%%%%%%%%%%%%%%%%%%
%%%%%%%%%%%%%%%%%%%%%%%%%%%%%%%%%%%%%%%%%%%%%%%%%%%%%%%%%%%%%%%%%%%%%%%%%%%%%%%%%%%%%%%%%%%%%%%%%%%%%%%%%%%%%%%%%%%

\section{Derivation of the Region in (\ref{region_SW})}\label{appen_SW_region_corr}

We prove that the admissible region for the SW problem with one-sided encoder cooperation is \eqref{region_SW}, and that \eqref{region_SW} is obtained from $\mathcal{R}_{\mathrm{WAK}}(P_{X_1,X_2,Y})$ stated in Theorem \ref{tm_bc_capacity}. To this end set $U=X_2$ and evaluate the rate bounds in \eqref{region_AK} to get
\begin{align}
R_{12}&\geq I(V;X_1|X_2)\nonumber\\
R_1&\geq H(X_1|X_2)-I(V;X_1|X_2)\nonumber\\
R_2&\geq H(X_2|X_1)\nonumber\\
R_1+R_2&\geq H(X_1,X_2).\label{region_ak_reduced}
\end{align}

%%%%%%%%%%%%%%%%%%%%%%%%%%%%%%%%%%%%%%%%%%%%%%%%%%%%%%%%%%%%%%%%%%%%%%%%%%%%%%%%%%%%%%%%%%%%%%%%%%%%%%%%%%%%%%%%%%
%%%%%%%%%%%%%%%%%%%%%%%%%%%%%%%%%%%%%%%%%%         Markov Graph      %%%%%%%%%%%%%%%%%%%%%%%%%%%%%%%%%%%%%%%%%%%%%

\begin{figure*}[!t]
\begin{center}
\begin{psfrags}
    \psfragscanon
    \psfrag{A}[][][0.8]{\ $M_1$}
    \psfrag{B}[][][0.8]{\ \ \ \ \ $M_2$}
    \psfrag{C}[][][0.8]{$\mspace{-5mu}X^{q-1}$}
    \psfrag{D}[][][0.8]{$X_q$}
    \psfrag{E}[][][0.8]{\ \ \ \ \ \ $X_{q+1}^n$}
    \psfrag{F}[][][0.8]{\ \ \ $Y_1^{q-1}$}
    \psfrag{G}[][][0.8]{\ $Y_2^{q-1}$}
    \psfrag{H}[][][0.8]{\ \ \ $Y_{1,q}$}
    \psfrag{I}[][][0.8]{\ $Y_{2,q}$}
    \psfrag{J}[][][0.8]{\ \ \ \ $Y_{1,q+1}^n$}
    \psfrag{K}[][][0.8]{\ $Y_{2,q+1}^n$}
    \psfrag{X}[][][0.8]{\ }
    \subfloat[]{\includegraphics[scale=0.44]{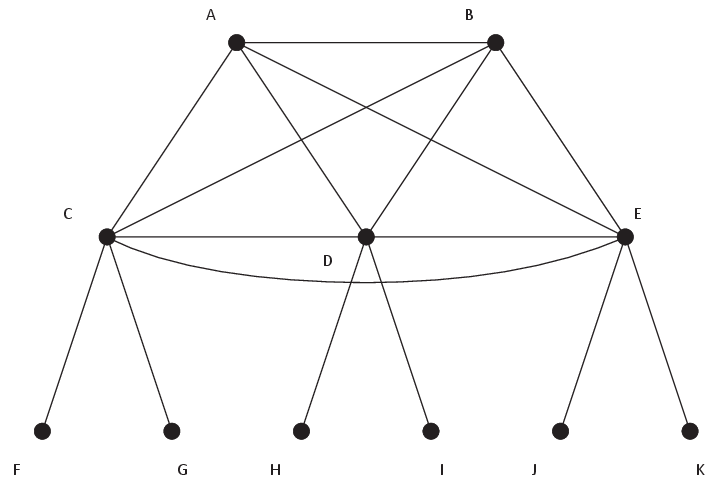}}\ \ \ \ \ \  
    \subfloat[]{\includegraphics[scale=0.44]{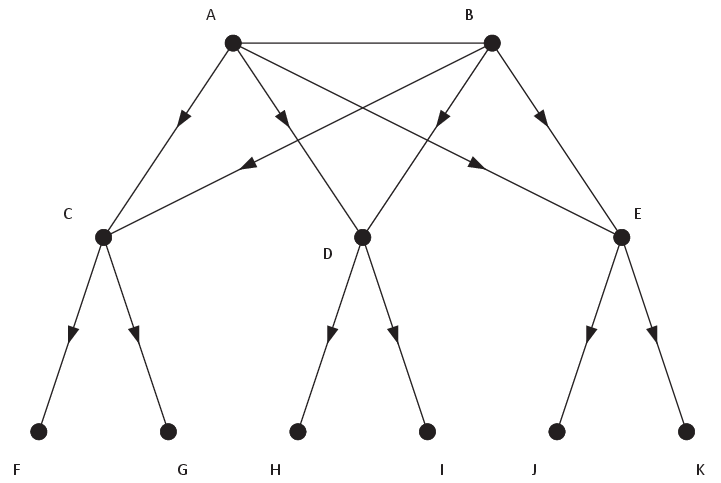}}\ \ \ \ \ \ 
    \subfloat[]{\includegraphics[scale=0.44]{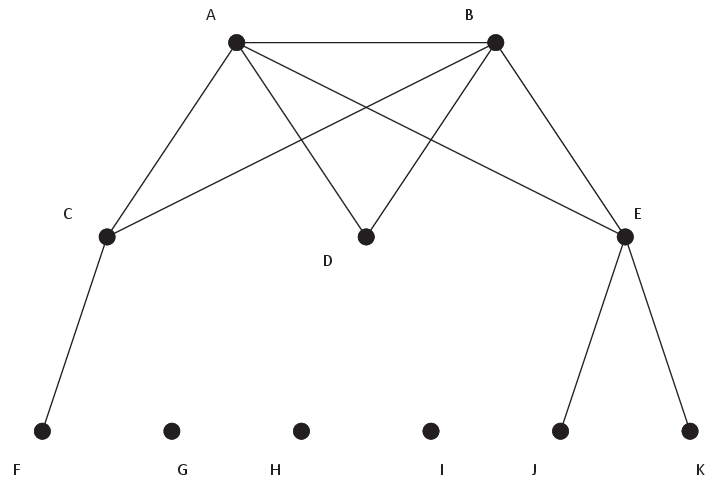}}
    \caption{(a) The undirected graph that corresponds to the PMF from \eqref{Markov_sdbc2_sufficient_temp}: the relation \eqref{Markov_sdbc2_sufficient2} holds because all paths from $Y_{2,q}$ to $(M_1,M_2,Y_1^n,Y_{2,q+1}^n)$ pass through $X_q$. (b) The FDG that stems from \eqref{Markov_sdbc2_sufficient_temp}: \eqref{Markov_sdbc2_sufficient2} follows since $\mathcal{C}=\big\{X_q\big\}$ d-separates $\mathcal{A}=\big\{Y_{2,q}\big\}$ from $\mathcal{B}=\big\{M_1,M_2,Y_1^n,Y_{2,q+1}^n\big\}$. (c) The undirected graph obtained from the FDG after the manipulations described in Definition \ref{DEF:D-sep}.} \label{fig_FDG}
\psfragscanoff
\end{psfrags}
\end{center}%\vspace{-5mm}
\end{figure*}
%%%%%%%%%%%%%%%%%%%%%%%%%%%%%%%%%%%%%%%%%%%%%%%%%%%%%%%%%%%%%%%%%%%%%%%%%%%%%%%%%%%%%%%%%%%%%%%%%%%%%%%%%%%%%%%%%%%

\par The structure of the region in \eqref{region_SW} implies $R_{12}\leq H(X_1|X_2)$. Thus, it suffices to show that for every $0\leq R_{12}\leq H(X_1|X_2)$ there is a random variable $V$ that admits the Markov property $V-X_1-X_2$ such that $I(V;X_1|X_2)=R_{12}$. Since $R_{12}\leq H(X_1|X_2)$, there is a real number $\gamma\geq 0$ such that
\begin{equation}
R_{12}= H(X_1|X_2)-\gamma.
\end{equation}
Set the auxiliary random variable $V\triangleq(\Theta,\widetilde{V})$, where $\Theta\sim\mbox{Ber}\left(\lambda\right)$, $\lambda\in[0,1]$, is a binary random variable independent of $(X_1,X_2)$ that takes values in $\mathcal{O}=\{\theta_1,\theta_2\}$, and
\begin{eqnarray}
\widetilde{V}=\begin{cases}
X_1\ ,& \Theta=\theta_1\\
0\ ,& \Theta=\theta_2
\end{cases}.\label{choice_v_sw}
\end{eqnarray}
Taking
\begin{equation}
\lambda=\frac{H(X_1|X_2)-\gamma}{H(X_1|X_2)}\label{lambda_sw}
\end{equation}
results in
\begin{align*}
I(V;X_1|X_2)&=\lambda I(X_1;X_1|X_2)+\bar{\lambda}I(0;X_1|X_2)\\
            &=H(X_1|X_2)-\gamma\\
            &=R_{12}
\end{align*}
and implies the achievability of \eqref{region_SW}.
%
%\par The optimality of (\ref{region_SW}) follows by the Cut-Set bound.
%
\par The converse follows by the generalized Cut-Set bound \cite[Theorem 1]{Cohen_Netwrok_SI2009} and characterizes \eqref{region_SW} as the admissible rate region for the SW problem with one-sided encoder cooperation.
%
%\par To prove the optimality of (\ref{region_SW}) we have the following converse. For the lower bound on $R_1$ we have
%\begin{align*}
%  nR_1&\geq H(T_1)\\
%      &= H(T_1,T_{12})-H(T_{12}|T_1)\\
%      &\stackrel{(a)}\geq H(T_1,T_{12}|X_2^n)-H(T_{12})\\
%      &\stackrel{(b)}\geq I(T_1,T_{12};X_1^n|X_2^n)-nR_{12}\\
%      &\stackrel{(c)}= H(X_1^n|X_2^n)-H(X_1^n|T_1,T_{12},X_2^n,T_2)-nR_{12}\\
%      &\stackrel{(d)}\geq H(X_1^n|X_2^n)-nR_{12}-n\epsilon_n\\
%      &\stackrel{(e)}= \sum_{i=1}^nH(X_{1,i}|X_{2,i})-nR_{12}-n\epsilon_n\\
%\end{align*}
%where:\\
%  (a) follows because conditioning reduces entropy;\\
%  (b) follows because $|\mathcal{T}_{12}|=2^{nR_{12}}$;\\
%  (c) follows because $T_2$ is a deterministic function of $(T_{12},X_2^n)$;\\
%  (d) follows by Fano's inequality according to which $H(X_1^n|T_1,T_2)\leq\epsilon_n$, where $\lim_{n\to\infty}\epsilon_n=0$.\\
%  (e) follows because the sequences $\big\{(X_{1,i},X_{2,i})\big\}_{i=1}^n$ are drawn in an i.i.d. manner.
%\ \\
%\par The lower bounds on $R_2$ and $R_1+R_2$ are established as for the case without cooperation.

%%%%%%%%%%%%%%%%%%%%%%%%%%%%%%%%%%%%%%%%%%%%%%%%%%%%%%%%%%%%%%%%%%%%%%%%%%%%%%%%%%%%%%%%%%%%%%%%%%%%%%%%%%%%%%%%%%%
%%%%%%%%%%%%%%%%%%%%%%%%%                   APPENDIX H - MARKOV PROOF               %%%%%%%%%%%%%%%%%%%%%%%%%%%%%%%%%
%%%%%%%%%%%%%%%%%%%%%%%%%%%%%%%%%%%%%%%%%%%%%%%%%%%%%%%%%%%%%%%%%%%%%%%%%%%%%%%%%%%%%%%%%%%%%%%%%%%%%%%%%%%%%%%%%%%

\section{Proof of the Markov Relation in (\ref{Markov_sdbc})}\label{APPEN:Markov_Proof}

We present two proofs for the Markov relation in \eqref{Markov_sdbc}, each based on a different graphical method. The first uses the sufficient condition via undirected graphs that was introduced in \cite{Permuter_steinberg_Helper2010}. The second approach relies on the notion of d-separation in functional dependence graphs (FDGs), for which we use the formulation from \cite{Kramer_FDG2003}.

\par By the definitions of the auxiliaries $V$ and $U$, it suffices to show that
\begin{equation}
(M_1,M_2,M_{12},Y_1^{t-1},Y_{2,t+1}^n,Y_{1,t})-X_t-Y_{2,t}\label{Markov_sdbc2_sufficient}
\end{equation}
is a Markov chain for every $t\in[1:n]$. In fact, we prove the stronger Markov relation
\begin{equation}
(M_1,M_2,Y_1^n,Y_{2,t+1}^n)-X_t-Y_{2,t}\label{Markov_sdbc2_sufficient2}
\end{equation}
from which \eqref{Markov_sdbc2_sufficient} follows because $M_{12}$ is determined by $Y_1^n$. Since the channel is SD, memoryless and without feedback, for every $(m_1,m_2)\in\mathcal{M}_1\times\mathcal{M}_2$, $(x^n,y^n_1,y^n_2)\in\mathcal{X}^n\times\mathcal{Y}_1^n\times\mathcal{Y}_2^n$ and $t\in[1:n]$, the structure of the joint PMF from \eqref{EQ:induced_bc_PMF} gives
\begin{align*}
&P(m_1,m_2,x^n,y_1^n,y_2^n)\\
&=P(m_1)P(m_2)P(x^n|m_1,m_2)P(y_1^{t-1}|x^{t-1})P(y_2^{t-1}|x^{t-1})\\
&\mspace{40mu}\times P(y_{1,t}|x_t)P(y_{2,t}|x_t)P(y_{1,t+1}^n|x_{t+1}^n)P(y_{2,t+1}^n|x_{t+1}^n).\numberthis\label{Markov_sdbc2_sufficient_temp}
\end{align*}

\noindent Given \eqref{Markov_sdbc2_sufficient_temp}, the Markov relation in \eqref{Markov_sdbc2_sufficient2} follows by using either of two subsequently explained methods.

\subsection{Via Undirected Graph}

Fig. \ref{fig_FDG}(a) shows the undirected graph that stems from the PMF in \eqref{Markov_sdbc2_sufficient_temp} with respect to the principles described in \cite{Permuter_steinberg_Helper2010}. Namely, the nodes of the graph correspond to the random variables in \eqref{Markov_sdbc2_sufficient_temp}. All the nodes that are associated with random variables that appear together in any of the terms in the factorization of \eqref{Markov_sdbc2_sufficient_temp} are connected by edges. For instance, the term $P(x^n|m_1,m_2)$ induces edges that connect the nodes of $M_1$, $M_2$, $X^{t-1}$, $X_t$ and $X_{t+1}^n$ with one another. The Markov chain in \eqref{Markov_sdbc2_sufficient2} follows from Fig. \ref{fig_FDG}(a), since all paths from $Y_{2,t}$ to $(M_1,M_2,Y_1^n,Y_{2,t+1}^n)$ pass through $X_t$.

\subsection{Via Functional Dependence Graph and d-Separation}

%%%%%%%%%%%%%%%%%%%%%%%%%%%%%%%%%%%%%%%%%%%%%%%%%%%%%%%%%%%%%%%%%%%%%%%%%%%%%%%%%%%%%%%%%%%%%%%%%%%%%%%%%%%%%%%%%%
%%%%%%%%%%%%%%%%%%%%%%%%%%%%%%%%%%%%%%%%%%         Markov Graph      %%%%%%%%%%%%%%%%%%%%%%%%%%%%%%%%%%%%%%%%%%%%%

%\begin{figure}[!t]
%\begin{center}
%\begin{psfrags}
%    \psfragscanon
%    \psfrag{A}[][][0.8]{\ $M_1$}
%    \psfrag{B}[][][0.8]{\ \ \ \ \ $M_2$}
%    \psfrag{C}[][][0.8]{\ \ \ \ $X^{q-1}$}
%    \psfrag{D}[][][0.8]{$X_q$}
%    \psfrag{E}[][][0.8]{\ \ \ \ \ \ $X_{q+1}^n$}
%    \psfrag{F}[][][0.8]{\ \ \ $Y_1^{q-1}$}
%    \psfrag{G}[][][0.8]{\ $Y_2^{q-1}$}
%    \psfrag{H}[][][0.8]{\ \ \ $Y_{1,q}$}
%    \psfrag{I}[][][0.8]{\ $Y_{2,q}$}
%    \psfrag{J}[][][0.8]{\ \ \ \ $Y_{1,q+1}^n$}
%    \psfrag{K}[][][0.8]{\ $Y_{2,q+1}^n$}
%    \psfrag{X}[][][0.8]{\ }
%\subfloat[]{\includegraphics[scale=0.6]{Markov_graph_FDG1.eps}}
%\subfloat[]{\includegraphics[scale=0.6]{Markov_graph_FDG2.eps}}
%\caption{(a) The FDG that stems from (\ref{Markov_sdbc2_sufficient_temp}): (\ref{Markov_sdbc2_sufficient2}) follows since $\mathcal{C}=\big\{X_q\big\}$ d-separates $\mathcal{A}=\big\{Y_{2,q}\big\}$ from $\mathcal{B}=\big\{M_1,M_2,Y_1^n,Y_{2,q+1}^n\big\}$. (b) The undirected graph obtained from the FDG after the manipulations described in Definition \ref{DEF:D-sep}.} \label{fig_FDG}
%\psfragscanoff
%\end{psfrags}
%\end{center}\vspace{-5mm}
%\end{figure}
%%%%%%%%%%%%%%%%%%%%%%%%%%%%%%%%%%%%%%%%%%%%%%%%%%%%%%%%%%%%%%%%%%%%%%%%%%%%%%%%%%%%%%%%%%%%%%%%%%%%%%%%%%%%%%%%%%%

Fig. \ref{fig_FDG}(b) shows the FDG induced by \eqref{Markov_sdbc2_sufficient_temp}. The structure of FDGs allows one to establish the conditional statistical independence of sets of random variables using the notion of d-separation.

\begin{definition}[d-separation \cite{Kramer_FDG2003}]\label{DEF:D-sep}
Let $\mathcal{A}$, $\mathcal{B}$ and $\mathcal{C}$ be disjoint subsets of the vertices of an FDG $\mathcal{G}$. $\mathcal{C}$ is said to d-separate $\mathcal{A}$ from $\mathcal{B}$ if there is no path between a vertex in $\mathcal{A}$ and a vertex in $\mathcal{B}$ after the following manipulations of the graph have been performed.
\begin{enumerate}
\item Consider the subgraph $\mathcal{G}_{\mathcal{ABC}}$ of $\mathcal{G}$ consisting of the vertices in $\mathcal{A}$, $\mathcal{B}$ and $\mathcal{C}$, as well as the edges and vertices encountered when moving backward one or more edges starting from any of the vertices in $\mathcal{A}$, $\mathcal{B}$ or $\mathcal{C}$.
\item In $\mathcal{G}_{\mathcal{ABC}}$, delete all edges coming out of the vertices in $\mathcal{C}$. Call the resulting graph $\mathcal{G}_{\mathcal{AB}|\mathcal{C}}$.
\item Remove the arrows on the remaining edges of $\mathcal{G}_{\mathcal{AB}|\mathcal{C}}$ to obtain an undirected graph.
\end{enumerate}
\end{definition}

The Markov relation from \eqref{Markov_sdbc2_sufficient_temp} follows by setting $\mathcal{A}=\big\{Y_{2,q}\big\}$, $\mathcal{B}=\big\{M_1,M_2,Y_1^n,Y_{2,q+1}^n\big\}$ and $\mathcal{C}=\big\{X_q\big\}$, and noting that $\mathcal{C}$ d-separates $\mathcal{A}$ from $\mathcal{B}$ \cite{Kramer_FDG2003}. To see this, in Fig. \ref{fig_FDG}(c) we show the undirected graph obtained from the FDG in Fig. \ref{fig_FDG}(b) by applying the manipulations described in Definition \ref{DEF:D-sep} with respect to the specified choices of $\mathcal{A}$, $\mathcal{B}$ and $\mathcal{C}$.

\par Neither of the methods is a special case of the other. While the first method (via undirected graphs) involves graphs with more edges, the derivation of the Markov relations using such graphs is more direct. The second method (via FDGs and d-separation) requires manipulating the original FDG. However, the FDG is typically simpler than its undirected counterpart.

\section*{Acknowledgment}
The authors would like to thank the associate editor and the anonymous reviewers for their constructive comments that helped simplify some of the proofs and improve the presentation of the paper.

\bibliographystyle{unsrt}
\bibliographystyle{IEEEtran}
\bibliography{ref}

\begin{IEEEbiographynophoto}{Ziv Goldfeld}
(S'13) received his B.Sc.\@ (summa cum laude) and M.Sc.\@ (summa cum laude) degrees in Electrical and Computer Engineering from the Ben-Gurion University, Israel, in 2012 and 2014, respectively. He is currently a student in the direct Ph.D. program for honor students in Electrical and Computer Engineering at that same institution.

Between 2003 and 2006, he served in the intelligence corps of the Israeli Defense Forces.

Ziv is a recipient of several awards, among them the Dean's List Award, the Basor Fellowship for honor students in the direct Ph.D. program, the Lev-Zion fellowship and the Minerva Short-Term Research Grant (MRG).
\end{IEEEbiographynophoto}

\begin{IEEEbiographynophoto}{Haim H. Permuter}
(M'08-SM'13) received his B.Sc.\@ (summa cum laude) and M.Sc.\@ (summa cum laude) degrees in Electrical and Computer Engineering from the Ben-Gurion University, Israel, in 1997 and 2003, respectively, and the Ph.D. degree in Electrical Engineering from Stanford University, California in 2008.

Between 1997 and 2004, he was an officer at a research and development unit of the Israeli Defense Forces. Since 2009 he is with the department of Electrical and Computer Engineering at Ben-Gurion University where he is currently an associate professor.

Prof. Permuter is a recipient of several awards, among them the Fullbright Fellowship, the Stanford Graduate Fellowship (SGF), Allon Fellowship, and and the U.S.-Israel Binational Science Foundation Bergmann Memorial Award. Haim is currently serving on the editorial board of the IEEE Transactions on Information Theory.
\end{IEEEbiographynophoto}

\begin{IEEEbiographynophoto}{Gerhard Kramer}
(S'91-M'94-SM'08-F'10) received the Dr. sc. techn. (Doktor der technischen Wissenschaften) degree from the Swiss Federal Institute of Technology (ETH), Zurich, in 1998.

From 1998 to 2000, he was with Endora Tech AG, Basel, Switzerland, as a Communications Engineering Consultant. From 2000 to 2008, he was with Bell Labs, Alcatel-Lucent, Murray Hill, NJ, as a Member of Technical Staff. He joined the University of Southern California (USC), Los Angeles, in 2009. Since 2010, he has been a Professor and Head of the Institute for Communications Engineering at the Technical University of Munich (TUM), Munich, Germany.

Dr. Kramer served as the 2013 President of the IEEE Information Theory Society. He has won several awards for his work and teaching, including an Alexander von Humboldt Professorship in 2010 and a Lecturer Award from the Student Association of the TUM Electrical and Computer Engineering Department in 2015. He has been a full member of the Bavarian Academy of Sciences and Humanities since 2015.
\end{IEEEbiographynophoto}

\end{document}